\documentclass{iopart}
\pdfoutput=1
\usepackage{harvard}
\usepackage{xspace}
\usepackage{graphicx}
\let\jnl=\rmfamily
\def\refe@jnl#1{{\jnl#1}}%
\newcommand\aj{\refe@jnl{AJ}}%
\newcommand\actaa{\refe@jnl{Acta Astron.}}%
\newcommand\araa{\refe@jnl{ARA\&A}}%
\newcommand\apj{\refe@jnl{ApJ}}%
\newcommand\apjl{\refe@jnl{ApJ}}%
\newcommand\apjs{\refe@jnl{ApJS}}%
\newcommand\ao{\refe@jnl{Appl.~Opt.}}%
\newcommand\apss{\refe@jnl{Ap\&SS}}%
\newcommand\aap{\refe@jnl{A\&A}}%
\newcommand\aapr{\refe@jnl{A\&A~Rev.}}%
\newcommand\aaps{\refe@jnl{A\&AS}}%
\newcommand\azh{\refe@jnl{AZh}}%
\newcommand\memras{\refe@jnl{MmRAS}}%
\newcommand\mnras{\refe@jnl{MNRAS}}%
\newcommand\na{\refe@jnl{New A}}%
\newcommand\nar{\refe@jnl{New A Rev.}}%
\newcommand\pra{\refe@jnl{Phys.~Rev.~A}}%
\newcommand\prb{\refe@jnl{Phys.~Rev.~B}}%
\newcommand\prc{\refe@jnl{Phys.~Rev.~C}}%
\newcommand\prd{\refe@jnl{Phys.~Rev.~D}}%
\newcommand\pre{\refe@jnl{Phys.~Rev.~E}}%
\newcommand\prl{\refe@jnl{Phys.~Rev.~Lett.}}%
\newcommand\pasa{\refe@jnl{PASA}}%
\newcommand\pasp{\refe@jnl{PASP}}%
\newcommand\pasj{\refe@jnl{PASJ}}%
\newcommand\skytel{\refe@jnl{S\&T}}%
\newcommand\solphys{\refe@jnl{Sol.~Phys.}}%
\newcommand\sovast{\refe@jnl{Soviet~Ast.}}%
\newcommand\ssr{\refe@jnl{Space~Sci.~Rev.}}%
\newcommand\nat{\refe@jnl{Nature}}%
\newcommand\iaucirc{\refe@jnl{IAU~Circ.}}%
\newcommand\aplett{\refe@jnl{Astrophys.~Lett.}}%
\newcommand\apspr{\refe@jnl{Astrophys.~Space~Phys.~Res.}}%
\newcommand\nphysa{\refe@jnl{Nucl.~Phys.~A}}%
\newcommand\physrep{\refe@jnl{Phys.~Rep.}}%
\newcommand\procspie{\refe@jnl{Proc.~SPIE}}%
          
\newcommand{\Al}{$^{26}$Al\xspace}
\newcommand{\about}{$\simeq$}
\newcommand{\degree}{$^{\circ}$}
\newcommand{\Fe}{$^{60}$Fe\xspace}

\newcommand{\Ti}{$^{44}$Ti\xspace}

\newcommand{\Msol}{M\ensuremath{_\odot}\xspace}

\begin{document}

\review[Nuclear Astrophysics]{Nuclear-Astrophysics Lessons from INTEGRAL}

\author{Roland Diehl}

\address{Max-Planck-Institut f\"ur extraterrestrische Physik, D-85741 Garching, Germany}
\ead{rod@mpe.mpg.de}

\begin{abstract}
Measurements of high-energy photons from cosmic sources of nuclear radiation through ESAÕs INTEGRAL mission have advanced our knowledge: New data with high spectral resolution showed that characteristic gamma-ray lines from radioactive decays occur throughout the Galaxy, in its interstellar medium and from sources. 
Although the number of detected sources and often the significance of the astrophysical results remain modest,  conclusions derived from this unique astronomical window of radiation originating from nuclear processes are important, complementing the widely-employed atomic-line based spectroscopy. 

We review the results and insights obtained in the past decade from gamma-ray line measurements of cosmic sources, in the context of their astrophysical questions.
\end{abstract}

\maketitle

\section{Introduction}\label{intro}

\begin{figure}[th] 
\centering
\includegraphics[width=0.95\textwidth]{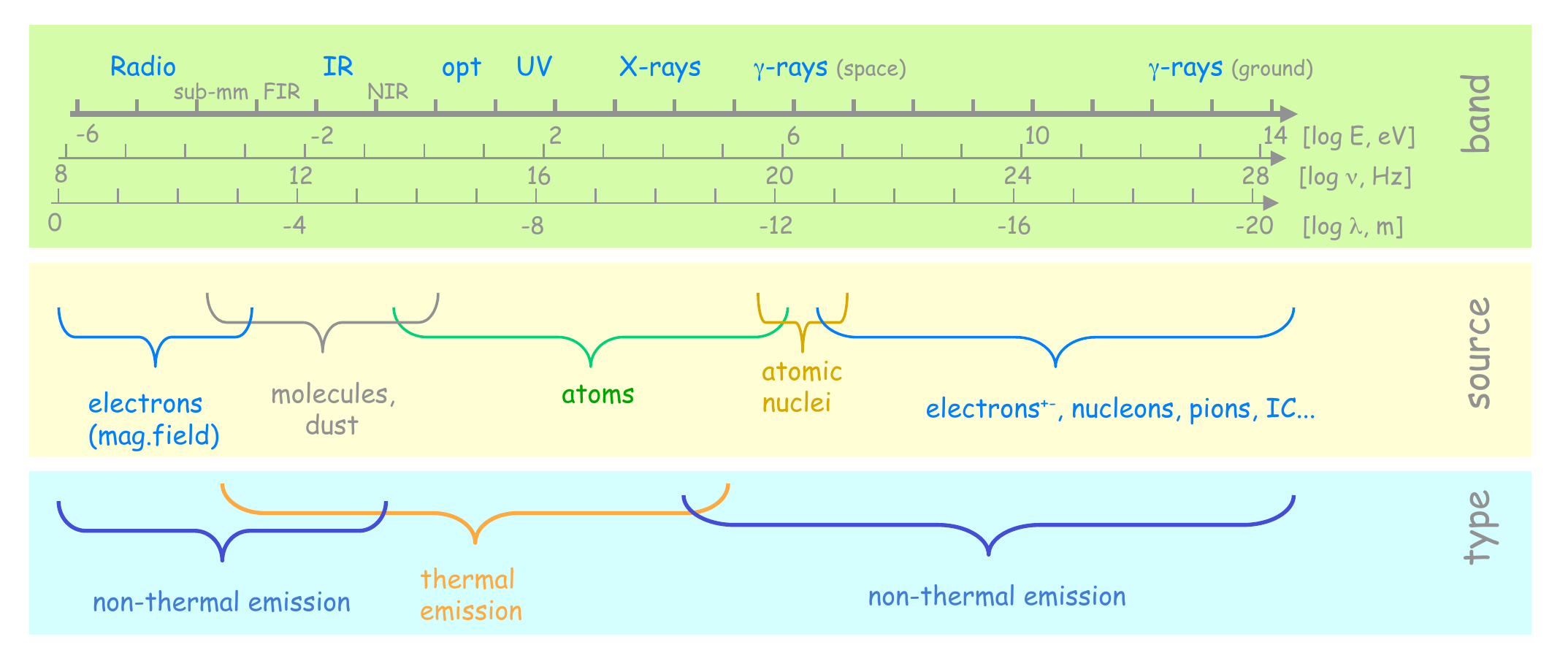}
\caption{The different  wavelength ranges within the electromagnetic spectrum contain specific information: Different sources of emission, and different emission processes are relevant.  
The nuclear radiation window is a small part only, located above the thermal-emission regime, and towards the high-energy and, at the low end of the (large) domain of gamma-rays. INTEGRAL instruments measure photon energies from 15~keV to \about~8000~keV.  \cite{2011LNP...812....3D}.}
\label{emspectrum.fig}
\end{figure}

Among the various windows within the electromagnetic spectrum (Fig.~\ref{emspectrum.fig}) accessible to different types of astronomical telescopes, the radiation originating from atomic nuclei is one of the least explored. The typical energies of such radiation range from  $\sim$100~keV to 10~MeV. Astronomers call such radiation from the nuclear-radiation window \emph{hard} (i.e. high-energy end) \emph{X-rays} to \emph{soft} (i.e. low-energy end) \emph{$\gamma$-rays}.

Characteristic lines in the \about~MeV window originate from transitions among the quantum states of nucleons within atomic nuclei, while spectroscopic signatures at lower photon energies are due to  transitions in the electron shells of atoms or molecules, or from molecular configuration changes (e.g. rotational state transitions). Atomic transitions incur changes in multipole moments of the electron configuration, which leads to emission or absorption of a photon with characteristic energy. Nuclear state transitions are predominantly mediated by the strong and weak interactions, which drive changes in the \emph{nuclear} multipole moment, analogously leading to emission or absorption of characteristic gamma-ray photons. Typical nuclear state energies and energy differences are measured in units of MeV, typically 3 or more orders of magnitude above energies of atomic transitions; the typical binding energy of a nucleon in a nucleus is \about~7--8~MeV. 

Nuclear-state transitions may be caused, e.g., by atomic collisions with energies exceeding \about~100~keV. More important are radioactive decays, however, which leave behind a daughter isotope in an excited state, thus preparing the state transition from which characteristic gamma-ray lines can be measured. 

\begin{figure}[th] 
\centering
\includegraphics[width=0.99\textwidth]{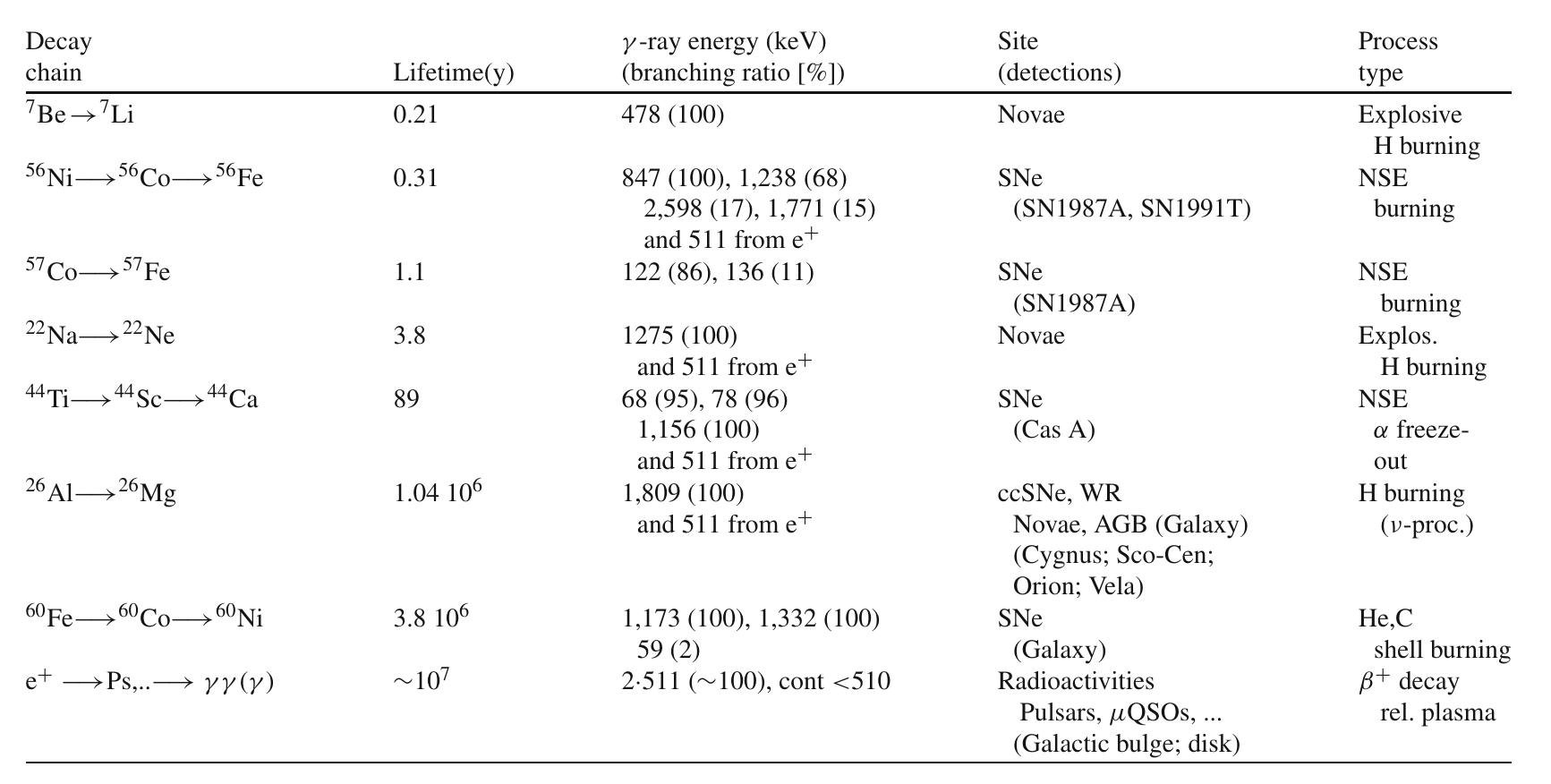}
\caption{The radioactive decay chains relevant for nucleosynthesis measurements, with their characteristic gamma-ray energies. This table is sorted by decay time.  \cite{2011LNP...812..345D}}
\label{lines.fig}
\end{figure}

Emission lines will be brightest for radioactive isotopes which are most abundant and which decay rapidly.  On the other hand, candidate sources such as novae, supernovae, and stellar cores  are dense and occulted by envelopes which are optically-thick to even penetrating gamma-rays ( absorption columns of a few g~cm$^{-2}$), and transparent only after weeks, typically. Therefore, candidate lines and isotopes  must originate from sufficiently-abundant and long-lived isotopes. The list of isotopes   (Fig.~\ref{lines.fig}) extends over decay times from weeks to 10$^6$y, and from a total of 17 gamma-ray lines at energies between 68~keV and 2.6~MeV, 9 lines have been detected so far. 

Astrophysical processes addressed in this nuclear-radiation window are related to \emph{high-energy astrophysics}. Objectives  are: Nucleosynthesis in stars and stellar explosions and associated radioactive decays of unstable isotopes; the processes leading to and characterizing cosmic rays, their production, and their interactions with interstellar gas; radiation from extremely violent objects characterized by extremes in energy density, such as the objects producing gamma-ray bursts.  

With INTEGRAL, massive-star nucleosynthesis has become accessible from stellar groups within our Galaxy, through gamma-rays from decays of long-lived (10$^6$y=1~My) \Al and \Fe. Core-collapse supernova interiors  create radioactive \Ti in addition to abundant $^{56}$Ni. Constraints on gamma-ray line fluxes exploit this new diagnostic for young supernova remnants. 
Surprisingly, none of the expected gamma-ray signatures from a nova, nor any other candidate nucleosynthesis source, such as an AGB star or SNIa, has been found. Nevertheless, the  implications of flux limits as obtained with INTEGRAL are significant for astrophysical models. 
Positrons are a plausible by-product of nucleosynthesis, in H-rich nuclear burning, and from equilibrium burning in hot environments such as supernovae. INTEGRALÕs all-sky mapping of positron annihilation gamma-rays presents a puzzle: emission is found at an unexpected location, in the GalaxyÕs bulge, while  the Galactic disk is low in brightness. Nucleosynthesis is part, but probably not all, of the explanation. 
Nuclear-emission signatures from other origins, such as de-excitation of excited nuclear energy levels, have only been seen in solar flares from the Sun. Other sources should be detectable, based on their expected rate of high-energy particle collisions. 

In the following Section we  expose the astrophysical issues for each of these sources, before we present the INTEGRAL mission (Section 3), and discuss results and findings (Sections 4,5).

\section{Astrophysical Issues and Nuclear Astrophysics}\label{intro_astro}

A thermal energy of 100~keV corresponds to a temperature of 10$^{9}$K (=1~GK). This is beyond the temperatures which are observed in even hottest cosmic plasma (e.g., accretion of matter onto compact objects such as neutron stars may heat plasma to temperatures of several keV at most). Therefore, such plasma almost never is \emph{thermalized}, and particle energy distributions (often following power laws in energy) must be convolved with transition cross sections to estimate emission brightness. MeV, or GK-scale, plasma temperatures are characteristic for the interiors of supernovae and massive stars, where matter is fully ionized, where nuclear reactions may change composition, and creation of particles such as electron-positron pairs and neutrinos becomes an important part of the energy budget. The equations describing composition and dynamics of matter thus become very different from the treatments of cosmic gas at lower energies, as equilibria and conservation laws must be formulated to include those processes. 
Collisional excitations of nuclear state transitions are at the transition regime towards relativistic\footnote{The electron rest mass of 511~keV implies that electrons with MeV kinetic energies are relativistic; MeV protons (proton rest mass 938.3~MeV)  
still would be non-relativistic at this energy.} particles, which implies their connection to the study of the origins of cosmic rays.

\subsection{Nucleosynthesis}
\begin{figure}[th] 
\centering
\includegraphics[width=0.79\textwidth]{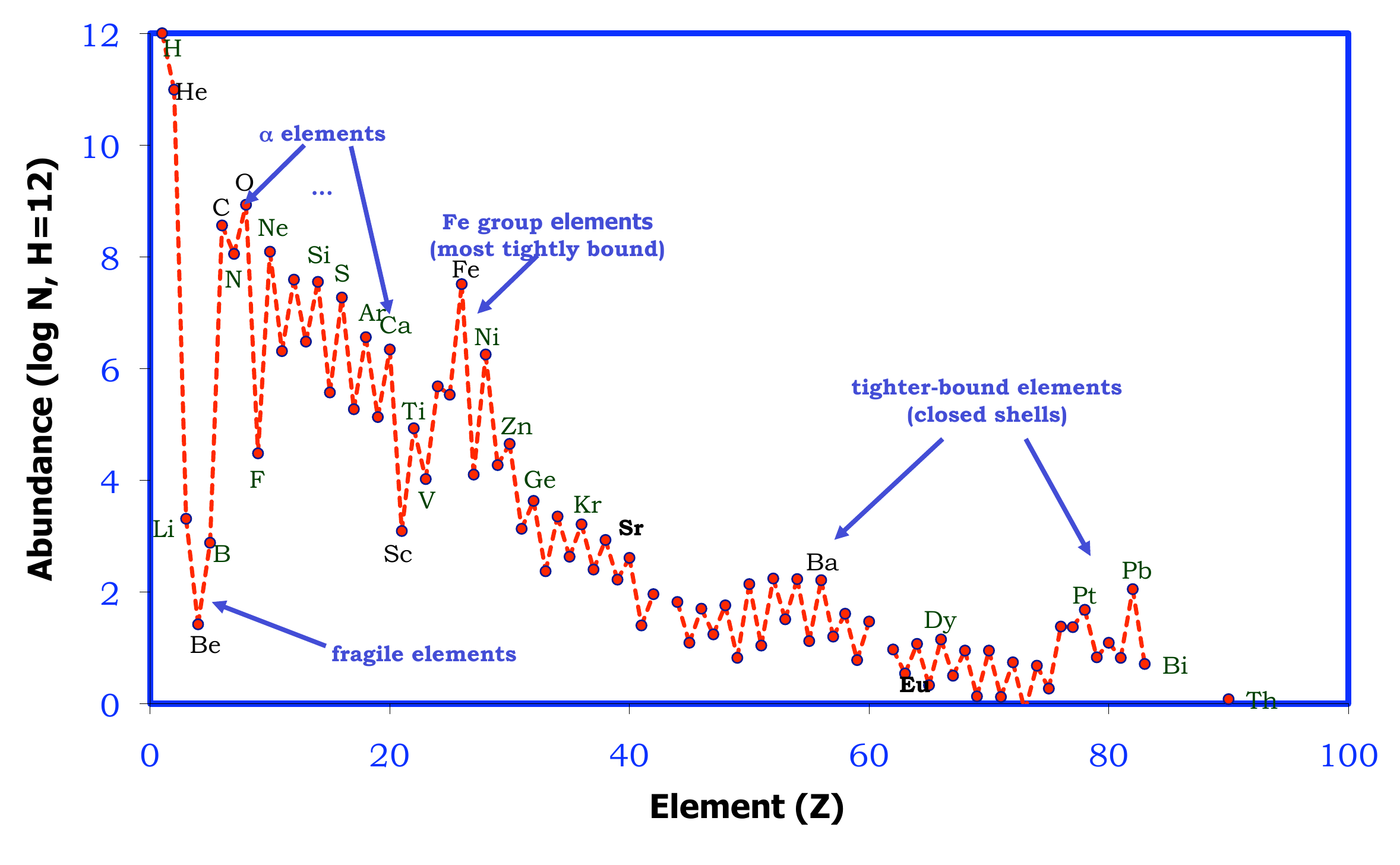}
\caption{The cosmic elemental abundances extend over 12 orders of magnitude. Apparent are three groups of elements, which suggest different nucleosynthesis sites: Light and primordial H and He, intermediate elements up to the iron-peak elements, and element heavier than the Fe group; additionally, odd/even abundance variations add another hint that nuclear structure and binding plays a prominent role. \cite{2011LNP...812..345D}}
\label{fig:abundances}
\end{figure}

The cosmic abundances of elements and isotopes cover a large dynamic range of about 12 orders of magnitude (Fig.~\ref{fig:abundances}). Abundance patterns and regularities provide strong hints on underlying nuclear-structure and cosmic-source processes, which determine abundances of elements and isotopes by groups. Primordial abundances as left behind after big bang nucleosynthesis include only a minor fraction \about~10$^{-7}$ of nuclei heavier than $^4$He. Nuclear fusion processes inside stars and their explosions, supplemented by nuclear interactions of interstellar gas, have contributed to transform initial hydrogen and helium into the current variety of \emph{metals} (i.e., all elements heavier than helium) with their \about~3000 isotopes. 

We understand that stars are stabilized by nuclear energy release against gravitational collapse. We describe stellar evolution as changes in structure and energy flow processes driven by successive nuclear burning stages towards the most-tightly bound nucleus of iron. Release of nuclear energy from those reactions occurs from the excess of nuclear binding energy in daughter nuclei. In an equilibrium situation, nuclear burning thus would lead to predominance of $^{56}$Ni, alongside with other, less tightly bound nuclei which are required to exist according to the nuclear Saha equation in nuclear statistical equilibrium to also satisfy the population of states of kinetic particle energies and phase space conservation. 

Thus, in principle, a local abundance maximum of iron group elements as observed seems plausible. Also, it appears plausible that in non-equilibrium conditions (i.e. either when equilibrium is approached, or when an equilibrium cannot be upheld and \emph{freezes out})  $\alpha$-multiple elements and isotopes with local nuclear-binding energy maxima obtain higher cosmic abundances than their neighbors on the nuclear chart, as observed. Furthermore, elements heavier than iron group elements should plausibly be formed through neutron capture reactions only, as the Coulomb barrier for charged-particle reactions increases with nuclear charge. Neutron capture properties of heavy elements, and in particular \emph{magic number} nuclei show up through local abundance maxima for \emph{rapid} and \emph{slow} \footnote{where the $\beta$-decay rate is the relevant measure, \emph{rapid} neutron capture being able to generate more neutron rich isotopes during burning (\emph{r-process}), while \emph{slow} neutron capture produces nuclei along the valley of stable isotopes only.} burning situations. The basics of cosmic nucleosynthesis appear understood \cite{1957RvMP...29..547B}.

Towards a description in terms of physics of the cosmic sites of nuclear burning, even 50 years after  \cite{1957RvMP...29..547B} we face issues that have not been solved up to now. Supernova light is understood as being powered mainly by $^{56}$Ni radioactive decay, as demonstrated by characteristic light curve and spectral-evolution data, but our understanding of how these explosions are seeded, initiated, and then develop towards observed remnants is deficient, and can not yet be calculated from stellar-evolution initial conditions following equations of physical processes. While the early explosion and light curve can be explained by some catastrophic event (releasing either bulk nuclear or gravitational energy), the ignition and unfolding details are still debated in terms of different progenitor star types for both of these explosion types \cite{2011PrPNP..66..309R,2005NatPh...1..147W}.  Thermonuclear supernovae probably arise from carbon burning ignited in degenerate-matter white dwarf stars; core-collapse of massive stars probably is turned into an explosion by the combined action of neutrino emission from the compact remnant and large-scale envelope oscillations. 
The variety of dynamics and spatial scales, and of relevant processes, continues to challenge computer modeling towards a consistent description throughout, although different phases and their physics have been clarified by comparing such studies to observational data.

Nuclear reactions are a key for the initiation and progress of explosions, and for the production of new elements and isotopes. They proceed in a highly dynamic environment and lasts for a second or less only. We are incapable of putting together nuclear reaction rate knowledge from physics experiments and nuclear structure theory with hydrodynamical descriptions of stellar explosions consistently to produce what is observed. Hence we must explore the observed products to learn about the conditions in the interiors of the cosmic sources of new isotopes. 

\begin{figure}[th] 
\centering
\includegraphics[width=0.48\textwidth]{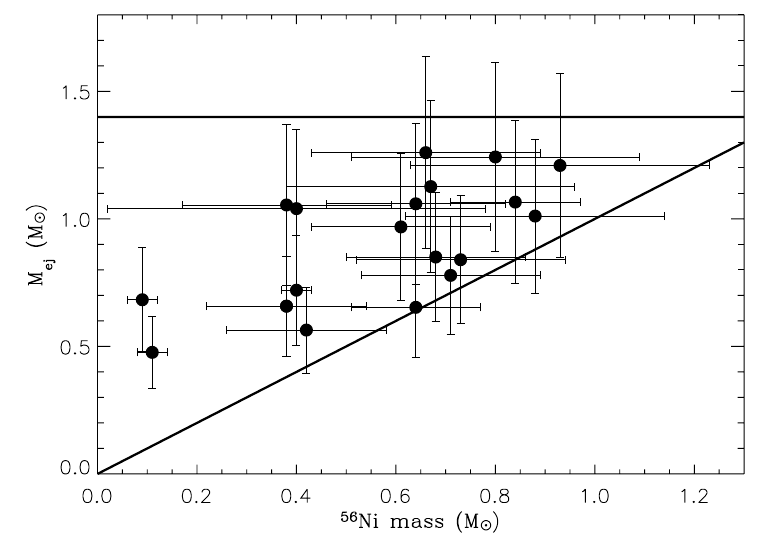}
\includegraphics[width=0.48\textwidth]{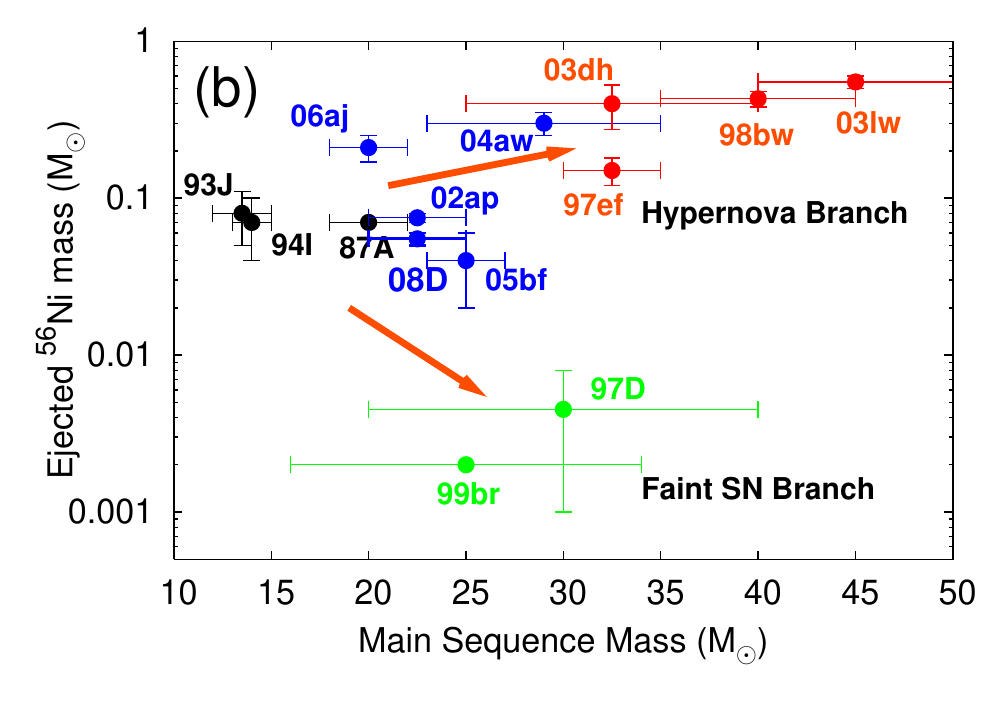}
\caption{The total amounts of $^{56}$Ni are often estimated from \emph{Arnett's rule}, assuming absence of other sources of supernova light. Variations are large, and not yet consistently reproduced by current models. \emph{Left:} SNIa (from \cite{2006A&A...450..241S}). \emph{ Right:} core-collapse supernovae (from \cite{2009IAUS..254..355N}).}
\label{fig_sn-ni-yields}
\end{figure}

One of the key constraints from observations is the broad range of inferred amounts of $^{56}$Ni (even for SNIa used as cosmic  ``standard candles''),  which must be described within the parameter range of models for each type of supernova (see Fig.~\ref{fig_sn-ni-yields}). 

Nucleosynthesis leads to evolving abundances on cosmic time scales, combining the yields of the variety of sources as time proceeds and stars and their explosions occur and change in detail. Descriptions of \emph{chemical evolution} have been put together, which follow the evolution within galaxies as stars form, evolve, and explode or lock up baryons. The resulting abundance ratios of observable elements can thus be described in some cases, but fail to reproduce observed ratios in some others. 
With high-resolution spectroscopy of photospheric abundances in stars, it has also been found that more-massive stars often show inconsistently deviant abundances: Carbon-rich stars present surprises \cite{2005ARA&A..43..531B}, nitrogen abundances are abnormal \cite{2005A&A...430..655S}, and lithium destruction appears as a puzzle \cite{2010IAUS..268..201S}, just to name a few. Even for hydrogen burning in the core of our Sun, a consistent description of structure (as observable in helio-seismology through sound wave recording of the extent of the convective region) and reaction rates (as constrained by neutrino fluxes from some reactions involved in hydrogen burning, including the effects of neutrino flavor oscillations) is not obtained \cite{2006ApJS..165..400B}. It is clear that turbulence and convection are not easy to describe on astrophysical scales, but discrepancies are larger than expected.
Even though here the effects of star formation and interstellar as well as galactic-scale and intergalactic gas flows add complexity, it is puzzling that such description works for some, and not other, element ratios. While the successes are encouraging in that the basic description of matter cycling from gas to stars may be realistic, the discrepancies cannot be related simply to nuclear-rate uncertainties or single object-class model deficiencies. Our knowledge about cosmic nucleosynthesis needs to be carried to the next level of precision, step by step. The ultimate goal would be to have a description of chemical evolution based on models of the various cosmic nucleosynthesis sources based on the physics of nuclear interactions, energy transport, and gas dynamics only.

Measurements which help to understand nucleosynthesis sources are currently dominated by spectroscopy and light curves in the UV/optical/IR. Here, spectral resolution and sensitivities are sufficient to obtain high quality data from stars in our own and nearby galaxies, and from supernova explosions in distant galaxies out to redshifts of \about~0.2 for core-collapse events \cite{2011arXiv1104.2274H}, and \about~0.6 for SNIa \cite{2008AJ....135.1598M}. 

\begin{figure}
\centering
\includegraphics[width=0.6\textwidth]{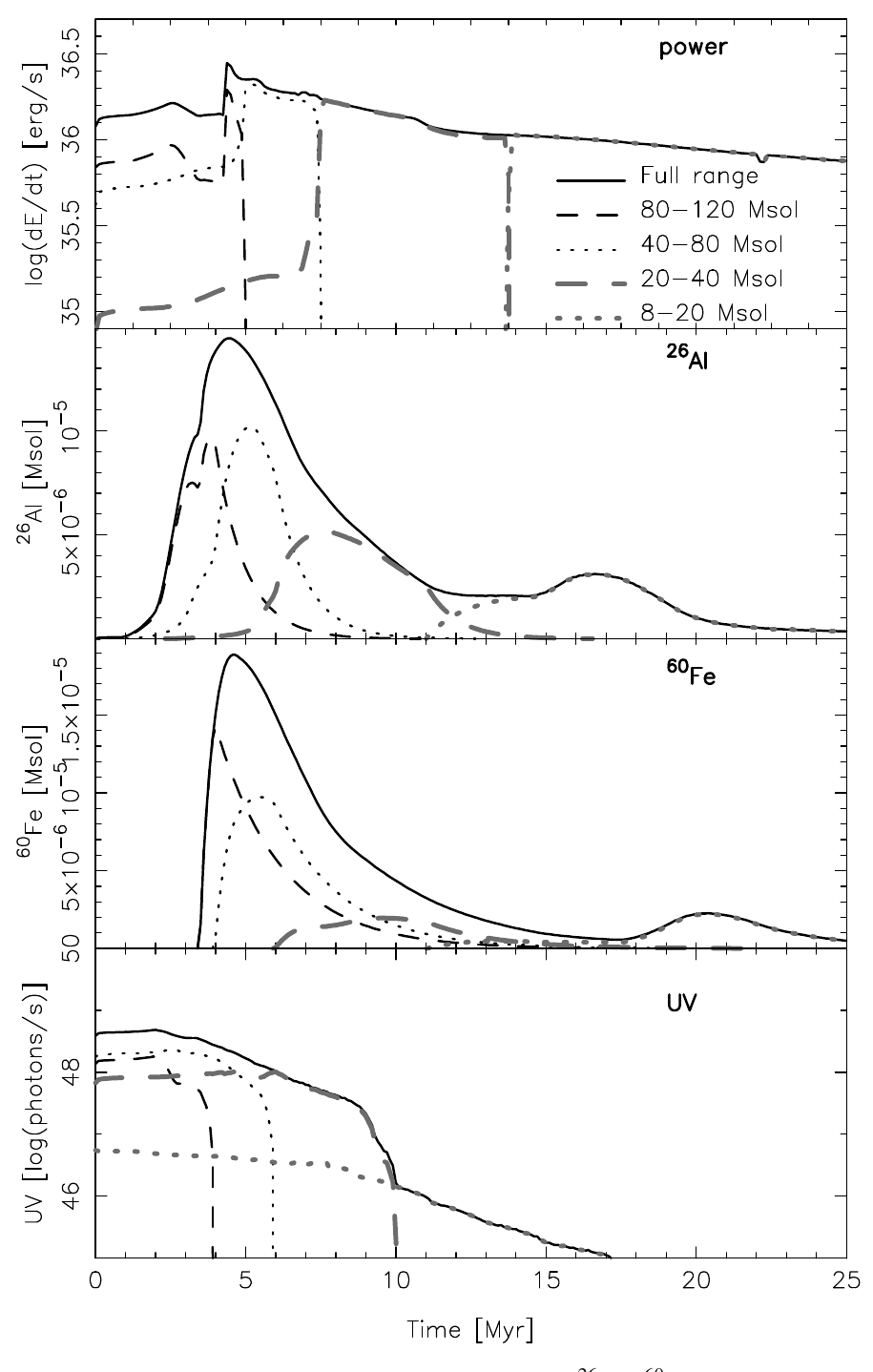}
\caption{The gamma-ray signal from interstellar decay of $^{26}$Al and $^{60}$Fe results from superimposed ejections of these isotopes by stellar winds and supernovae, as driven by mass-dependent stellar evolution. Here we show the time profile of interstellar radioactivities created by a coeval group of stars, as resulting from population synthesis, employing stellar evolution models with a standard mass distribution of stars (from \cite{2009A&A...504..531V}). Also shown (top and bottom panels) are the kinetic energy ejected through winds and supernovae, and the ionizing ultraviolet radiation, respectively.}
\label{fig:popsyn_26Al60Fe}
\end{figure}

Gamma-rays add to this in two important ways: 
Short-lived radioactive nuclei such as $^{56}$Ni and $^{44}$Ti are products of nucleosynthesis in the deep interiors of supernova explosions. Their measurement with gamma-ray telescopes is capable of penetrating deep into the interiors of these explosions, which are otherwise only accessible through neutrinos; even though the gamma-rays also cannot penetrate supernova envelopes, these radioactive isotopes carry the information about the environment of their formation, unaffected by the violent expansion of the supernova explosion.
Somewhat longer-lived radioactive nuclei also produce a gamma-ray signal from cumulative nucleosynthesis sources dominated by massive and short-lived stars. $^{26}$Al gamma-rays had shown that nucleosynthesis is currently ongoing in the Galaxy. INTEGRAL observations also recorded gamma-rays from $^{60}$Fe decay, and allow us to set constraints on nucleosynthesis as it occurs in the complex interiors of massive and short-lived stars, as they evolve towards their supernova. 
The ejection of these radioactive nuclei from massive stars and their supernovae occurs over times of 3--20~My after birth (see Fig.~\ref{fig:popsyn_26Al60Fe}). As such stars mostly are formed in groups or clusters, the gamma-ray signal samples such groups of stars, and hence a population of stars of different mass and evolution time scales. Both the study of such coeval groups and the total Galactic massive-star population have been made possible with gamma-ray telescopes, complementing other studies which exploit nucleosynthesis admixtures of the photosphere surfaces of such stars.

\subsection{Cosmic Rays and Particle Acceleration}
The observed energy spectrum of relativistic particles in interplanetary and interstellar space is characterized by a steep power law, which extends over a very large energy range from \about~MeV to 10$^{21}$~eV, with a typical power law index (slope) of -2.7. The existence of \emph{cosmic rays} throughout those energies is proof that cosmic environments or sources are capable of accelerating charged particles to those extreme and highly relativistic energies. This exceeds capabilities of terrestrial acceleration facilities such as the LHC by at least four orders of magnitude. Cosmic rays mostly consist of atomic nuclei and protons, with a lepton (i.e. electrons and positrons) fraction of \about~1\%. 

Nuclear radiation should be a useful tool to study cosmic-ray origins and propagation, due to the supra-thermal energies involved. In particular, high-energy collisions with ambient interstellar gas produce excited nuclei and/or spallation products which are radioactive, thus leading to characteristic nuclear emission. Spallation reaction cross sections for high-energy collisions with energies exceeding \about~1~MeV/nucleon are of the order tens of mb, and only depend weakly on energy (i.e., they peak at several MeV/n and fall off beyond several tens of MeV/n). Collisional excitation cross sections are up to one order of magnitude higher, i.e. up to hundreds of mb. 

With a typical cosmic-ray energy content of interstellar space of \about~1~eV~cm$^{-3}$ and gas densities of the order 1--10$^4$~cm$^{-3}$, nuclear radiation luminosities of the ISM are expected to be still undetectable by INTEGRAL. 
But special situations could arise where cosmic ray acceleration sites such as supernova remnants or superbubbles are located close to molecular clouds. This may be the case in the inner Galaxy, or in the Orion region. Estimates of nuclear-line emission for the Orion region have obtained intensities up to 10$^{-4}$~ph~cm$^{-2}$s$^{-1}$ for the brightest line (de-excitation of $^{*12}$C at 4.4~MeV) for optimistic models \cite{1994A&A...283L...1B,1998A&A...331..726P} stimulated by COMPTEL data \cite{1994A&A...281L...5B}, as opposed to the \about three orders of magnitude lower  4~10$^{-6}$ ~ph~cm$^{-2}$s$^{-1}$~rad$^{-1}$ for the agreed standard estimate of for such emission \cite{2002ApJS..141..523K,1979ApJS...40..487R}. 

It turned out that the early reports of COMPTEL detection of C and O line emission from the Orion region were probably contaminated by an instrumental artifact \cite{1999ApL&C..38..349B}. Therefore, those optimistic nuclear line flux estimates are generally not held up, and hopes for detecting such emission even from the entire inner Galaxy remained low (see \cite{1997AIPC..410.1074B} for the COMPTEL result). 

Acceleration of charged particles towards relativistic energies occurs during solar flares in the Sun. This process had been at the roots of gamma-ray spectroscopy from cosmic sources \cite{1971SSRv...12..486C}. 
Although this is only one of the candidate processes for acceleration of particles to cosmic-ray energies, measurements of characteristic gamma-ray lines from solar flares have provided a rich database for the study of this magnetic-field reconfiguration-related acceleration process. 
The \about~11-year solar cycle had shown in its 23$^{rd}$ cycle a double-peaked maximum of activity in 2000 and 2002, decaying towards a minimum in 2009, with a predicted cycle-24 maximum in 2013. 
Hence, both RHESSI and INTEGRAL were able to register valuable solar-flare gamma-ray spectra, as discussed below.

\subsection{Other Extreme Cosmic Environments or Sources}
Energies sufficient for nuclear interactions can be obtained also in extreme cosmic sources, where \emph{extreme} would be the local energy density as compared to the highest energies which typically characterize cosmic sources. Comparing scales, hydrostatically stabilized stars obtain surface temperatures of several 10,000~K, while accretion onto compact objects generates temperatures of keV energies (corresponding to 10$^7$~K), and nuclear burning generates 10$^9$K-environments.
MeV-level temperatures of order 10$^{10}$K can only be confined within a source by extremes of overlying matter, such as massive stars or accreting black holes, and are characteristic for explosive events otherwise. 

Surface layers of neutron stars undergo nuclear burning during \emph{Type~I X-ray bursts}; the overlying neutron-star atmosphere shows a photosphere of these events of keV temperatures at most.   
One may expect that Type-I X-ray burst emission could also include a line features from neutron capture (at 2.23~MeV), which could convey important information from its expected gravitational redshift. 

Above a threshold energy of 1.022~MeV, pair creation becomes possible, and opens another degree of freedom for plasma states, with a candidate feature in the positron annihilation line at 511~keV energy (at rest). Pair plasma is expected to exist in accreting compact stars and black holes, especially when a jet is launched, such as in microquasars or active galactic nuclei. Also pulsars and magnetars with particle acceleration up to GeV/TeV energies are expected to generate pair plasma, and annihilation line features would be of great interest. None of these have been reported so far.

Gamma-ray bursts are understood to arise from relativistic fireballs of energy, generated by a relativistically-expanding jet, which may be launched from accretion onto a newly-forming black hole. The emitted gamma-ray spectrum at the surface is probably characterized by thermalized emission from all degrees of freedom for matter and radiation here, including pair plasma and synchrotron emission. It is found as featureless and characterized by power law shapes, with a \emph{break} near \about~300~keV. 

By analogy to quasar absorption line studies of interstellar clouds and galaxies along the line of sight to the background source emitting a bright continuum spectrum, matter between such extreme sources emitting gamma-rays above the nuclear-excitation threshold and the observing telescopes should leave its imprints on the observed spectrum through nuclear-line absorption. This could turn out to be a new astronomical tool to study interstellar gas around such cosmic extremes, providing access to fully-ionized gas that is transparent in all atomic lines \cite{2011ExA...tmp..116G}.

\section{INTEGRAL and other Nuclear Radiation Telescopes}\label{intro_instrum}
\begin{figure} 
\centering
\includegraphics[width=0.7\textwidth]{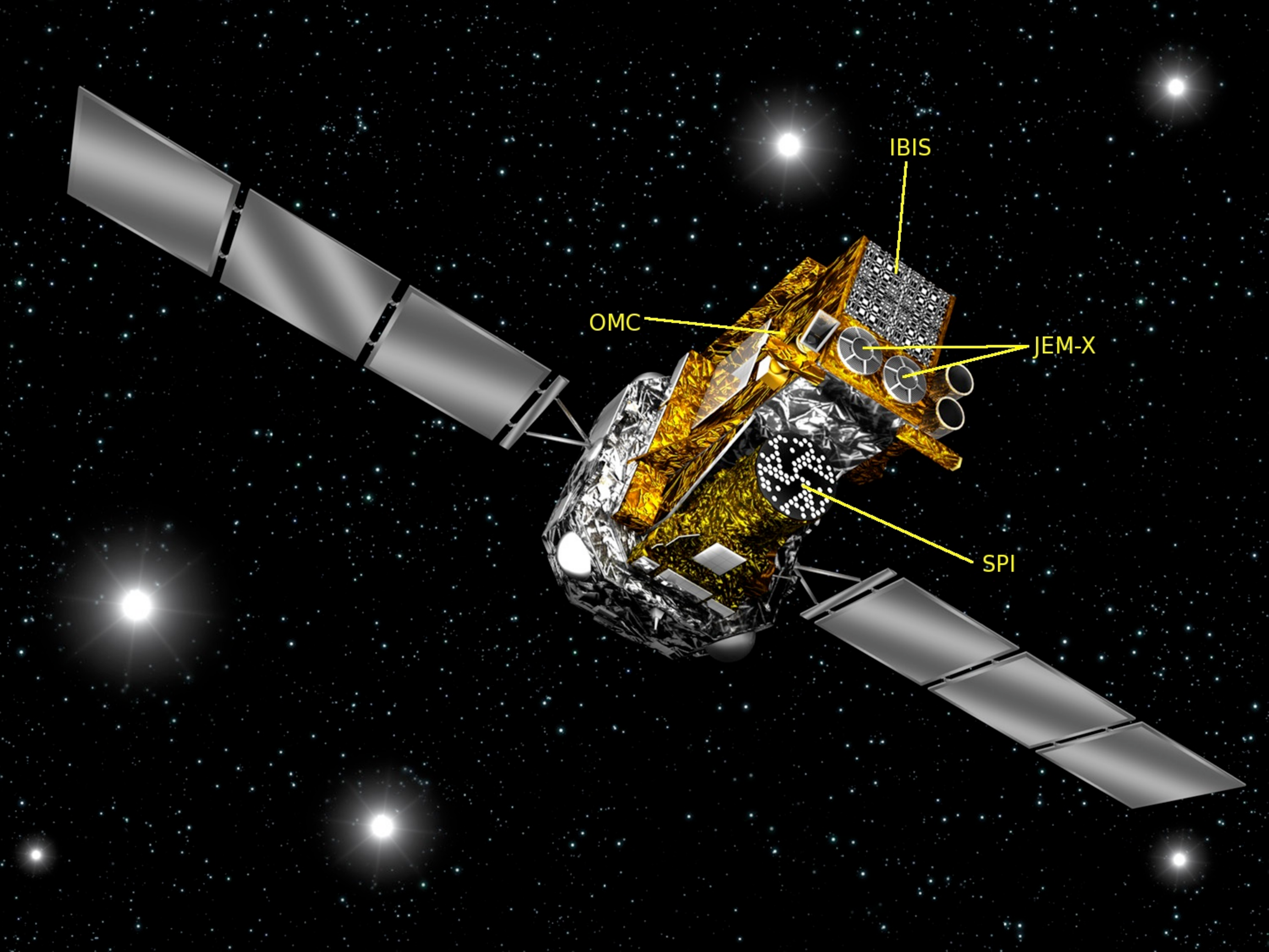}
\includegraphics[width=0.29\textwidth]{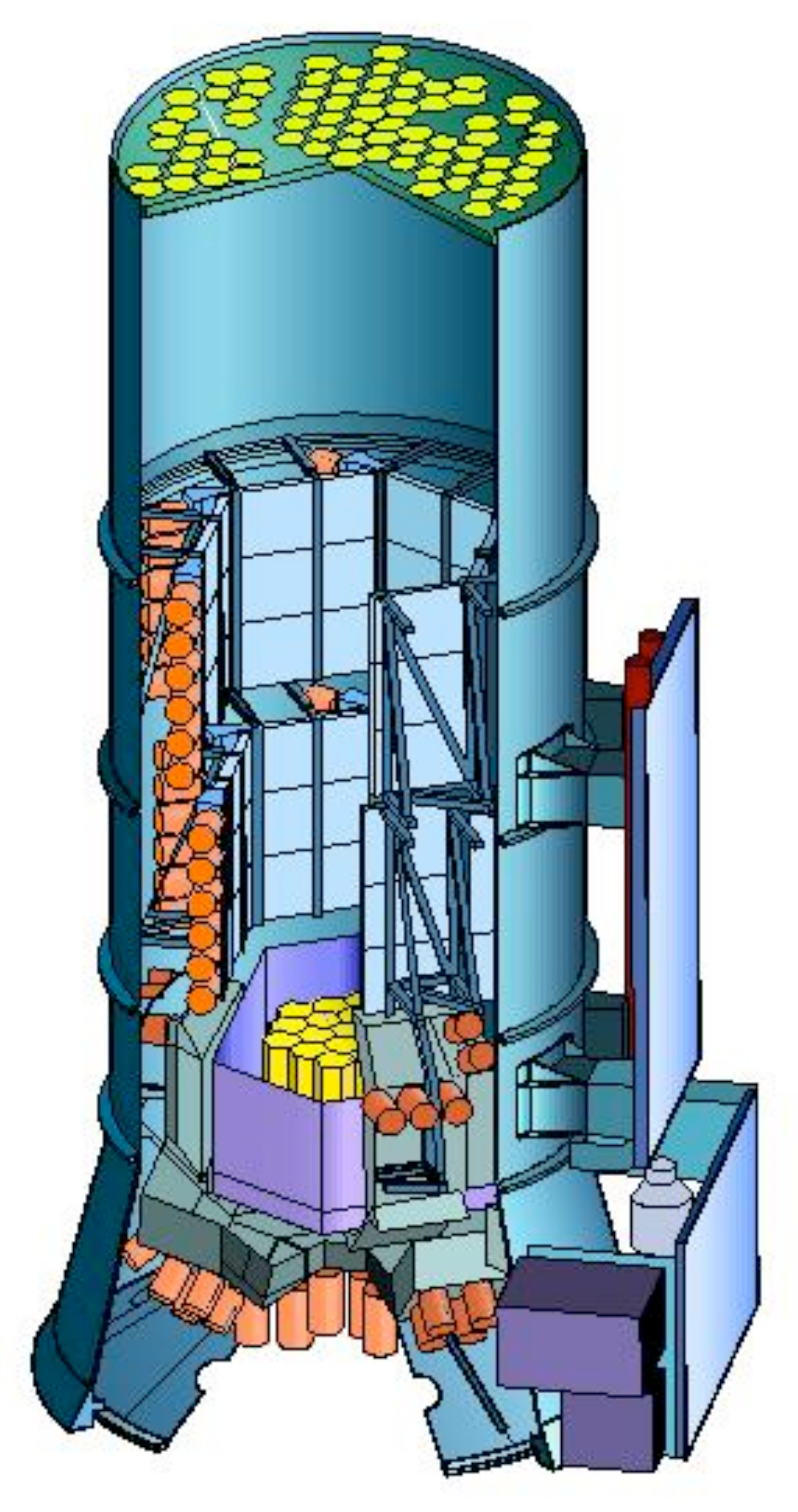}
\caption{{\it Left:} Artist impression of the INTEGRAL spacecraft. Dimensions are (5x2.8x3.2) m. The deployed solar panels are 16~m across. The mass is 4 t (at launch) including 2 t of payload, which consists of two main telescopes, the imager IBIS and the spectrometer SPI, and X-ray and optical monitors JEM-X and OMC. The coded aperture masks for SPI and IBIS are shown for illustration. (Copyright ESA, C. Carreau). {\it Right:} Schematic illustration of the SPI spectrometer. Dimensions (height, diameter) are 2.26x0.78~m. }
\label{integral.fig}
\end{figure}

\subsection{INTEGRAL instruments and mission}
The European Space Agency's (ESA) INTErnational Gamma-Ray Astrophysics Laboratory (INTEGRAL) satellite platform (figure \ref{integral.fig}) hosts two main instruments, the SPI spectrometer \cite{2003A&A...411L..63V} and the IBIS imager \cite{2003A&A...411L.131U}. Both instruments employ coded masks in their apertures for imaging,  multi-element detector planes capture the shadowgram from cosmic X- and $\gamma$-ray sources that is cast by the mask.
The field of view of the main instruments are of similar size of order 10\degree  (IBIS 9x9\degree, SPI 16x16\degree, fully-coded parts, corner-to-corner). Two monitor instruments complement the INTEGRAL instrumentation, the JEM-X coded-mask X-ray telescope addressing lower energies with a wider field of view, and the OMC optical monitor camera for simultaneous optical exposure of the target sky regions. Additionally, the IREM charged-particle monitor detector system records the cosmic-ray irradiation, which is an important source of instrumental backgrounds and disturbances. 

 \emph{Dithering} the satellite pointing direction around a celestial target direction in \about~2\degree steps helps to add more modulation to the sky signal, so it can be separated from the (\about~constant) instrumental background. 
The background should not vary within \about~2\degree attitude changes; any other background variations should either be traced by charged particle detectors (IREM, or the SPI anti coincidence system) if due to charged-particle events such as flares or other space weather, or else be slow compared to the dithering variations. The 3-day orbital period was chosen to provide slow background variations.
Typically, the satellite pointing is changed every \about~1800~s, and observations way vary from tens of ks to Ms, depending on science target. 

For nuclear astrophysics, the SPI instrument  is most relevant, and will be in focus for  this paper. It features Ge detectors with high collection area in the \emph{nuclear radiation window}: SPI records photons in the energy range 15--8000~keV, with a  detection area of 250~cm$^2$ and a nominal sensitivity of \about~3~10$^{-5}$~ph~cm$^{-2}$s$^{-1}$ (3$\sigma$, 1~Ms observing time; this is somewhat energy dependent due to instrumental background, e.g. at 511 keV 2.8~10$^{-5}$~ph~cm$^{-2}$s$^{-1}$, about a factor~2 better in the few-hundred keV range, and falling off towards higher energy with decreasing detection efficiency to below the given value above \about~2~MeV).  SPI's main asset is the fine energy resolution of typically E/$\delta$E=500 or 3~keV at 1300--1800~keV, due to high-purity n-type Ge semiconductor detectors operated at \about~80~K and \about~4~kV, with annealings maintaining this spectral performance \cite{2003A&A...411L..91R}. This allows spectroscopy of nuclear lines through unambiguous identification of line energies. Line shape analysis allows determination of source kinematics from Doppler-modified line properties, enabling astrophysical studies of velocities down to the 100~km~s$^{-1}$ regime. The large field of view  is defined by the hexagonal arrangements of the 19-element Ge detector camera (70~mm deep, 55~mm wide detectors, covering a densely-packed 268~mm wide camera plane) and the 127-element tungsten mask with 6~cm wide elements. Partial coding of the sky signal occurs also outside this field of view and can be used for stronger sources. Moreover, the \emph{ditherings} of pointing direction during an observation in a standard pattern of 5x5 pointings separated by 2.1\degree extends the effective field of an observation by another \about~10\degree. The intrinsic imaging resolution of SPI has been determined from calibrations on ground an on the Crab nebular as 2.7\degree (FWHM of the response to a point source). The imager instrument is much superior in resolution (12'), which is however achieved at lower end of the energy range: 3$\sigma$ sensitivities for a gamma-ray line at 1~MeV for a 1~Ms observation are 2.5~10$^{-5}$ and 3.8~10$^{-4}$ph~cm$^{-2}$s$^{-1}$ for SPI and IBIS, respectively; SPI's imaging resolution is still the best-achieved so far for nuclear (\about~MeV) emission (COMPTEL: 3.8\degree FWHM, \cite{1993ApJS...86..657S}).      

The INTEGRAL satellite  was launched into space in October 2002 from Baikonur (Kasachstan). A highly-elliptical orbit had been chosen (perigee at launch 7000~km, apogee 150000~km, orbital period 3 days) to avoid the radiation belts near Earth with its high flux of trapped charged particles, and to have longer times of stable background conditions for such instruments and their particular sensitivity to cosmic-ray irradiation. INTEGRAL is a \emph{medium-sized mission} for ESA, at a more modest cost level  than the  \emph{cornerstone missions}  XMM-Newton or Herschel, which all were part of ESA's  \emph{Horizon~2000} program. 

The INTEGRAL mission \cite{2003A&A...411L...1W} was planned for a duration of 3 years, with possible extension by another 2 years. After its scientific success and excellent performance, ESA has been deciding every 2 years upon further extensions of the mission, evaluating costs and merits of continued operations, while considering the status of both the mission hardware and the astrophysical questions. Currently, the INTEGRAL mission is approved till the end of 2014, with its next 2-year extension to be assessed in late 2012. 
The total cost of INTEGRAL as launched was approximately 700~million Euro, contributed by ESA for the satellite and project plus mission infrastructure, by Russia for the launching rocket and facilities, and by various ESA member countries for the instruments (such as mainly France, Germany, and Italy for the two main instruments of INTEGRAL). Operations expenses are typically 1\% of these total costs, per year. 

INTEGRAL data analysis is somewhat more complex than common for astronomical telescopes. This arises for two main reasons: (1) The instrumental response of the telescope to celestial sources is not as sharply-defined, imaging resolution to gamma-ray sources is on the degree scale only; moreover, the tails and wings of the instrument response function are significant, and lead to cross-talk of sources over significant distances of order 10\degree.  (2) Instrumental background at MeV energies is enormous and dominates the total signal; time variability adds challenges, and \emph{empty field} references are not easy to obtain due to the instrumental response characteristics.
For the SPI telescope, thus, data are fitted iteratively with parametrized models for the sky emission and instrumental backgrounds. The optimization of amplitude parameters describing a celestial source's emission (image in broader bands, or highly-resolved spectra for adopted spatial distribution models) leads to the scientific result of interest. 
Further details may be found in \cite{2003A&A...411L.117D,2003A&A...411L.127S,2007arXiv0712.1668K,2011ApJ...739...29B}.

\subsection{Similar and complementing astronomical Instruments}
Within INTEGRAL's mission time, other observing facilities in the field of high-energy astrophysics became available: The \emph{Fermi} gamma-ray satellite was launched June 2008, and includes instrumentation optimized for MeV-range energies through its  \emph{Gamma-Ray Burst Monitor}. At modest energy resolution (10\% FWHM), the general nuclear spectroscopy potential of GBM's NaI and BGO detectors is limited, while being excellent to measure transient sources and search for nuclear signatures herein. 
The SWIFT satellite launched in November 2004 currently provides the best  all-sky coverage searching for transients in hard X-rays; the coded-mask Burst Alert Telescope (BAT) with its 2~sr field of view extends up to 150~keV in energy, providing arcmin location accuracy, with ~3--4~keV energy resolution near 100~keV. The NuSTAR mission pushes the technology of a focusing hard X-ray mirrors towards the nuclear radiation window, operating between 6 and 79 keV, thus including radioactive decay of $^{44}$Ti (one of the isotopes of interest in Table~\ref{lines.fig}). The $^{44}$Ti lines at 68 and 78~keV are the lowest-energy lines known from any cosmic radioactive decay. NuSTAR was launched in June 2012, and will map $^{44}$Ti emission within supernova remnants, with 50~arcsec spatial resolution within its 8.4x8.4 arcmin field of view.


Nucleosynthesis is also studied by elemental abundance measurements through photospheric absorption or emission lines, for all species where isotopic composition is dominated by a single isotope, or otherwise nuclear origins are clear. Additionally, radio line features from isotope-discriminating molecular rotation bands, and isotopic signatures in IR and optical from heavy elements have become accessible in recent years. Moreover, presolar grains which are included in meteorites are being studied in the laboratory, with great isotopic abundance detail and precision \cite{2004ARA&A..42...39C}. They contribute a sample probably directly from nucleosynthesis sources , although grain formation and transport adds uncertainties. 

Particle acceleration is also studied through direct cosmic-ray  measurements near Earth (e.g. \cite{2009NuPhS.188..296A}). Abundances of isotopes generated by spallation along their propagation in the galaxy, such as from elements Li, Be, and B, but also radioactive isotopes such as $^{59}$Ni, provide constraints on cosmic-ray propagation from its sources throughout the Galaxy \cite{2005NuPhA.758..201I,2001SSRv...99...27M}. Diffuse and high-energy gamma-ray emission is characterized by broadband emission processes such as inverse-Compton scattering on starlight, but also pion and Bremsstrahlung emission from cosmic-ray interactions with interstellar gas. Modeling the gamma-ray emission intensities and spectrum from transport of cosmic rays through the Galaxy has proven to be a valuable tool to constrain the overall cosmic-ray theories \cite{2007ARNPS..57..285S}. But as spectral features are less clear, both hadronic and leptonic acceleration models are held plausible, and  the question of cosmic-ray origins is still open.

\section{Understanding Cosmic Nucleosynthesis Sources}\label{sources}
\subsection{Nucleosynthesis and gamma-rays}
Radioactive ejecta from cosmic sources of nucleosynthesis generate different types of gamma-ray sources: In the case of short-lived ($\leq $My) radioactive isotopes, they decay within the exploding source during its expansion, or, alternatively for longer-lived radioactive isotopes and annihilating positrons,  they produce an extended, diffuse emission. 
Early observations of the sky in gamma-rays had shown the existence of such lines, and brightest were the positron annihilation line at 511 keV \cite{1975ApJ...201..593H}, 
and the 1809~keV line from $^{26}$Al \cite{1982ApJ...262..742M}. 
Supernova 1987A in the LMC then led to first direct proof of supernova light origin from radioactive $^{56}$Ni decay \cite{1988Natur.331..416M}. 
This had prepared the ground for a first gamma-ray line sky survey with the Compton Gamma-Ray Observatory (CGRO; 1991-2000) with its OSSE and COMPTEL instruments. CGRO measurements first detected supernova radioactivity from $^{57}$Co in SN1987A \cite{1992ApJ...399L.137K}  and from $^{44}$Ti in Cas A \cite{1994A&A...284L...1I}. Diffuse radioactivity in $^{26}$Al was mapped over the Galaxy \cite{1995A&A...298..445D}, and positron emission gamma-rays were seen from the inner, bulge region of our Galaxy as an extended source \cite{1997ApJ...491..725P}. 
Yet, CGRO had a modest spectral resolution of order 10\%, inadequate for any gamma-ray spectroscopy in the sense of identifying new lines or constraining kinematics of source regions through line shape measurements; balloon-borne measurements had indicated the power of such measurements \cite{1990ApJ...351L..41T,1989Natur.339..122T} with solid state detectors operated at cryogenic temperatures. The GRSE instrument had originally been foreseen as one of CGRO's detectors to address nuclear-line spectroscopy; this instrument was removed from the observatory as cost overruns in CGRO development and launch preparations had made such trims necessary. 
INTEGRAL's spectrometer SPI was set out to perform such gamma-ray spectroscopy within the ESA Mission program. With a spectral resolution of 2--3~keV in the MeV regime, SPI is the highest spectral-resolution instrument ever operated for such observations, and will remain so for many years; no MeV-gamma-ray  spectroscopy mission is currently at the horizon of any of the international space agencies' programs. INTEGRAL data will establish the nuclear-astrophysics legacy database for (at least) the current generation of astrophysicists.

INTEGRAL's early-mission program included a core program part \cite{2003A&A...411L...1W}, where the plane of the Galaxy was surveyed, and regions of particular interest for nucleosynthesis studies were observed, such as the Cygnus and Vela regions, and the Cas A supernova remnant. The inner Galaxy itself obtained substantial additional exposure from monitoring programs on its rich population of transient X-ray sources. Altogether, at this time, INTEGRAL's sky exposure covers all regions of candidate sources of  gamma-rays from nucleosynthesis, although sensitivities in particular outside the inner-galaxy region are not always sufficient to constrain nucleosynthesis.

\subsection{Supernovae of Type Ia}\label{sources_snia}
Understanding the origins and explosion physics of supernovae of Type Ia is motivated by two main reasons. SNIa are bulk producers of Fe-group elements and intermediate mass elements, and therefore key drivers of the metal enrichment of the universe through nucleosynthesis. Secondly, SNIa play a prominent role in today's physical cosmology, as they are used as standardizable candles with known  brightness, for tracing cosmological expansion \cite{2011ARNPS..61..251G}. 

It is commonly agreed that explosions of white dwarfs, the compact inner-core remnants of the evolution of stars with masses below \about~9~\Msol, lead to SNIa (see\cite{2011LNP...812..233I,2011PrPNP..66..309R} for recent reviews): This naturally explains the absence of hydrogen features in spectra (which observationally defines supernovae of type I), as the stellar envelope is shed in the giant phase and the core has been processed through nuclear fusion towards carbon and oxygen for stars in the mass range 0.5--9~\Msol. The nuclear energy which can be liberated from conversion of a white dwarf's carbon and oxygen to iron exceeds the gravitational binding energy of such stars (\about~10$^{49}$erg) by at least an order of magnitude. It was long thought that the upper limit ( Chandrasekhar mass, \about~1.4~\Msol) for white dwarf masses is responsible for the homogeneity of the explosions in that only white dwarfs reaching that limit would generate SNIa; but this has been put in doubt as observational diversity cannot be denied any more, and several observational categories of SNIa are distinguished \cite{2011MNRAS.412.1441L}. Currently-considered candidate models, therefore, are the variety of (i) explosions triggered in white dwarf central regions as the Chandrasekhar mass limit is reached, (ii) explosions triggered by an off-center event such as a He-shell flash on white dwarfs of a broader mass range, and (iii) explosions resulting from merging of two white dwarfs as it may terminate the evolution of a binary system. More complex scenarios of binary evolution with recurrent pulsations culminating in a supernova are also discussed \cite{2011PrPNP..66..309R}. In all cases, the runaway nuclear carbon burning will provide the energy which disrupts the white dwarf. But the nuclear C-burning flame might propagate in different density regimes from outer layers (\about~10$^4$g~cm$^{-3}$) to central regions (\about~10$^{9}$g~cm$^{-3}$), and hence very differently, due to the extremely steep temperature sensitivity of carbon burning reactions. It is unclear how close to nuclear statistical equilibrium burning (NSE) conditions will settle, within the explosion time scale of \about~100~ms before the nuclear burning ceases due to the explosive expansion of the fuel. Three-dimensional effects will be important, and determine how instabilities grow and determine the nuclear-burning flame front structure, which has typical dimensions of mm to cm; if laminar burning would prevail, the flame could not proceed fast enough to outrun thermal expansion and would fizzle, therefore turbulent burning is likely in SNIa. As a result, the SNIa nuclear ashes may range from almost-pure $^{56}$Ni as predicted for deflagration models with NSE burning conditions to $^{56}$Ni amounts below 0.1~\Msol and mostly intermediate-mass element ashes resulting from partial burning of double-degenerate merger or detonation models \cite{2007Sci...315..825M}. Obviously, this outcome has profound implications for the cosmic enrichment of galaxies with heavy elements, as SNIa are one of the major sources providing freshly-synthesized new nuclei into interstellar space, and are the main providers of Fe-group nuclei. 

The brightness standard derives from an empirical calibration based on a relation between absolute brightness and light curve shape, as brighter events have been found to rise and fall more slowly than the up to a factor \about~10 fainter ones \cite{1993ApJ...413L.105P,1999AJ....118.1766P} (the \emph{Phillips relation}). For use in cosmological studies it is essential that such brightness normalization is independent of cosmic time (redshift). Yet, recently the existence of substantial variety among SNIa has been established \cite{2005ApJ...623.1011B,2010Natur.466...82M,2011arXiv1110.5809W,2011ApJ...731L..11N}.
Constraints on dark energy predominantly rely on the absence of evolutionary biases in this brightness normalization \cite{2006ApJ...648..884R}. For determination of dark-energy properties such as the equation-of-state parameter $w$, a precision level of percent must be achieved on statistical, but more so on systematic variations. Models should properly account for biases such as could arise from the progenitor's metallicity, i.e. the carbon to oxygen (or even admixtures of neon) ratio in white dwarfs at the end of stellar evolution within a binary system. Concerns arise from an apparent intrinsic reddening variation with luminosity in SNIa events \cite{2011AJ....142..156S} \footnote{This is obtained when reddening is used as a free parameter to minimize dispersion of SNIa brightness in the Hubble diagram. This however yields lower reddening values than would be plausible from galactic reddening, and thus hints on an intrinsic effect.}. This requires an understanding of the composition of the outer SNIa layers which are responsible for the reddening, and which also contain partly unburnt material from the progenitor star. Hence, such effects can only be understood if the dynamics of the explosion and the extent of nuclear burning herein can be understood in its event-to-event variations (or, expressed in another way, in viewing-angle variations). 

The brightness of SNIa explosions in its initial phase reflects mainly the sudden deposition of a large amount of energy near the center of an exploding compact star:  \about~10$^{51}$erg of total energy can be estimated from the velocities measured in ejecta, \about~10$^9$cm~s$^{-1}$. The initially radiation-dominated sphere converts this thermal energy of the runaway nuclear burning in a dense and compact star ($\rho\sim$10$^7$g~cm$^{-3}$, R$\sim$10$^8$cm) into kinetic energy. The stellar matter expands  almost adiabatically and cools, leading to a luminosity maximum at \about~30 days after explosion in optical emission. Thereafter, brightness fades rapidly as the remnant cools, while also the photosphere recedes due to its expansion. The decrease of supernova light after maximum brightness then clearly slows down, over a time scale of months. This requires an additional source of energy beyond the initial explosion, which, according to common belief, is due to MeV photons and positron emitted in the radioactive decay of $^{56}$Ni, and down-scattered in energy in the expanding envelope.  Energy deposition efficiency reduces in the diluting remnant, and thus the decrease in brightness does not directly translate into radioactive decay, but require photon transport modeling in the expanding supernova for $^{56}$Ni gamma-rays as they down-scatter in energy from \about~MeV to eV energies as observed in optical/NIR light curves.  Current estimates of $^{56}$Ni masses make use of such radiation propagation models \cite{2007ApJ...662..487W}. Although such models have reached a remarkable sophistication and consistency, their intrinsic dependency on the complexities of radiation transport and on the density structure of the expanding envelope remains a concern, possibly incurring  systematic distortions. 

These arguments underline the need for a physical model of this supernova type. 
Measurements of the total $^{56}$Ni content are a key, and can be performed through gamma-rays in a way that is rather independent of the density structure of the expanding supernova. Moreover, as the expanding supernova becomes transparent to gamma-rays on a time scale of \about~100~days, the rise of radioactive gamma-ray luminosity, as well as gamma-ray line shapes and line-to-continuum ratio convey information about the spatial distribution of $^{56}$Ni and the envelope structure. Typically, the gamma-ray luminosity maximum is expected at 70--90~days after explosion, much later than the luminosity maximum in thermal (optical to infrared emission) supernova light.

\begin{figure}
\centering
\includegraphics[width=0.68\textwidth]{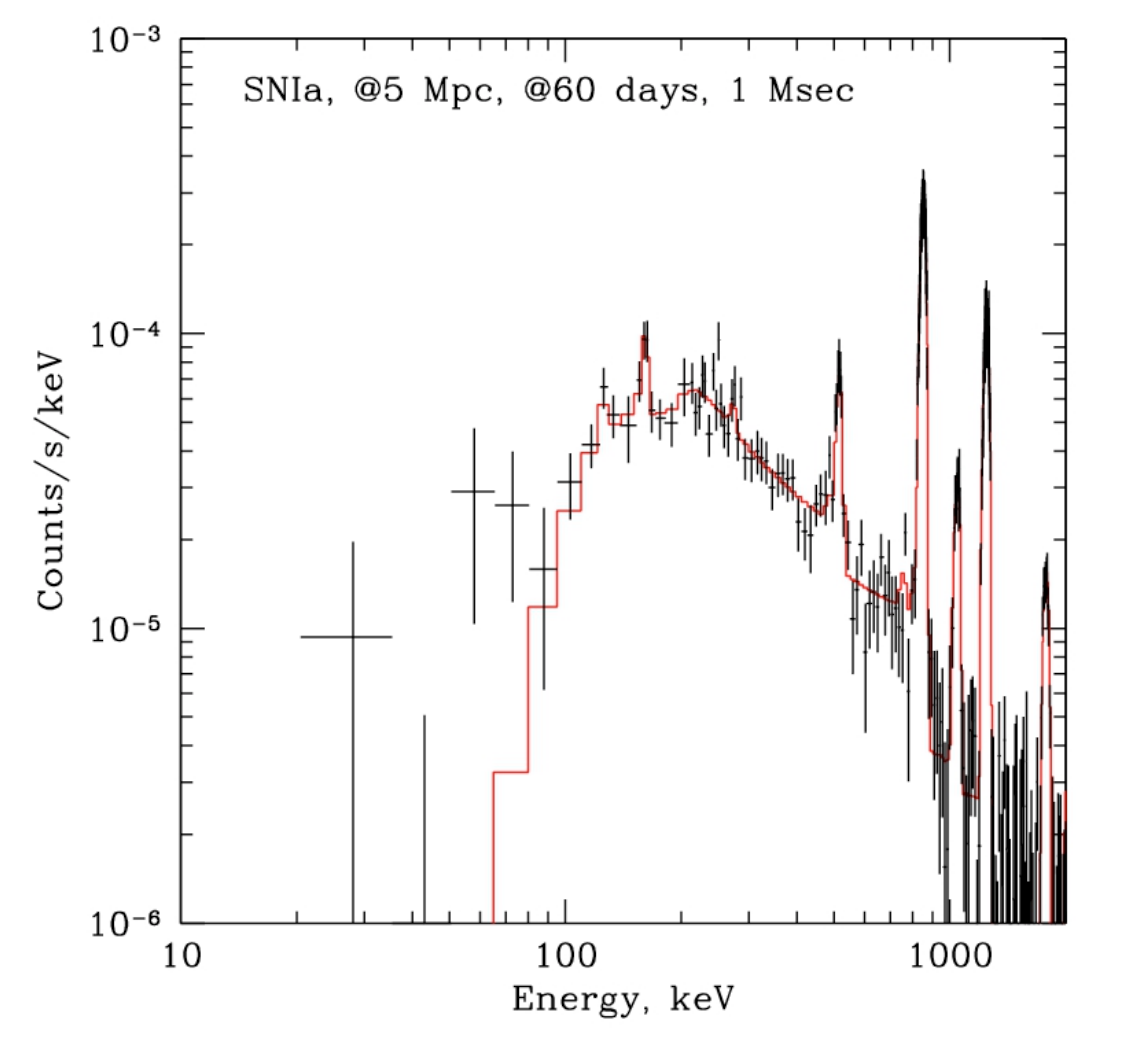}
\caption{The gamma-ray signal expected in the SPI instrument from a nearby supernova of type Ia 60 days after the explosion with 1Ms integration time, assuming background already eliminated. The characteristic broad lines from the decay of $^{56}$Co are clearly seen on top of Comptonized continuum emission. This simulation assumes a homogeneously-mixed explosion of a 1.4~\Msol white dwarf producing 0.7~\Msol of $^{56}$Ni with outer-ejecta velocity 10$^4$~km~s$^{-1}$, and a distance of 5 Mpc. Figure by E. Churazov, from \cite{2011SSRv..161..149W}.}
\label{fig:SNIa_sim}
\end{figure}

Hope for an accurate measurement of the amount of $^{56}$Ni which powers a SNIa light curve (canonically about 0.7~\Msol  ($\pm$0.3) \cite{2006A&A...450..241S}) through direct radioactive-decay gamma-rays had rested on the opportunity of a sufficiently-nearby event, closer than about 5~Mpc. Such events should occur once every 2--3 years \cite{2008NewAR..52..377I}, although estimated numbers vary by factors \about~2 depending on the assumptions about the supernova rate in this more-heterogeneous population of nearby galaxies.  
The CGRO mission in 1991--2000 had two \emph{realistic opportunities} to measure characteristic gamma-rays from a SNIa: SN1991T occurred early in the mission, yet was quickly recognized as an anomalous event with exceptionally-high brightness and $^{56}$Ni yield. The hint for the expected gamma-ray lines supports this, in spite of its relatively small distance of 13 Mpc \cite{1997AIPC..410.1084M}. The second event, SN1998bu at \about~17~Mpc, seemed much fainter, and no gamma-ray emission could be detected \cite{2002A&A...394..517G}. Therefore, even after nine mission years, the gamma-ray view of SNIa still was in its infancy.

Fig.~\ref{fig:SNIa_sim} shows how, ideally, the SPI instrument aboard INTEGRAL should see such a signal from a SNIa at 5~Mpc distance. The radioactive-decay gamma-ray lines are broadened, and overlay a broad continuum which is caused by Compton scatterings of these characteristic line gamma-rays in the envelope.  The variety of diagnostic measurements is capable to decode details and differences among explosion models \cite{2008MNRAS.385.1681S}.   
So far in INTEGRAL's mission,  only one SNIa, SN2003gs at 16~Mpc distance, had been considered a candidate for INTEGRAL observations, but no signal could be found. No other SNIa has been nearer and also observable from solar-array and celestial-visibility constraints until 2011. 

\begin{figure}
\centering
\includegraphics[width=0.68\textwidth]{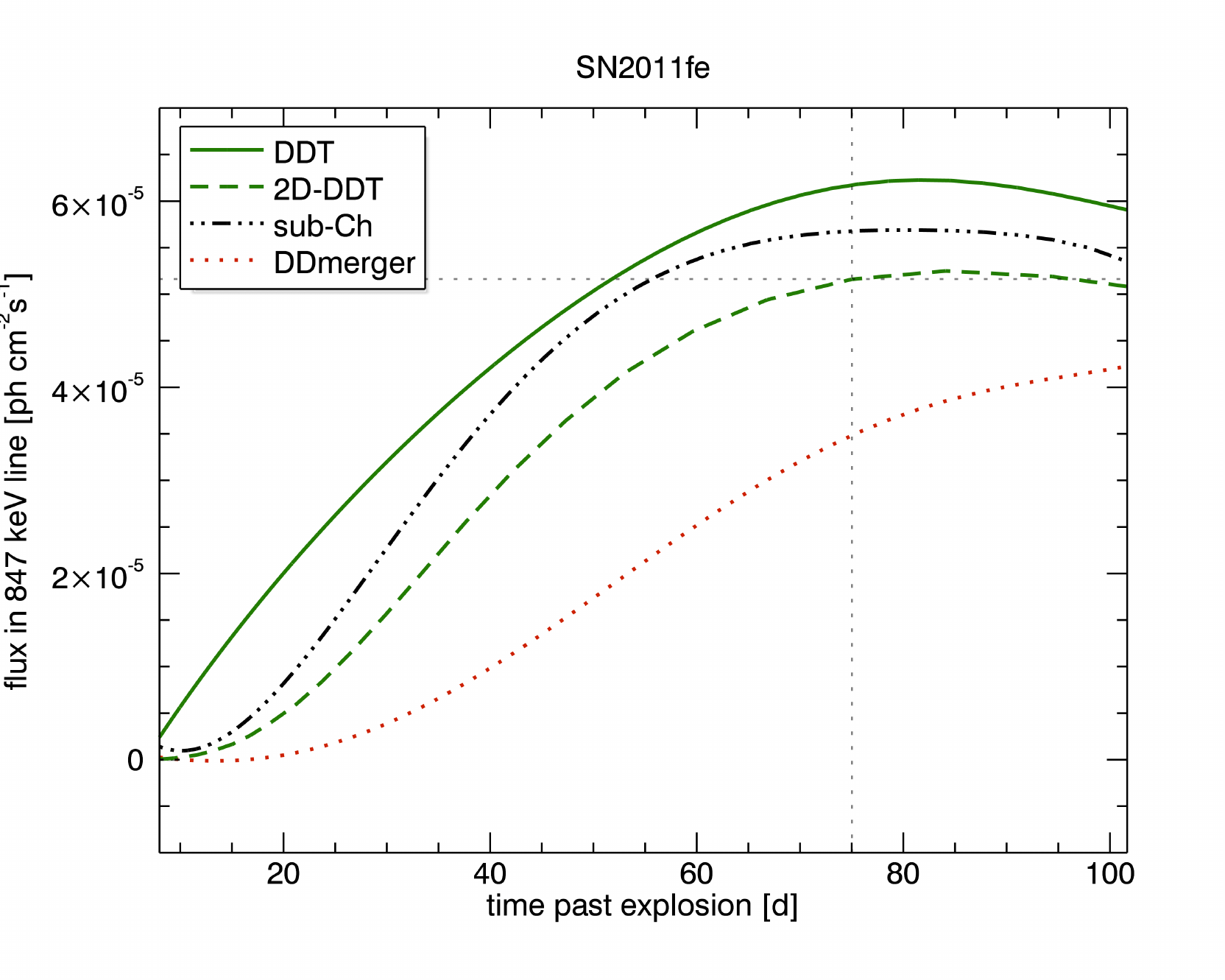}
\caption{The gamma-ray line emission at 847 keV from decay of $^{56}$Co rises towards a maximum at about 90 days after explosion, according to several model variants for the case of SN2011fe at 6.4~Mpc distance (S. Sim 2011, private communication and \cite{2012arXiv1203.4839R}). INTEGRAL  observed SN2011fe during days 6-20 and 45-89. The SPI narrow-line sensitivity of 3~10$^{-5}$ph~cm$^{-2}$s$^{-1}$ degrades with line width, to about 3~10$^{-4}$ph~cm$^{-2}$s$^{-1}$ if lines would be 50~keV wide.}
\label{fig:SN2011fe_847}
\end{figure}

SN2011fe occurred on August 24, 2010, in the nearby galaxy M101/NGC5457 \cite{2011Natur.480..344N}, and may be a promising candidate to observe gamma-ray emission from radioactive decays directly. At a distance of 6.4 Mpc and with lines likely to be significantly broadened, it still is marginal, considering the instrumental-background level. But as nearby SNIa are rare, INTEGRAL observed this event for 4~Ms total. One of the observations (1~Ms) was scheduled early after the explosion, in order to search for $^{56}$Ni which could have been produced on the outer surface and appear early, if SN2011fe would be triggered by a He flash on a white dwarf's surface \cite{2010ApJ...714L..52S}. A second observation of 2+1~Ms was scheduled towards the expected maximum gamma-ray brightness, where a line detection is most probable, and then could discriminate among candidate models \cite{2011PrPNP..66..309R,2012arXiv1203.4839R} (see Fig.~\ref{fig:SN2011fe_847}). Shown in this figure are currently-favored model classes of a delayed detonation (DDT, shown for a 1- and 2-dimensional treatment), sub-Chandrasekhar, and (double-degenerate) white-dwarf merger models. Particular interest in this supernova was stimulated by apparent absence of a companion star, thus making a double-degenerate merger model more likely for this event \cite{2011Natur.480..348L}.  At later times after the gamma-ray maximum, the SN as a whole is transparent to gamma-rays, and explosion type differences cannot be constrained, as occultation differences among models with different explosion morphology vanish, and only the total $^{56}$Ni mass determines the gamma-ray brightness. Detailed analysis of the observations is currently underway. First inspections did not show any of the candidate lines from the $^{56}$Ni decay chain in the early observations \cite{2011ATel.3683....1I}. 

After the COMPTEL mission, and with INTEGRAL another nine years of a suitable gamma-ray telescope in operation, SNIa gamma-rays still have not contributed to help understand SNIa. We need a significant advance in sensitivity below 10$^{-6}$~ph~cm$^{-2}$s$^{-1}$ as proposed by some new mission concepts. Also, luck with sufficiently nearby events, within few Mpc, is needed. SN2011fe, or a fortunate future event may prove and exploit INTEGRAL's potential herein during its extended mission.  

\subsection{Core-Collapse Supernovae}\label{sources_ccsn}
Massive stars with initial masses above 8--10~\Msol end their evolution through gravitational collapse, as their energy reservoir of releasing nuclear binding energy through fusion reactions is exhausted \cite{2007PhR...442...38J,2011LNP...812..153T}. At this point in their evolution, the large stellar cores which are stabilized by electron pressure of a degenerate Fermi gas of electrons finally become unstable, when these electrons reach higher energies and can thus be captured by nuclei in weak transitions. 
As this reduces degeneracy pressure, the ensuing gravitational collapse occurs on a time scale of a second, and is only moderated by decomposition of infalling nuclei into nucleons and $\alpha$~particles (consuming again the nuclear energy that had powered the stability of the star in previous phases of its evolution), and by beta decays. 

\begin{figure}
\centering
\includegraphics[width=0.88\textwidth]{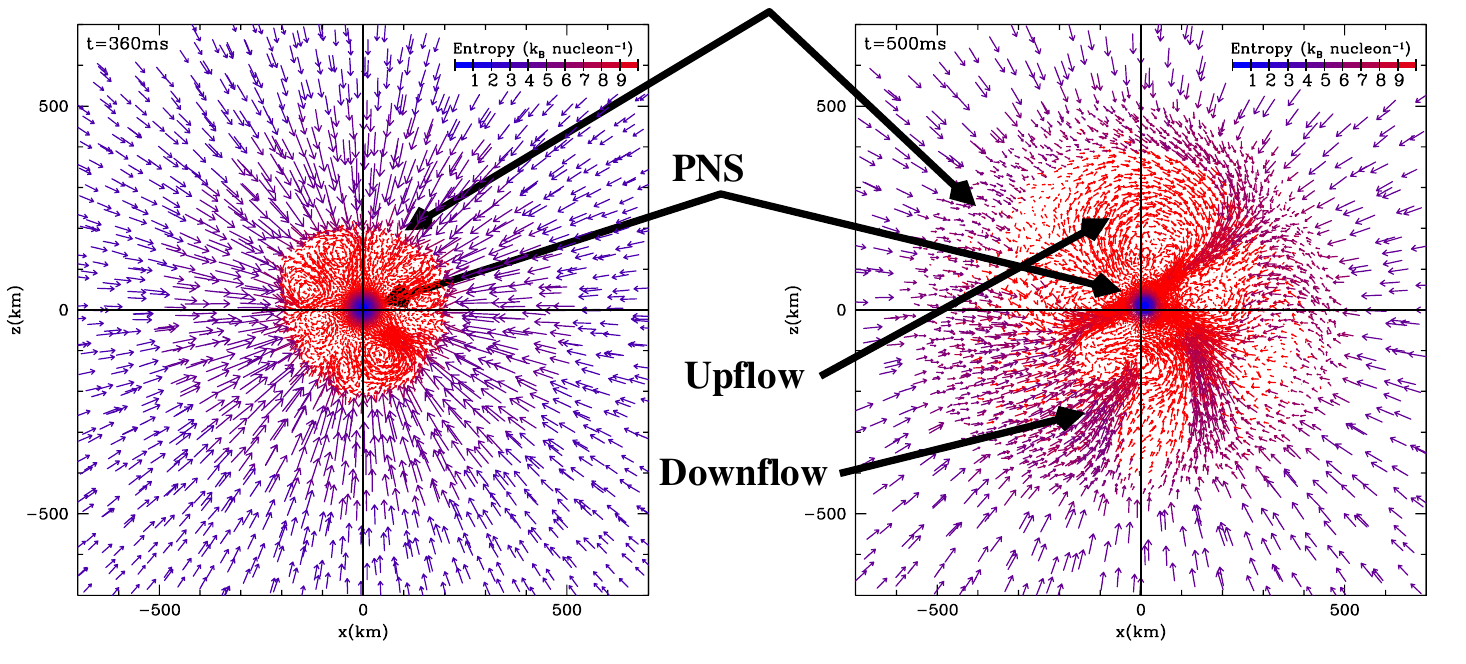}
\caption{The inner region of a gravitational-collapse supernova is governed by in falling matter from the imploding core. The core-bounce shock develops into an accretion shock, instabilities develop and lead to inward and outward gas flows, as the supernova explosion develops. (from \cite{2008nuco.confE.101F}).}
\label{fig:ccSN_inner}
\end{figure}

Once infalling material reaches the inner core and mounts up to densities corresponding to those of matter in atomic nuclei (\about~10$^{13}$g~cm$^{-3}$), this \emph{core bounce} creates a shock above the newly-forming proto-neutron star. A shock region forms at radii around 120~km, where infalling matter is decomposed  into nucleons and $\alpha$~particles more efficiently. This also consumes the shock-region's energy, thereby becoming insufficient for exploding the outer parts of the star. But here matter is still dense enough to be opaque to neutrinos, which are copiously produced in the proto-neutron star as matter neutronizes from electron captures and inverse $\beta$-decays,. Additionally $\nu$'s are produced as the neutron star cools, emission of thermal $\nu$ at typical thermal energies of \about~10~MeV results. Thus, neutrino scatterings (elastic and inelastic) deposit additional energy in this inner region, they \emph{heat} the \emph{gain region}\footnote{As nuclear reactions occur herein, these regions also cool very efficiently through both radiation and neutrinos. The gain region is defined as the inner part where $\nu$-heating dominates all cooling processes.}. Now an unstable situation is set up, since a hot inner region underlies colder/cooling material. Instabilities develop and will be characteristic for the dynamics. This results in destruction of spherical symmetry, causing violent turbulence and possibly large-scale asymmetric inwards as well as outwards directed gas flows (see an example from core-collapse simulations in Fig.~\ref{fig:ccSN_inner}; \cite{2008nuco.confE.101F,2010ApJ...714.1371H}). Infalling material from the star may thus be channeled to low altitudes near the neutron star, at the same time hot nuclear-burning blobs may be swirled outward towards the stellar envelope. Numerical simulations of various types and groups have shown this consistently.
 
At this point, modeling of all relevant physics is difficult, and dynamics probably is driven by secondary mechanisms of energy transfers from the violent interior of the gravitational collapse to the infalling envelope. These neutrino interactions have long been thought to cause the supernova explosion, once modeled with sufficient accuracy and including angular dependences of relativistic scatterings. Studies showed that explosions may occur, specifically for less-massive stars near 10~\Msol, yet in many numerical models of gravitational collapses actually the accretion shock stalled and did not result in an explosion. It has then been shown that mechanical energy could be transferred more directly from the inner turbulent core towards the infalling envelope through acoustic waves, thus could excite large-scale \emph{acoustic} oscillations, which may be sufficient to explode the envelope; this would occur at late times, up to seconds after the onset of the gravitational collapse and the core bounce\footnote{For this reason, it had been overlooked in earlier simulations, which were often terminated about one second after bounce when dynamics seemed settled.}. 
Clearly, core-collapse supernova explosions occur from an unstable and largely variable initial state after the gravitational collapse. Residual rotational energy of the core, magnetic fields, neutrino interactions, hydrodynamic instabilities and their evolution, and nuclear reactions, all play a role in how a collapse turns into an explosion. It is not understood how this will occur in stars of different masses, rotational, and magnetic state; observations can help, and most prominently observations which investigate the products of nuclear burning, as the burning rates are all very sensitive to environmental conditions. 

Nuclear reactions are expected to occur in the dense shock region approaching nuclear statistical equilibrium (NSE), hence producing Fe-group elements and also substantial amounts of radioactive $^{56}$Ni. At the same time, in regions nearer to the neutron star, neutrino interactions with nuclei occur in the \emph{$\nu$-process} and mainly liberate nucleons from nuclei, thus stimulating proton and neutron capture reactions of remaining nuclei. Further out, decomposition of in falling matter will also provide abundant free nucleons and $\alpha$-particles, and thus add an \emph{$\alpha$-rich} flavor to nuclear burning. Once the explosion sets in, material will expand and cool, and nuclear burning will \emph{freeze out}. 

As this onset of the explosion arises from an unstable situation and probably is triggered by rather minor changes in energy transport, non-spherical behavior may lead to an explosion with clumps and jets, as observed in several prominent supernova remnants. Under such conditions, it has been shown that in particular $^{44}$Ti ejection would vary by more than an order of magnitude. \cite{2006A&A...450.1037T,2010ApJS..191...66M}.
This is probably the most-uncertain part of core-collapse supernova nucleosynthesis. Large spreads in environmental parameters of density and temperature, and probably even more importantly for nuclear processing, in the neutron-to-proton ratio are likely to occur. Several theoretical studies have shown such diversity of outcomes, for production of intermediate-mass elements \cite{2011ApJ...741...78M}. Moreover, exotic isotopic mixtures could result from inner neutron-rich and also proton-rich nucleosynthesis, from rapid neutron capture on iron-group seed nuclei (\emph{r-process products}) in the neutrino-driven wind region behind the shock. The conditions and dynamics in this region are also highly uncertain, but plausibly related to core-collapse supernovae because observed in metal-poor stars which probably were formed from interstellar gas which had been enriched by one or few nucleosynthesis events only; also for proton-rich heavy-element isotopes, this region is a plausible origin \cite{2006ApJ...637..415F,2011arXiv1112.4651A,2011ApJ...729...46W}\cite{2009ApJ...694L..49F,2009A&A...494..829P}. 
  
Finally, the explosion will drive a shock through the outer envelope beyond the core, before these parts could take notice of the inner collapse, and some \emph{explosive nucleosynthesis} will occur in those shock-heated regions.  This explosive nucleosynthesis is characterized by short nuclear burning times which lead to large deviations from equilibrium and hydrostatic nuclear burning patterns.  
Altogether, one expects that abundant oxygen, silicon, and iron-group nuclei will be ejected, mixed with small but important contributions of heavy element products from the neutrino-driven wind zone.

Radioactive ejecta which could be observed with gamma-ray spectrometers include $^{56}$Ni and $^{44}$Ti, at estimated amounts of \about~0.1 and \about~few~10$^{-4}$~\Msol, respectively. Although both isotopes should be produced in core collapse events which form neutron star remnants and hence eject freshly-produced isotopes, only the ejection of $^{56}$Ni is certain, as it powers the supernova light (although amounts vary by almost three order of magnitude \cite{2011Ap&SS.336..129N}), while $^{44}$Ti ejection is much less clear \cite{2006A&A...450.1037T}. 

Early one-dimensional models of Type II or Type~Ib supernovae obtain typical $^{44}$Ti yields of 5~10$^{-5}$~\Msol, with large (factors 2--4) variations with progenitor mass \cite{1996ApJ...464..332T}. Analyzing the impacts of deviations from spherical symmetry and the entropy-dependency of $\alpha$-rich freeze-out nucleosynthesis, an overall enhancement of $^{44}$Ti relative to $^{56}$Ni of a factor of \about~5 was considered plausible  \cite{1997ApJ...486.1026N}; but it remains to be shown that this holds for a realistic and detailed supernova model. Recently, three-dimensional simulations of a nucleosynthesis network under supernova-interior conditions were explored \cite{2011ApJ...741...78M}. These studies investigated how the forward/outgoing and the reverse shock would modify density and temperature in these inner regions, and possibly lead to intermittent NSE, QSE-group, and freeze-out conditions. They obtain mass fraction variations over 6--8 orders of magnitude for $^{44}$Ti production, and thus reinforce that the specific supernova explosion trajectory in the $Y_e, \rho, T$ phase space  will determine the ejected amount of $^{44}$Ti. 
Measurements of $^{44}$Ti gamma-rays from all accessible young supernova remnants are the tool to clarify nucleosynthesis in these inner supernova regions.

The search for young SNR in $^{44}$Ti gamma-ray emission encompasses spectroscopy in three lines which occur in the radioactive decay of $^{44}$Ti. The first stage of its decay chain had been re-assessed in 1995--2006 to have a half-life of 59~y ($\pm$0.3~y), it occurs predominantly (99.3\%) through capture of an electron into an excited $^{44}$Sc level, which de-excites by emission of the excitation-level energy through two X-rays at 68 and 78~keV. $^{44}$Sc itself $\beta$-decays with half life 3.7~h to excited $^{44}$Ca, which in almost all cases de-excites through emission of the characteristic 1157~keV $\gamma$-ray line. 
The second stage of this decay chain may be considered prompt, in view of the long-lifetime decay to initial $^{44}$Ti. 
While the two X-ray lines are best suited for X-ray telescopes when optimized for high-energy X-rays (such as IBIS and SPI  on INTEGRAL, and NuSTAR), the gamma-ray line is left to gamma-ray telescopes such as COMPTEL aboard CGRO and SPI on INTEGRAL.

The radioactive decay times of the $^{56}$Ni decay chain, but also of $^{44}$Ti decay, are sufficiently short to ensure bright gamma-ray emission from each individual supernova event, and supernova rates are low enough to make source confusion or overlaps unlikely even for degree-scale resolution as typical for gamma-ray telescopes. Nearby and recent supernovae are thus the target us such studies. Historic supernovae in our own Galaxy are (in order of approximate age) SNR G1.9+0.3 (\about~100~y), SNR Cas A (\about~340~y), SNR Kepler (probably Type Ia; \about~410~y), SNR Tycho  (probably Type Ia; \about~440~y), SNR G41.1-0.3 (\about~600 or 16000~y), SNR 3C~58 (\about~630~y), Crab (\about~958~y), and SNR 1006/Lupus (probably Type Ia; \about~1006~y). Then, SN1987A in the Large Magellanic Cloud (LMC) is of course a prominent object of study outside our own Galaxy, but sufficiently nearby to be a candidate for $^{44}$Ti gamma-ray search.

In both the $^{56}$Ni and the $^{44}$Ti decay chain,  electron captures are the main decay channels. This could lead to profound distortions of radioactive decay properties, as electrons need to be close to the nucleus for the decay to occur. An atomic shell needs to be occupied at least in the inner shell, and fully-ionized atoms would not decay. The effect of a fully-ionized supernova remnant would be to store the $^{44}$Ti decay energy until electron captures become possible, the exponentially-declining line intensity setting in at a later time. This has been analyzed to potentially result in underestimated $^{44}$Ti yields from a young supernova by up to a factor 2, depending on ionization structure of the remnant, and hence on how much of the ejecta have been subject to the reverse shock already \cite{1999A&A...346..831M}. Clumping of inner ejecta as well as asphericities of the explosion will be important here.

\subsubsection{Cas A}
The radioactivity afterglow of the Cas A supernova has become a key test case for core-collapse constraints, as this SNR has been observed with a variety of instruments throughout the electromagnetic spectrum. The distance to Cas A has been settled to 3.4~kpc (+0.3/-0.1~kpc) \cite{1995ApJ...440..706R}. The remnant appears as a 5' diameter filled shell in line and continuum emission from radio through IR and optical to X-ray bands; this diameter corresponds to \about~5~pc at this distance. Its explosion is dated to a precision of \about~few years as 1671, based on proper motion extrapolations of optical knots and Flamsted's reported optical supernova (see \cite{2004NewAR..48...61V,1997NuPhA.621...83H}). At present, the outer shock of the explosion blast wave is clearly seen in radio and X-ray bands, while the reverse shock that is expected cannot be clearly identified, but supposedly now has propagated through \about 2/3 of the remnant already, re-heating the remnant gas to X-ray emission temperatures. Although this remnant age is well beyond the 89~y decay lifetime of $^{44}$Ti, these $^{44}$Ti radioactivity observations are an important complement in the study of the Cas A supernova explosion. 

\begin{figure}
\centering
\includegraphics[width=0.88\textwidth]{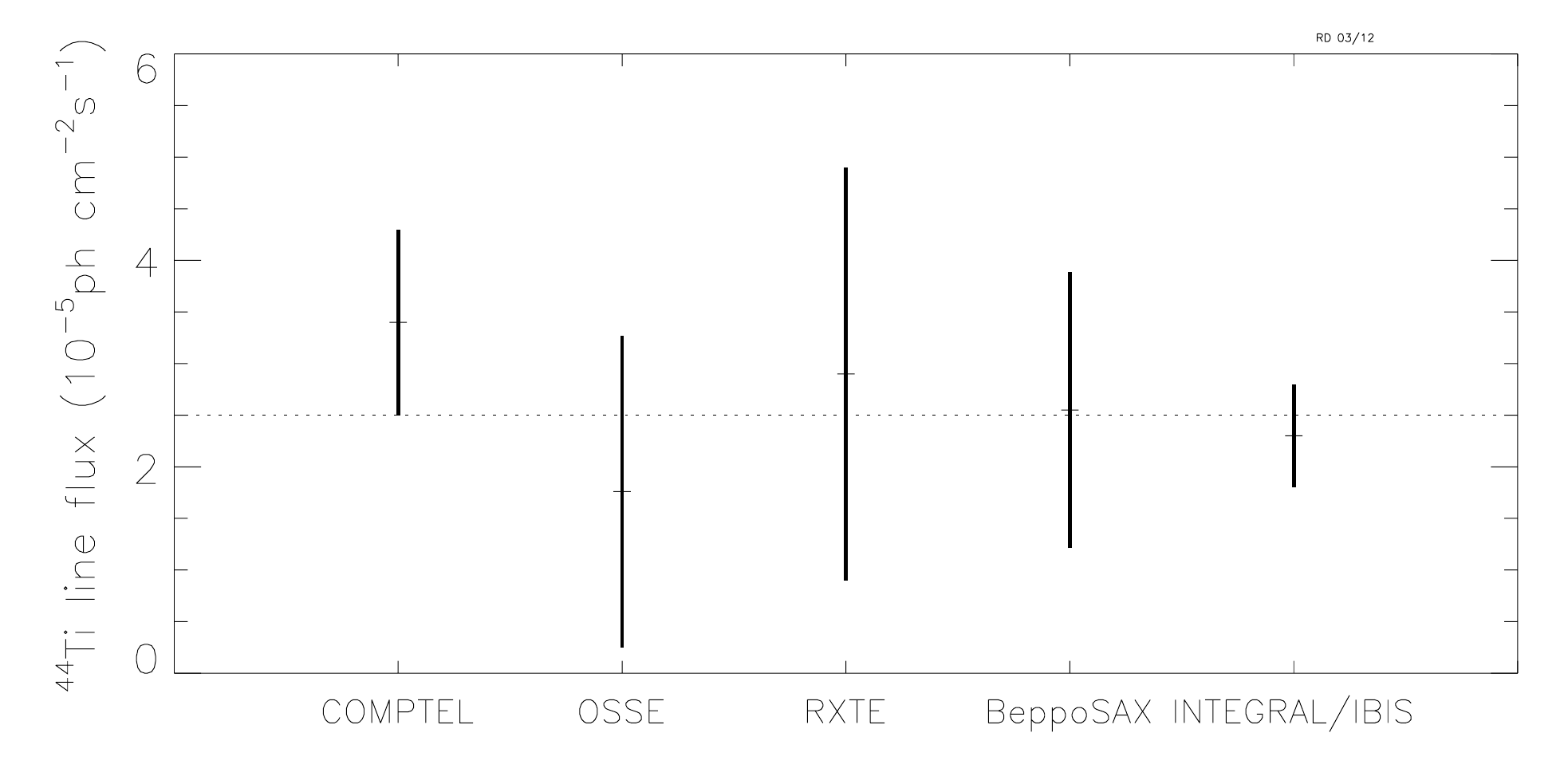}
\caption{The Cas A $^{44}$Ti fluxes as reported from gamma-ray telescopes which could measure either the low-energy (68 and 78~keV) or high-energy (1157 keV) lines from $^{44}$Ti decay. A flux value of \about~2.5~10$^{-5}$ph~cm$^{-2}$s$^{-1}$ (dotted line) seems consolidated. }
\label{fig:CasA_44Tiflux}
\end{figure}

Observations with the COMPTEL telescope had discovered $^{44}$Ti  from Cas A in 1994 \cite{1994A&A...284L...1I}. The flux measured in the 1157~keV line was  consolidated with more observations, and was assessed as (3.5~$\pm$1.2)~10$^{-5}$ph~cm$^{-2}$s$^{-1}$ \cite{2000AIPC..510...54S,1997A&A...324..683D}, corresponding to a $^{44}$Ti mass of \about~1--2~10$^{-4}$~\Msol. Earlier analysis reported a higher flux, and corrections for underlying continuum also turned out to complicate the line flux interpretation.

INTEGRAL has spent substantial exposure on the Cas A region,  a key target already for the early-mission core program. From 4.5~Ms of exposure accumulated with core and dedicated open-program time, the currently most-precise flux measurement from $^{44}$Ti decay has been obtained with IBIS, with a value of  (2.3$\pm$0.5)~10$^{-5}$ph~cm$^{-2}$s$^{-1}$ \cite{2006ApJ...647L..41R}. SPI data at the 68 and 78~keV lines from $^{44}$Ti decay have been analyzed as well. After an initial report of a detection consistent with expectations \cite{2007ESASP.622..105M}, it was concluded later that background variations were too large for a sufficiently-sensitive measurement \cite{2009A&A...502..131M}. Fig. \ref{fig:CasA_44Tiflux} summarizes the $^{44}$Ti line flux measurements for the Cas A supernova.  This has been translated into an amount of $^{44}$Ti at the time of explosion of 1.6~10$^{-4}$~\Msol, with an uncertainty of ($^{+0.6}$/$_{-0.3}$)~\Msol. 
It is clear that Cas A's $^{44}$Ti ejection was substantially (\about a factor three) above predictions from models.

From observations of this SNR at other wavelengths, we have learned much detail about the Cas A supernova explosion. Optical emission from fast-moving and O-rich but H-deficient knots suggests that the explosion itself originated from a Wolf-Rayet star, and was irregular in shape,  including jet-like features. X-ray emission from the shocked gas within the remnant has shown that intermediate-mass elements from C and O through Ne, Mg, S, Si, and Ca are all seen, consistent with the explosion of an onion-like structured massive star with its characteristic composition before collapse, O being the dominant element; an O mass of 1-3~\Msol for a total ejecta mass of 2--4~\Msol was inferred \cite{2002A&A...381.1039W,2004NewAR..48...61V}.  But unlike other SNR, Cas A does not show emission lines from the intermediate-mass elements such as Ne and Mg at the brightness expected from massive-star nucleosynthesis yields; this might be due to the reverse shock not yet reaching such ejecta, but detection of Si line emission suggests that ejecta are already shocked quite far towards inside. Hence, the explosion morphology of Cas A certainly does not reflect a spherical explosion of an onion-shell type massive star. 

The progenitor star has been estimated from absence of H emission to have lost its envelope already due to a Wolf-Rayet phase. This lead  to an estimate of a progenitor mass of 22--25~\Msol  \cite{2004NewAR..48...61V}. Also, the surroundings of the explosion then may be inferred to be structured in density by  the progenitor star's mass loss and wind, which was found consistent with compositions as seen in the outer part of the remnant. 

Supernova light  scattered by interstellar clouds has added another interesting observable for the Cas A supernova. Observing time variable emission from the general direction of Cas A, this may be traced back to emission variability in the Cas A object, when interpreted as re-radiated (dust) emission from Cas A. The scattered/reflected light thus probes past luminosity of Cas a, from the light propagation delay between source and reflector, and also different viewing aspects of the source. A suitable sample of interstellar clouds thus enables comparisons of echoes from supernova light emitted in different directions, and thus probes the symmetry of the explosion. Initial light echo data obtained from the Spitzer mission had been interpreted to reflect X-ray activity of the compact remnant star \cite{2005Sci...308.1604K}. According to \cite{2011ApJ...732....3R}, the central source is rather the supernova itself, i.e. the UV flash of the shock break out, which gives a view of the explosion from different viewing angles, and confirms asymmetry of the explosion itself.

While INTEGRAL's IBIS measurement confirms an ejected $^{44}$Ti amount of 1-2~10$^{-4}$~\Msol \cite{2006ApJ...647L..41R}, the non-detection of the 1157~keV decay line by INTEGRAL's SPI instrument \cite{2009A&A...502..131M} suggests that $^{44}$Ti-rich ejecta move at velocities above 500~km~s$^{-1}$ \cite{2009A&A...502..131M}. As Doppler broadening increases with photon energy, this makes the high-energy decay line much broader than SPI's instrumental resolution, hence the signal-to-background ratio for this line degrades and may escape detection, while the 68 and 78~keV lines have been measured from this same event.  It will be interesting to see how NuSTAR will map out the $^{44}$Ti emission morphology for Cas A, with its high imaging resolution, to  teach us more on supernova asymmetries and clumping of inner ejecta for the case of Cas A. 

\subsubsection{SN1987A}
Supernova SN1987A had been the target of early gamma-ray observations with several balloon and satellite missions. The detection of $^{56}$Co decay lines with the Gamma-Ray Spectrometer scintillation detector instrument on the Solar Maximum Mission was a surprise, as it had been expected that the supernova envelope would become transparent to inner nucleosynthesis products much later. Therefore, this detection suggested that inner ejecta were actually mixed and inhomogeneous in the envelope, and blobs from inner regions which may be enriched with $^{56}$Ni could thus appear earlier. 

SN1987A has been followed through the years, and the decline of the light curve in IR through optical emission could be traced with high precision. Already after few years, it became clear that the decline had slowed down, consistent with late or long-lived radioactivity energy input such as expected from decay of $^{44}$Ti. The inferred amount from the light curve has been determined at this time as (1.5 $\pm$1.0)~10$^{-4}$~\Msol \cite{2002NewAR..46..487F}. Using infrared lines in the 23--27~$\mu$m band, the supernova interior was investigated through ISO observations at 3425 and 3999 days post explosion, the light curve was modeled adopting different amounts of $^{56}$Ni and $^{44}$Ti and solving for the temperatures and ionization levels in an expanding remnant with adopted expansion and density, to obtain an upper limit of  1.1~10$^{-4}$~\Msol.  The larger value above was confirmed \cite{2011A&A...530A..45J} with a value of (1.4 $\pm$0.5)~10$^{-4}$~\Msol, from modeling the SN spectrum in detail throughout the NIR-optical-UV range. This is somewhat less than the value determined for Cas A, but still above the values typically resulting from supernova models. On the other hand, an alternative recent analysis explains the late SN87A light curve mainly through $^{55}$Co or $^{60}$Co decays with limits to $^{44}$Ti contributions of \about10$^{-5}$~\Msol only \cite{Seitenzahl87A2012}.

\begin{figure}
\centering
\includegraphics[width=0.68\textwidth]{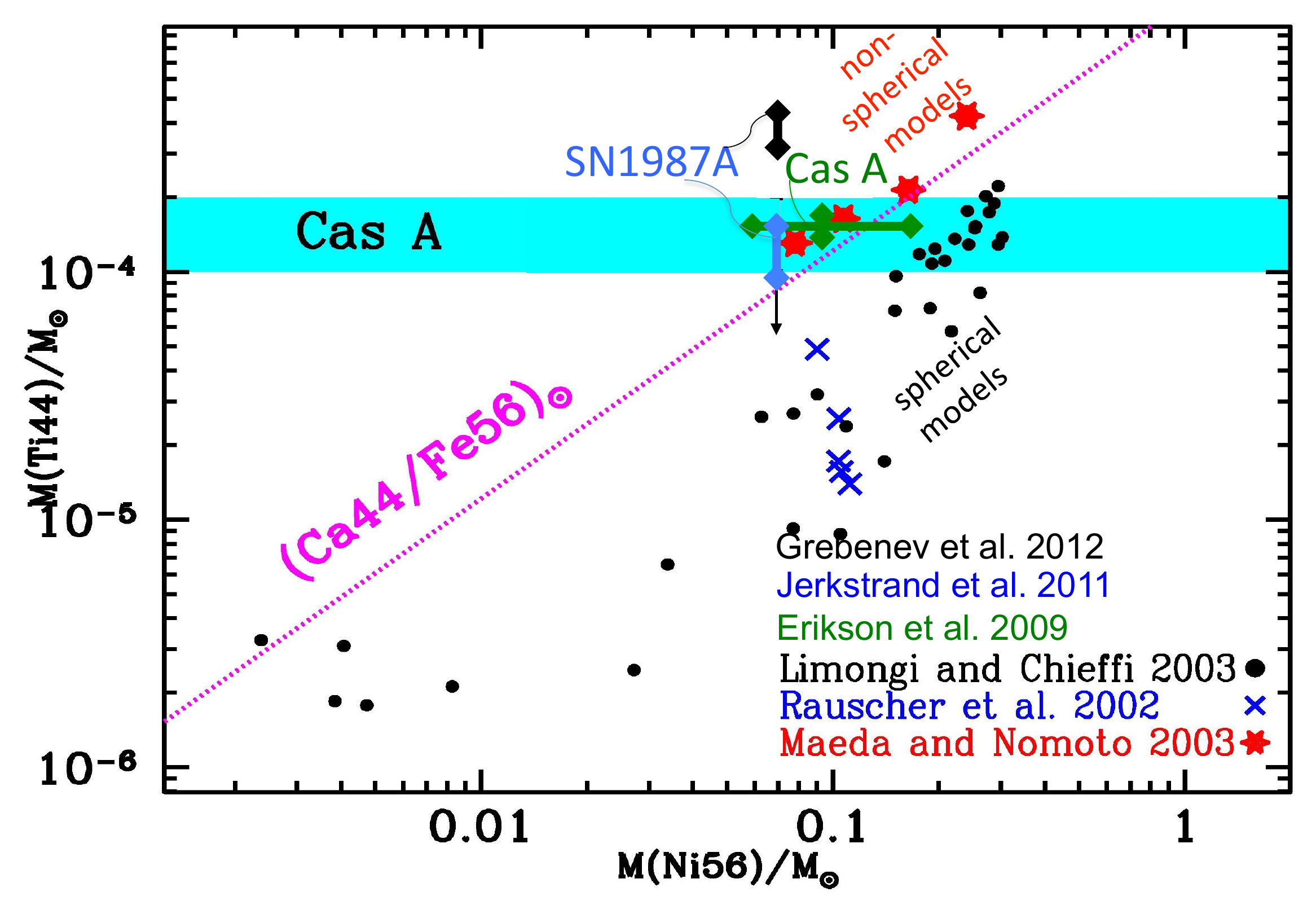}
\caption{The ratio of radioactive isotopes $^{56}$Ni and $^{44}$Ti is a sensitive probe of the effective location of the \emph{mass cut}, the separation between ejected material and the neutron-star remnant of a core-collapse supernova \cite{1996ApJ...464..332T}. Measurements for Cas A and SN1987A appear consistent with the ratio as inferred from solar abundances of the stable daughter isotopes. Models assuming spherical symmetry (black dots) fall below observations in this representation, while non-spherical explosion models indicate better agreement (adapted after \cite{2004ESASP.552...15P}).}
\label{fig:Ni-Ti-ratio}
\end{figure}

INTEGRAL had observed the LMC region with SN1987A early in the mission for \about~1.75~Ms, with no signal found at the location of the supernova. However with additional observations in 2010/2011, a detection of SN1987A in an energy band conforming to the 68 and 78~keV lines from $^{44}$Ti decay with IBIS was first reported in 2011 \cite{GrebenevSN87A}. This provides an important second case, where for a well-observed supernova radioactivity in  $^{56}$Ni and $^{44}$Ti can be compared with model predictions for different assumptions of explosion symmetry (see Fig.~\ref{fig:Ni-Ti-ratio}; see also \cite{2011A&A...530A..45J,2002NewAR..46..487F,2010A&A...517A..51K}). 

\subsubsection{Other young SNR, and $^{44}$Ti from supernovae}
One would expect to detect $^{44}$Ti emission from several objects along the plane of the Galaxy \cite{2006A&A...450.1037T}, if $^{44}$Ti ejection were a typical property of core-collapse supernovae, at a typical yield as constrained by the solar $^{44}$Ca/$^{56}$Fe abundance ratio and core-collapse supernova rate to be 10$^{-4}$~\Msol. In COMPTEL's survey \cite{1997A&A...324..683D,1999ApL&C..38..383I},  just Cas A turned out to be the single clearly detected source \cite{2000AIPC..510...54S}, and no source was found in the inner Galaxy where e.g. $^{26}$Al has shown massive stars to exist and explode. 
INTEGRAL  has also surveyed the plane of our Galaxy, obtaining comparable or higher sensitivity for the inner Galaxy. The IBIS search for $^{44}$Ti point sources  did not find any source either, beyond Cas A  \cite{2006NewAR..50..540R}. This is surprising and shows that $^{44}$Ti ejection is not a typical property of core-collapse supernovae. 
In their analysis, \cite{2006A&A...450.1037T} note the discrepancy of model yields with the solar abundance of $^{44}$Ca. Arguing that no source of $^{44}$Ca could avoid the $^{44}$Ti progenitor, they therefore scaled model yields up by a factor 3 in their simulation of supernovae brightness in $^{44}$Ti, finding several above instrumental limits. This is also seen in Fig.~\ref{fig:Ni-Ti-ratio}, showing that the solar $^{44}$Ca/$^{56}$Fe ratio sets a constraint for the $^{44}$Ti/$^{56}$Ni ratio ejected from supernovae. For Cas A and SN1987A, from which we have independent constraints for $^{44}$Ti and $^{56}$Ni, these isotopic ratios fall onto the same solar-abundance constraint line. However, spherical models (black dots) fall short of $^{44}$Ti production by about a factor 3. When the impact of deviations from spherical symmetry are estimated, $^{44}$Ti yields can be increased by factors (see the red star symbols in Fig~\ref{fig:Ni-Ti-ratio}). 
As discussed above, we expect large event-to-event variations. Therefore, it seems more appropriate to abandon this \emph{average-yield} assumption, and rather investigate on the one hand event by event, including all information about peculiarities of each supernova, and on the other hand the consistency of expected variations and rates with solar abundances. 
In non-spherical explosions with clumps and jets, $^{44}$Ti ejection could vary by more than an order of magnitude, as supported by explorations of the parameter space in density and temperature and their impact on nuclear reaction networks as they produce $^{44}$Ti \cite{2010ApJS..191...66M,2011ApJ...741...78M}.

\subsection{Diffuse Line Emission and Massive Stars}\label{sources_stars}

\begin{figure}
\centering
\includegraphics[width=0.68\textwidth]{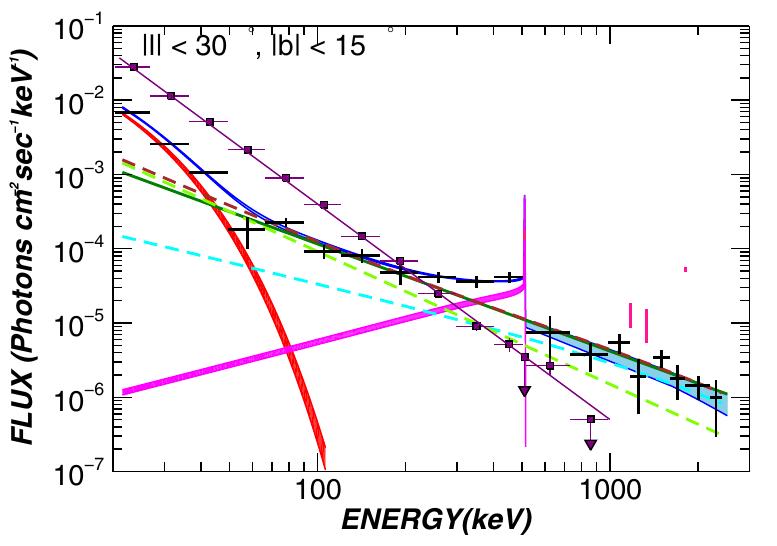}
\caption{INTEGRAL data integrated along the Galactic plane, showing continuum emission and line features from positron annihilation, \Al, and $^{60}$Fe radioactive decay (adapted from \cite{2011ApJ...739...29B}).}
\label{fig:galridge_spec}
\end{figure}

The INTEGRAL survey of the Galaxy confirms the extended, large-scale emission along the entire plane of the Galaxy, which had been imaged earlier using COMPTEL data \cite{1994A&A...292...82S}. Its spectrum is dominated by a bright, power-law-type continuum, due to interactions of cosmic ray electrons with interstellar radiation and matter  (figure \ref{fig:galridge_spec},  \cite{2011ApJ...739...29B}. Superimposed are line features from positron annihilation (together with its lower-end continuum) at 511 keV, and from \Al at 1808.63~keV and $^{60}$Fe at 1773 and 1332~keV. 
The large-scale nature of \Al emission had been found from COMPTEL observations \cite{1995A&A...298..445D,1999A&A...344...68K}, and had been attributed as mainly due to massive-star related sources \cite{1996PhR...267....1P}. For $^{60}$Fe and for positron annihilation, no corresponding COMPTEL imaging results are available. 

\begin{figure}
\centering
\includegraphics[width=0.68\textwidth]{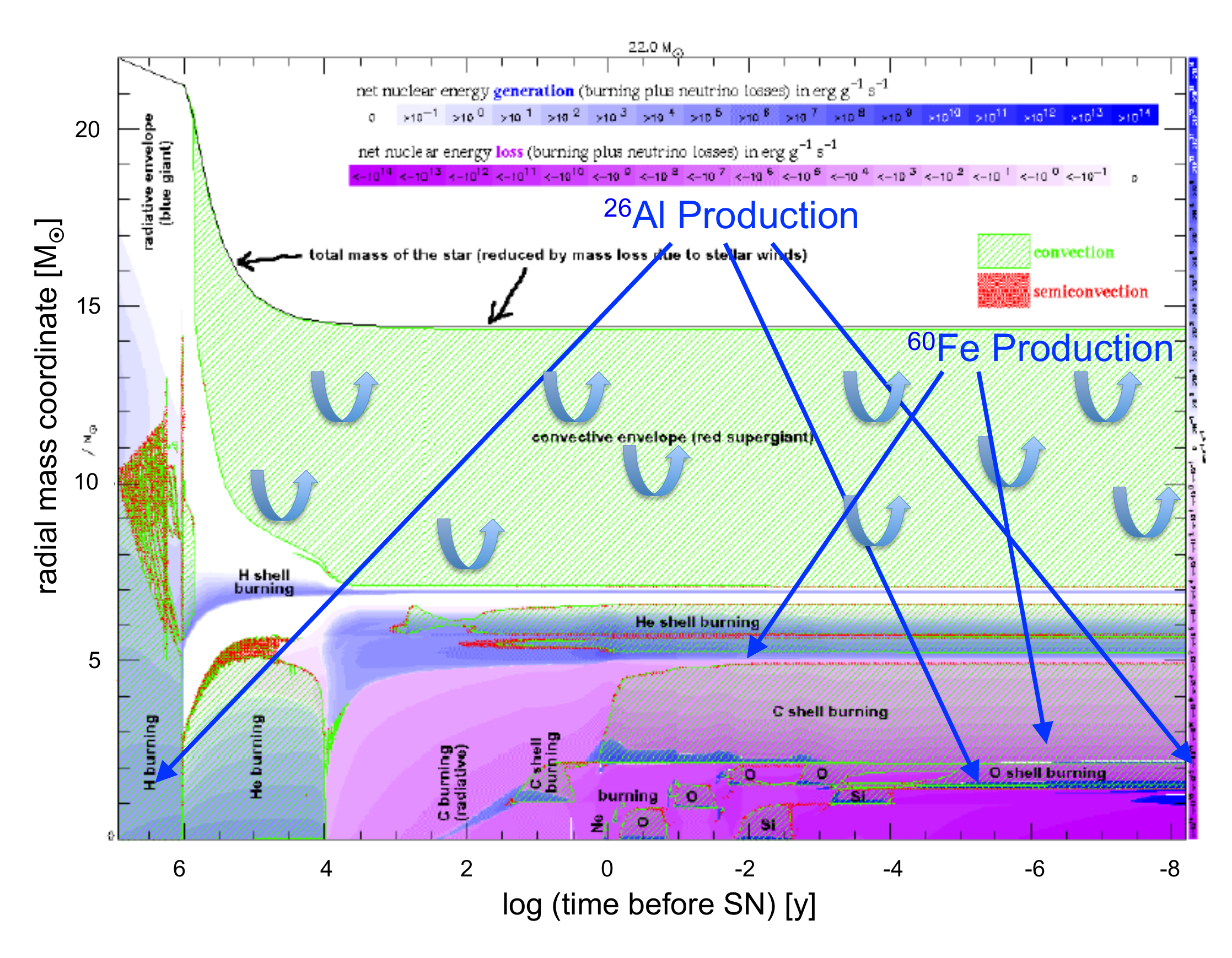}
\caption{The interior structure of massive stars is complex towards their later evolution: Intermittent shell burning occurs, simultaneously and in addition to core nuclear burning. This \emph{Kippenhahn diagram} shows the evolution of a massive-star interior towards the supernova in a logarithmic time scale (adapted from \cite{2003NuPhA.718..159H}, for a star of initial mass 22~\Msol). The production sites of $^{26}$Al and $^{60}$Fe are indicated, as well as the mass loss from stellar wind in the Wolf-Rayet phase. }
\label{fig:Al-Fe_nucleosynthesis}
\end{figure}

Massive-star interiors and supernovae are candidate sources of long-lived radioactive isotopes $^{26}$Al ($\tau\sim$~1.04~My) and $^{60}$Fe ($\tau\sim$~3.8~My)  \cite{2011LNP...812..153T}. The evolution of massive stars accelerates towards later burning stages, as burning temperatures increase and neutrino losses become important. Fig.~\ref{fig:Al-Fe_nucleosynthesis} shows  the complex interior structure of massive stars, with nuclear-burning and convective regions indicated (hatched). After initial core hydrogen burning (main sequence phase), hydrogen burning proceeds in a shell until the supernova occurs. All regions further inside are not coupled to the convective envelope any more, from which at later stages a strong wind leads to substantial mass loss of the star. This wind may include nucleosynthesis products from the early main sequence phase. Within the star, at later phases, intermittent shell burning occurs simultaneously and in addition to core nuclear burning.  The production sites of $^{26}$Al and $^{60}$Fe are indicated (arrows): $^{26}$Al is produced in the hydrogen-burning phases from initial $^{25}$Mg, and later in the carbon and neon burning core and shell as proton-release reactions allow further processing of remaining $^{25}$Mg, plus during the supernova explosion as explosive Ne/C burning. Only $^{26}$Al from core hydrogen burning is mixed into the envelope and thus ejected during the intense stellar-wind phase. $^{60}$Fe is expected to be synthesized from neutron capture reactions on $^{54}$Fe as neutron release reactions such as $^{13}$C($\alpha$,n) and $^{22}$Ne($\alpha$,n) are activated; but here, a subtle balance of neutron capture and convective transport away from neutron-rich regions is required to ensure that $^{60}$Fe is not destroyed by further neutron captures. All $^{26}$Al and $^{60}$Fe produced within the star after the main sequence phase is released only with the supernova explosion, together with additional contributions from explosive burning in the supernova itself. 

Both $^{26}$Al and $^{60}$Fe are long-lived compared to the stellar evolution of their massive star sources, which proceeds on the time scale of \about~10$^6$y (Fig.~\ref{fig:Al-Fe_nucleosynthesis}). Therefore, decay of these isotopes in interstellar space as produced by the most-rapidly evolving, most-massive stars will occur simultaneously with further ejection by lower-mass stars with correspondingly slower evolution towards their supernova. If we consider the likely formation of a \emph{group} of massive stars from a dense molecular-cloud core, their concerted release of radioactive isotopes occurs over a period of \about~3 to 20~My after star formation \cite{2009A&A...504..531V} (see Fig.~\ref{fig:popsyn_26Al60Fe}). The measurement of $^{26}$Al and $^{60}$Fe radioactivity gamma-rays, therefore, is a tool to verify our \emph{stellar-mass averaged} predicted yields from models of massive-star evolution and nucleosynthesis. In particular, since the same massive stars are plausible producers of both those isotopes, yet from different regions and epochs inside the stars, the measurement of the ratio of $^{26}$Al to $^{60}$Fe emission provides a valuable tool, as systematic uncertainties in the stellar populations themselves (richness, distance) cancel in such ratio.  

\begin{figure}
\centering
\includegraphics[width=0.48\textwidth]{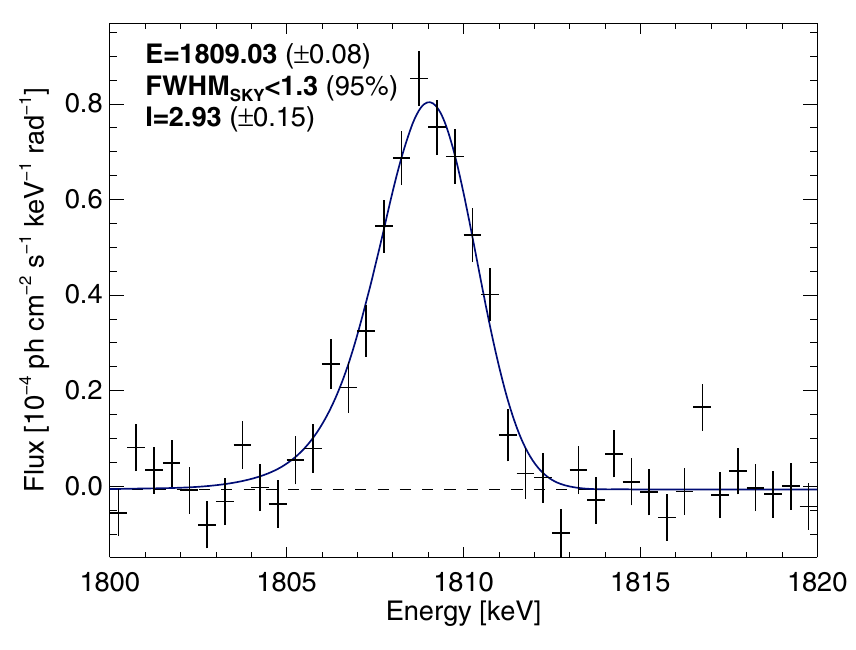}
\includegraphics[width=0.48\textwidth]{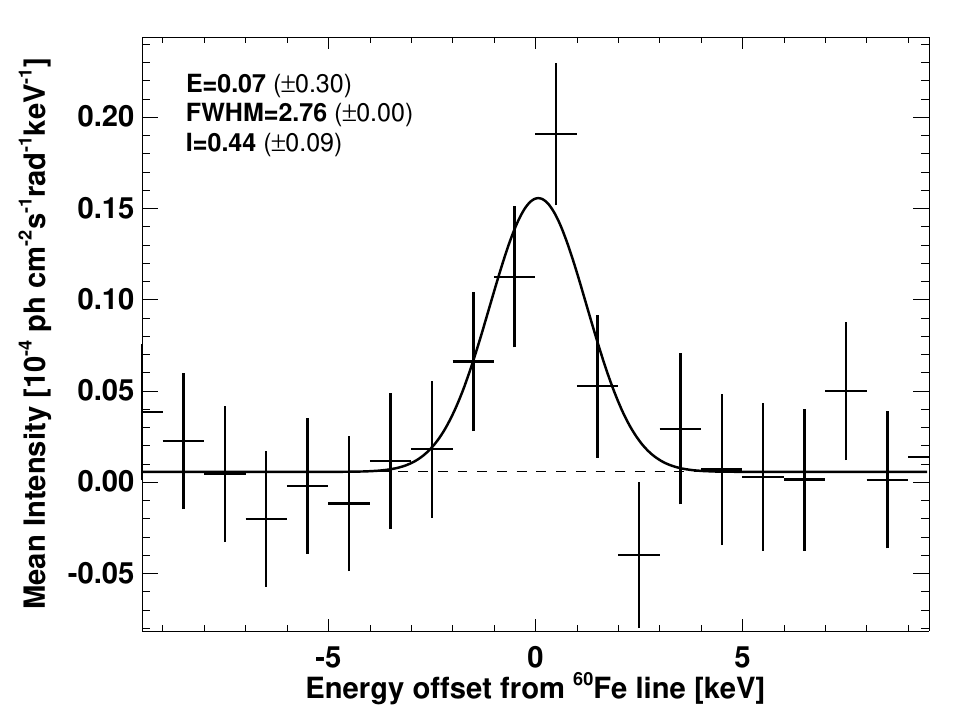}
\caption{The spectra measured by SPI from the entire plane of the Galaxy for $^{26}$Al \emph{(left)} \cite{2006A&A...449.1025D}, and $^{60}$Fe \emph{(right)} \cite{2007A&A...469.1005W}.}
\label{fig:Al-Fe_spectraSPI}
\end{figure}

Gamma-rays from the decays of  $^{26}$Al and $^{60}$Fe  isotopes have been measured with SPI/INTEGRAL from the Galaxy at large \cite{2006A&A...449.1025D,2009A&A...496..713W,2007A&A...469.1005W} Fig.~\ref{fig:Al-Fe_spectraSPI}. In the inner Galactic ridge, a flux at 1808.63~keV of (2.63~$\pm$0.2)~10$^{-4}$ph~cm$^{-2}$s$^{-1}$rad$^{-1}$ has been derived \cite{2010A&A...522A..51D}. The $^{26}$Al line is a very bright, diffuse and extended signal from the Galaxy at large, detected with SPI at a significance above 28$\sigma$ \cite{2009A&A...496..713W}.

\begin{figure}
\centering
\includegraphics[width=0.78\textwidth]{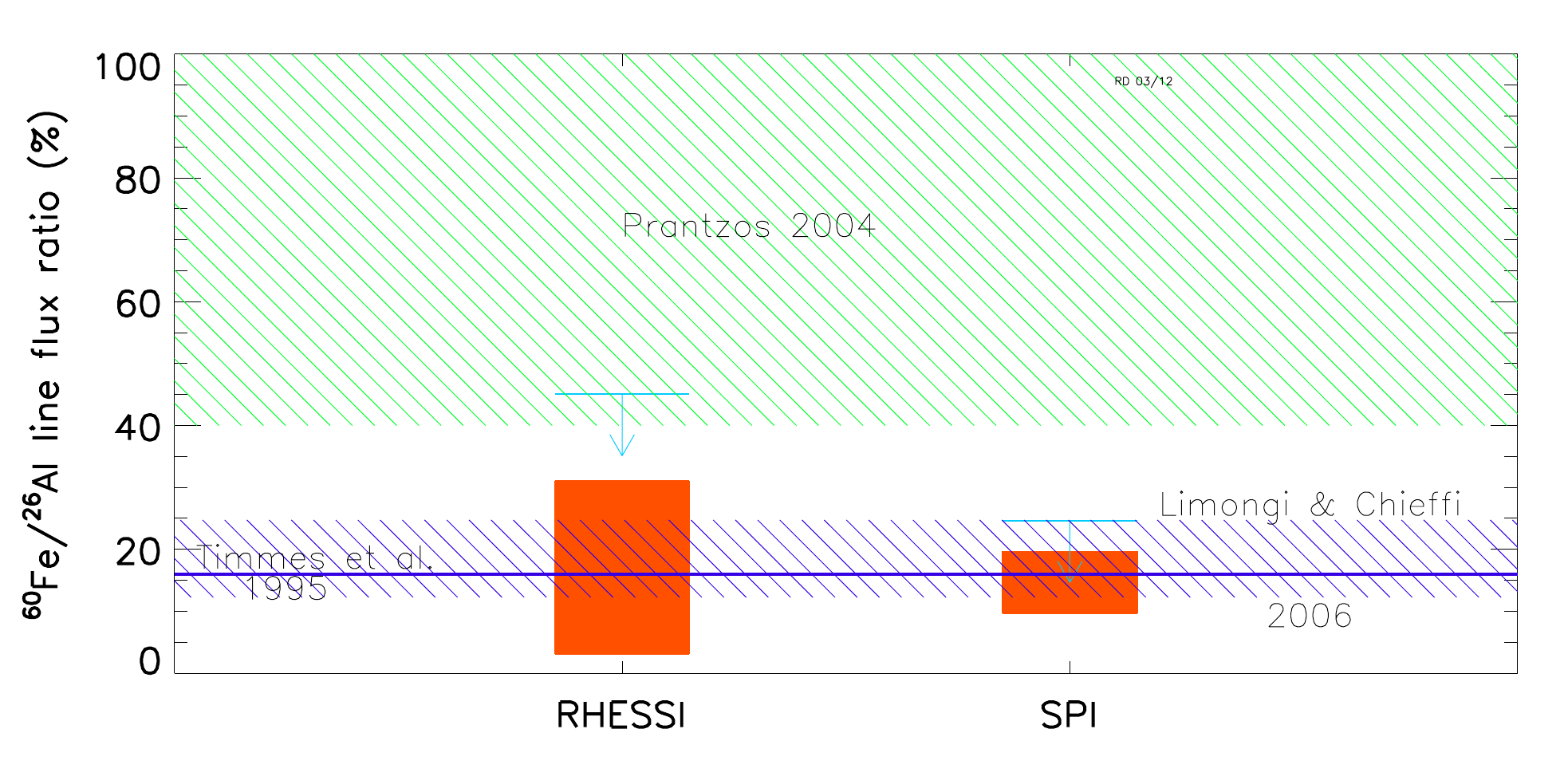}
\caption{The measured values for $^{60}$Fe/$^{26}$Al gamma-ray intensity ratios reported from RHESSI \cite{2004ESASP.552...45S} and INTEGRAL \cite{2007A&A...469.1005W} are consistent with values reported from earlier and most-recent massive-star nucleosynthesis models \cite{1995ApJ...449..204T,2007PhR...442..269W,2006ApJ...647..483L}. An intermediate assessment of model yields \cite{2004A&A...420.1033P} had assessed a much brighter model-predicted $^{60}$Fe emission than seen by instruments.  Both instruments report rather marginal detections (therefore, 2$\sigma$ upper limits are also shown in this graph).}
\label{fig:Al-Fe_ratio}
\end{figure}

The $^{60}$Fe gamma-ray line emission occurs in a cascade of two gamma-ray lines, at 1172.9 and 1332.5~keV, with approximately-equal intensities. In none of these lines, separately analyzed in SPI single-detector and multiple-detector events, a convincing celestial gamma-ray line signal could be seen. Combining line intensities in both these lines, the spectrum shown in Fig.~\ref{fig:Al-Fe_spectraSPI} has been derived, which shows $^{60}$Fe emission from the sky with a combined significance of \about~5$\sigma$. Clearly, the total gamma-ray brightness in $^{60}$Fe decay is much lower than the one seen from $^{26}$Al. The measured gamma-ray brightness ratio is \about~15\%, with an uncertainty of \about~5\% (figure~\ref{fig:Al-Fe_ratio}). 

  In steady state, this brightness ratio constrains massive-star interiors globally, as those same sources produce each of these isotopes in different inner regions and at different times of stellar evolution -- yields from models for these object types should come out consistent for all their nucleosynthesis products \cite{2011LNP...812..345D}. Particularly interesting would be the $^{60}$Fe/$^{26}$Al ratio for source populations of specific ages, as the ratio varies significantly due to wind-released $^{26}$Al before any core-collapse supernova would eject $^{60}$Fe and more $^{26}$Al \cite{2009A&A...504..531V}. $^{60}$Fe is exclusively released in supernovae, although predominantly produced in the late shell-burning phase before the collapse of the core. Only the Cygnus region appears within INTEGRAL's sensitivity range for this, however \cite{2010A&A...511A..86M}. 

From nucleosynthesis models of massive stars and supernovae, it appears that more $^{60}$Fe would be expected, although currently-predicted yields are still in agreement with the observations, within quoted uncertainties. Note that $^{60}$Fe nucleosynthesis may also occur in a rare subtype of SNIa explosions at high yields of \Msol, although a single bright source being responsible for observed $^{60}$Fe gamma-rays is unlikely. $^{60}$Fe yields are particularly high (as compared to $^{26}$Al yields) in massive-star models. So, in spite of significant uncertainties of the measurement (see below), the $^{60}$Fe/$^{26}$Al ratio seems lower than expected. 
Possible causes could be suppression of $^{60}$Fe by more-efficient neutron capture reactions, leading to neutron-richer Fe isotopes. Alternatively, $\beta$-decay of $^{59}$Fe could occur at a higher rate than inferred from current nuclear theory. Both reaction types are being investigated in laboratory experiments at the present time. Note that the $\beta$-decay lifetime of the $^{60}$Fe isotope had been re-determined in 2008 \cite{2009PhRvL.103g2502R}, and found to be substantially longer, with (exponential) lifetime increased from 2.15 to 3.8~My (T$_{1/2}$ from 1.5 to 2.6~My).
Another cause of lower $^{60}$Fe yields could be deviations from the standard initial-mass distribution for the high-mass end of massive stars $\geq$60~\Msol. Those stars are rare and short-lived, and therefore observational determinations of the power-law slope of the IMF are more uncertain. 
And thirdly, convective regions in high-mass stars are bounded by physical processes of buoyancy and dynamics of stellar gas at changing composition. Convection thus is very complex, as shown by high-resolution three-dimensional hydrodynamics simulations. Approximations which are required in full stellar-evolution models, such as mixing lengths, overshoot, and semi convection parameters, could be somewhat different than currently adopted in these models. 
It appears that the bound of \about~15\% for the $^{60}$Fe/$^{26}$Al gamma-ray intensity ratio provides a significant observational bound, to be met by massive-star models, independent of absolute yield uncertainties (figure~\ref{fig:Al-Fe_ratio}).

The total mass of $^{26}$Al in the Galaxy had been inferred from earlier COMPTEL measurements to be 2--3~\Msol \cite{1996PhR...267....1P}. In such a mass determination, one must adopt a spatial source distribution to resolve the distance uncertainty, when converting a measured gamma-ray flux into a quantity of isotopes present in the Galaxy. Generally, large-scale models of Galactic structure or sources have been used, such as exponential-profile disk, azimuthally-symmetric nested rings, or spiral-arm models inferred from other observations. Yet, the irregularity of the large-scale emission as seen by COMPTEL from Galactic $^{26}$Al already suggested that the massive-star population in the Galaxy may be more clumpy than such large-scale models describe. Identified localized $^{26}$Al emission from the Cygnus region \cite{2009A&A...506..703M} supports this view, and also the detection of $^{26}$Al emission from the very-nearby Scorpius-Centaurus association \cite{2010A&A...522A..51D} (see below). Accounting for such localized enhancements, with more-advanced INTEGRAL measurements the Galactic mass of $^{26}$Al has been reported at somewhat lower values around 2\Msol, and ranging from 1.5 to 3.6~\Msol within uncertainties.

Using this Galaxy-wide amount of $^{26}$Al together with $^{26}$Al yields for massive stars across the entire mass range, and an initial-mass distribution, one can derive the total population of stars which corresponds to measured $^{26}$Al gamma-rays \cite{2006Natur.439...45D}. As massive stars above \about~8-10~\Msol all are believed to end their evolution as core-collapse supernovae, this corresponds to a determination of the \emph{Galactic} rate of core-collapse supernovae. We obtain a value of 1.54 ($\pm$0.89) supernovae per century, or a supernova every 65 years\footnote{This is somewhat lower than the value published in \cite{2006Natur.439...45D} from $^{26}$Al, due to accounting for foreground emission now attributed to the nearby Scorpius-Centaurus sources, rather than distant Galactic $^{26}$Al.}.  
The importance of this measurement of the Galactic supernova rate derives from the underlying method. Using penetrating gamma-rays from radioactivities ejected by supernova related sources throughout the current Galaxy, this approach does not suffer from corrections for occulted sources, such as e.g. a supernova rate determination based on O and B star counts does. Other measurements inherently assume negligible possible differences between our Galaxy and other galaxies, as they measure core-collapse supernova related objects in Milky-Way analogue galaxies which are seen face-on and hence free from occultation-bias of distant parts of our own Galaxy. It is interesting that the $^{26}$Al-based supernova rate determination agrees with measurements undertaken with those alternative approaches, and falls into the lower-value part among all those measurements (see detailed discussion in \cite{2006Natur.439...45D}).

\begin{figure}
\centering
\includegraphics[width=0.98\textwidth]{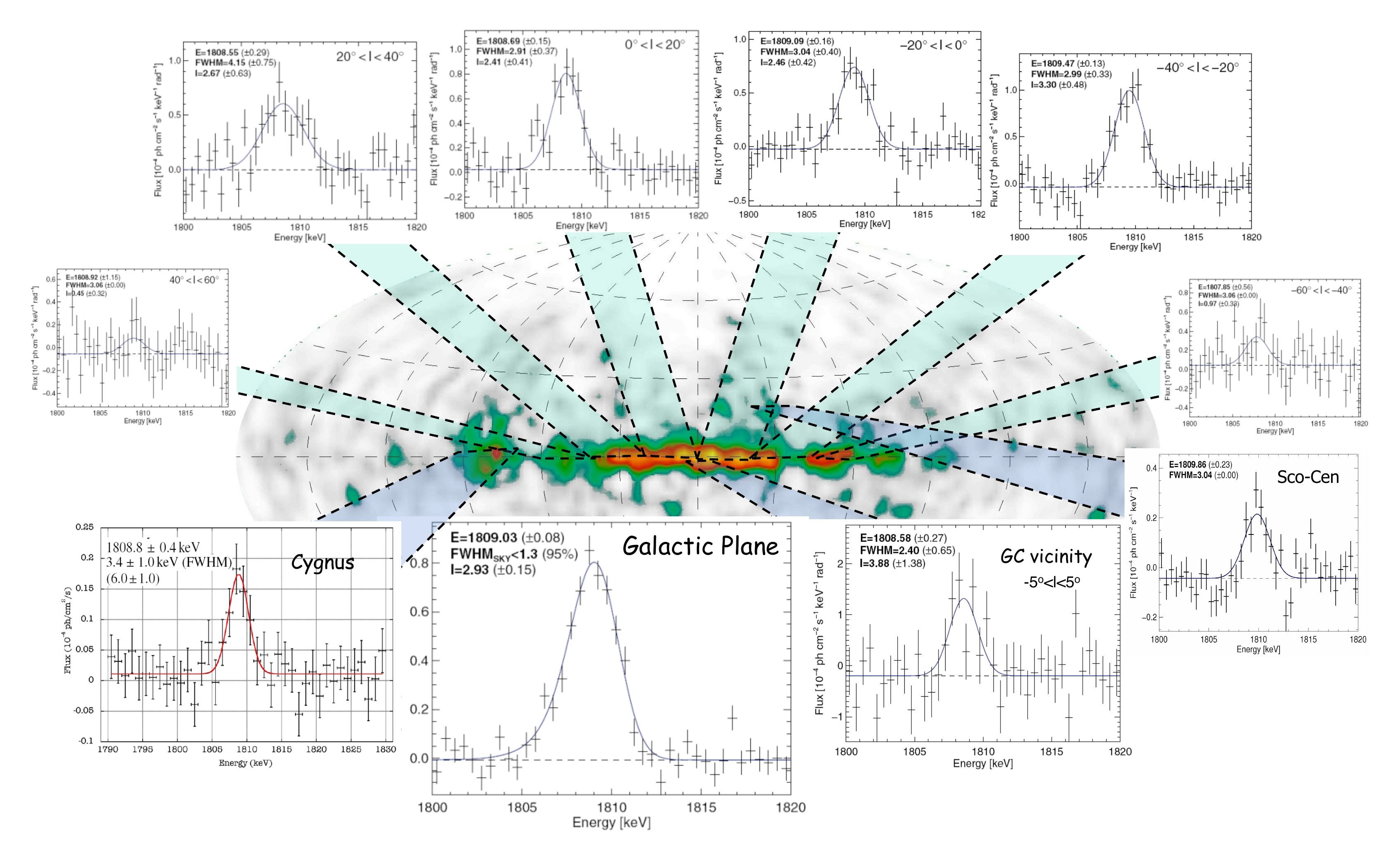}
\caption{The spectra showing the \Al gamma-ray line (E$_{lab}$~1808.63~keV) as measured by SPI from different regions along the plane of the Galaxy for $^{26}$Al \cite{2009A&A...496..713W}. }
\label{fig:Al_region_spectraSPI}
\end{figure}

$^{26}$Al is bright enough to also be seen from localized regions along the Galactic plane hosting many massive stars (Fig.~\ref{fig:Al_region_spectraSPI}). For 
a specific source region, distance uncertainties can often be resolved from other observations, and, moreover, the stellar population is determined through star catalogues. Therefore, a  comparison of predicted versus observed amounts of \Al can be made (see Fig.~\ref{fig:popsyn_26Al60Fe} and \ref{Fig_26AlCyg_pred-obs}), and provide a more specific test for massive-star models than can be obtained from Galaxy-wide analysis as discussed above. With deeper exposure, this became possible in the late INTEGRAL mission for the Cygnus \cite{2010A&A...511A..86M}, Carina \cite{2012A&A...539A..66V}, and Scorpius-Centaurus \cite{2010A&A...522A..51D} regions;  for other candidate locations of massive-star groups INTEGRAL's sensitivity is insufficient, due to their fainter \Al emission.

\begin{figure}
\centering
\includegraphics[width=0.78\textwidth]{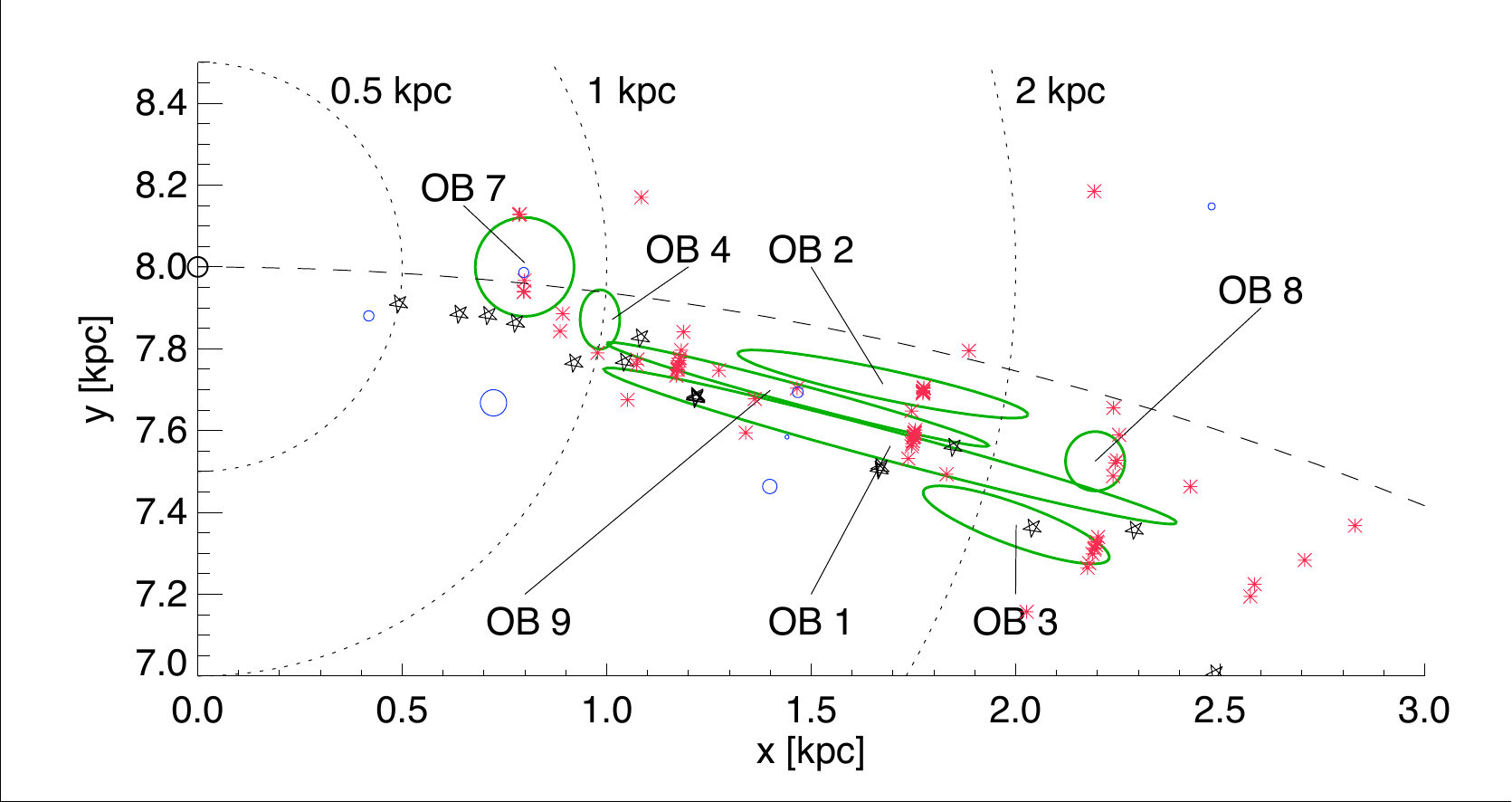}
\caption{The Cygnus region with its candidate sources of \Al  \cite{2002NewAR..46..535P}.  Ellipses reflect the uncertainties in the distances to the OB associations. Note that the richness of each association is subject to revisions, as new stellar searches proceed to deal with occultation by foreground interstellar clouds. }
\label{fig7:cygnusOB}
\end{figure}

\subsubsection*{$^{26}$Al in the Cygnus Region.}    
The COMPTEL \Al gamma-ray map showed emission from the Cygnus region at Galactic longitudes around  30\degree\ as the clearest and most-prominent feature beyond the bright ridge in the inner Galaxy.
Along this line of sight, there are 6 prominent OB associations at distances ranging from 0.7 to 2.5~kpc \cite{2002NewAR..46..535P}, plus about a dozen open clusters, some associated to those OB associations (Fig.~\ref{fig7:cygnusOB}). Their ages range from 2.5 to 7.5~My \cite{2002A&A...390..945K}. It appears that the Cygnus OB2 association dominates by far the stellar census of this \emph{Cygnus complex}. Possibly associations Cyg OB1, 2, and 9 are related to OB2 and may originate from the same parental molecular cloud \cite{2002A&A...390..945K}. Cyg OB2 may even be considered the most prominent case in our Galaxy of the extremely-rich superclusters, which appear prominent in other galaxies but are hard to recognize within our own Galaxy; about 120 stars in the high-mass range (20--120~\Msol) have been identified to relate to Cyg OB2; the other associations typically are ten times smaller. The age and distance of Cyg OB2 is 2.5~My and 1.57~kpc, respectively. 

Within the Cygnus region, star formation has probably occurred for more than 10~My, and has created a rich population of stars, as evident in the now-observed relatively-young OB associations. The interstellar medium in the Cygnus complex is very heterogeneous and filamentary, with hot cavities bounded by dense remains of the parental molecular clouds \cite{2002AstL...28..223L,2008A&A...486..453C}.
Thus, the stellar population is somewhat uncertain, leading to corresponding uncertainty in expected nucleosynthesis products \cite{2008A&A...486..453C,2010A&A...511A..86M}. 

Because of its young age, in Cyg OB2 stellar evolution even for the most-massive stars should still not be completed, and contributions from core-collapse supernovae to \Al production should be small or absent. Instead, Wolf-Rayet-wind ejected \Al from hydrostatic nucleosynthesis may be assumed to dominate, currently originating from Cyg~OB2 stars. In that case \Al gamma-rays from the Cygnus region could disentangle the different \Al production phases and regions within the same massive stars: In galactic-averaged analysis, one assumes a \emph{steady state} situation of \Al decay and production, such that the complete age range of stars is represented and contributes to \Al production with its time-averaged numbers of stars per age interval and their characteristic \Al ejection from either process (hydrostatic, or late-shell burning plus explosive. The age of currently-ejecting massive star groups suggests that \Al production is predominantly due to Wolf-Rayet wind ejection. 

\begin{figure}  
\centering 
\includegraphics[width=0.65\textwidth]{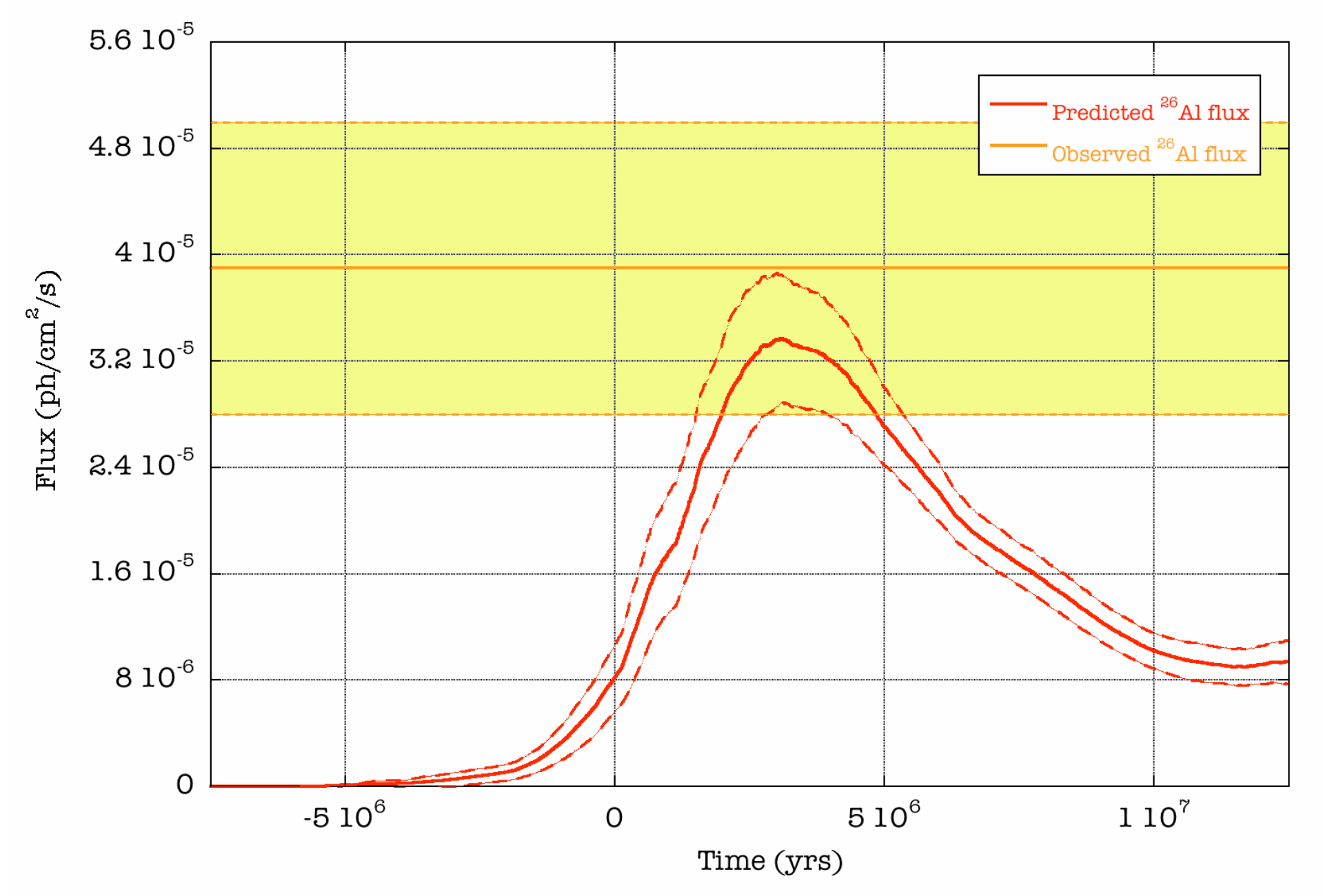}
\caption{The time history of \Al production in the Cygnus complex, as compared to the gamma-ray observations. The horizontally-shaded area presents the range given by the \Al gamma-ray data, the dashed lines bracket the uncertainty range of predictions from recent massive-star models through population synthesis. Expectations from such populations synthesis are on the low side of observed \Al gamma-rays, and predict the main \Al ejection still to come within the next \about~3~My.  Predictions are for sub-solar metallicity as appropriate for the Cygnus region itself. (see  \cite{2010A&A...511A..86M} for more details). 
}
\label{fig7:al_cygnus} 
\end{figure}

The \Al emission towards the Cygnus direction has been measured with INTEGRAL \cite{2004ESASP.552...33K,2009A&A...506..703M}. In the longitude interval [70\degree,96\degree] the total flux is $\sim$6~10$^{-5}$ph~cm$^{-2}$s$^{-1}$ from the line of sight towards the Cygnus OB associations. This has been decomposed into a large-scale galactic background, plus a contribution from the \emph{Cygnus complex} of  $\sim$3.9~10$^{-5}$ph~cm$^{-2}$s$^{-1}$ \cite{2009A&A...506..703M}. This recent discrimination of the \Al attributed to the Cygnus complex proper allows a more-specific comparison to expectations from population synthesis of the otherwise-determined stellar populations, as shown in Fig.~\ref{fig7:al_cygnus}.  From the observed stars, accumulation of the expected \Al production of massive-star groups as they evolve has always shown a problem of seeing 2--3 times more \Al gamma-rays than predicted. While much of this discrepancy was due to the occultation of stars behind molecular clouds \cite{2002A&A...390..945K}, some under-prediction ($\sim$25\% for solar metallicity) remained, though still within uncertainties of both observation data and models \cite{2010A&A...511A..86M}. But as metallicity is one of the critical factors for Wolf-Rayet-phase contributions of \Al, it is interesting that a prediction employing the metallicity value determined for the Cygnus region itself shows an under-prediction by a factor \about~4 (see Fig.~\ref{fig7:al_cygnus} at time=0) \cite{2010A&A...511A..86M}.

For a young and active region of massive-star action, one may plausibly assume that the interstellar medium would be peculiar and probably more dynamic than in a large-scale average. With the fine spectroscopic resolution of the INTEGRAL measurements, therefore initial hints for a broadened \Al gamma-ray line were tantalizing. With better data, it turns out that the \Al line seen from the Cygnus region is compatible with the laboratory energy (i.e. no bulk motion exceeding tens of km~s$^{-1}$) and with instrumental line width (i.e. no excessive Doppler broadening beyond $\sim$200~km~s$^{-1}$ \cite{2009A&A...506..703M}. Note that \Al ejection from Wolf Rayet winds would be with $\sim$1500~km~s$^{-1}$, decelerating as circum-stellar gas is swept up.

\begin{figure}  
\centering 
\includegraphics[width=0.8\textwidth]{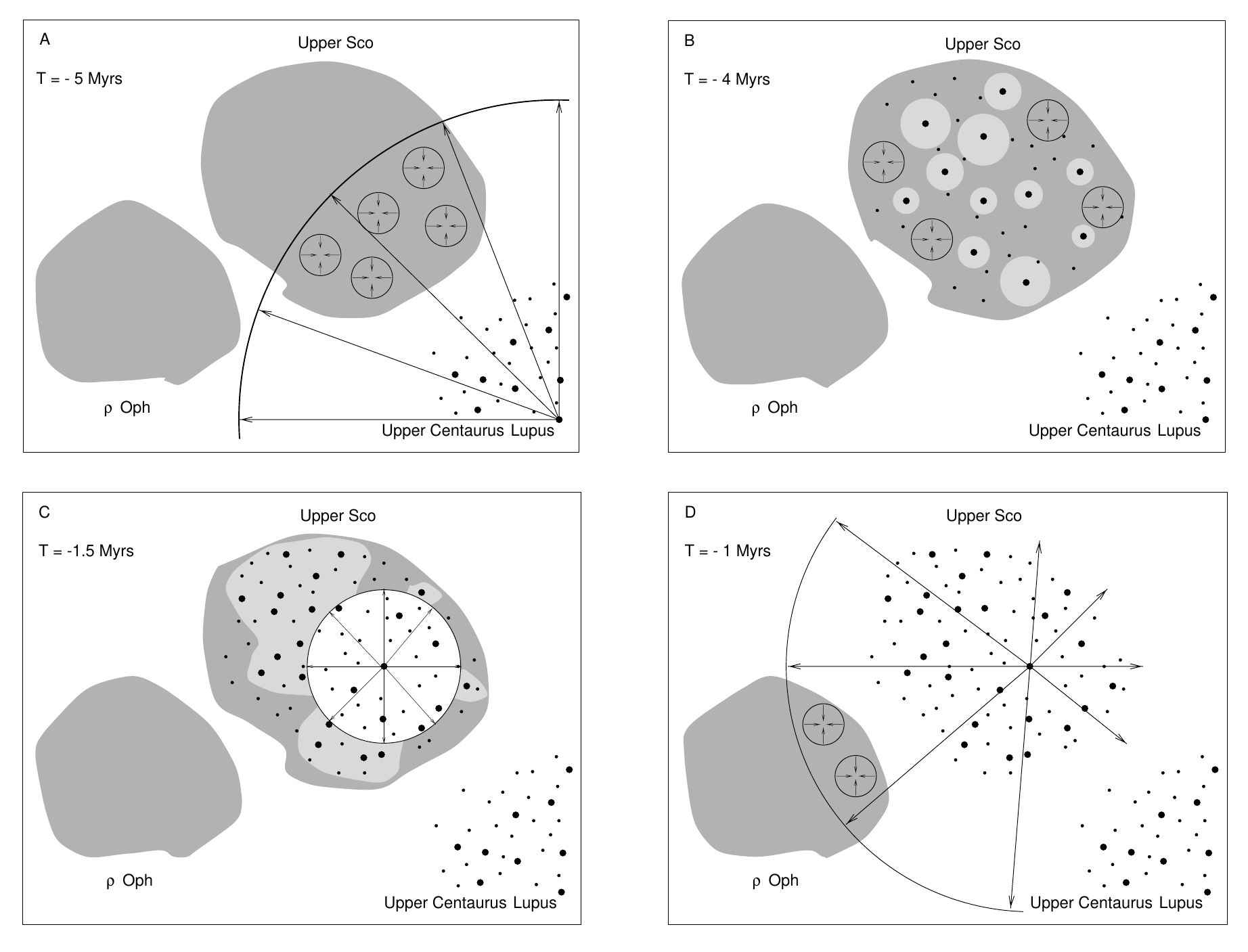}
\caption{The objects in the Sco-Cen region, as they could have evolved in a scenario of \emph{triggered star formation} (from  \cite{2002AJ....124..404P}). }
\label{fig:triggeredSF_sco-cen} 
\end{figure}   

\subsubsection*{$^{26}$Al in the Sco-Cen Region}    
The immediate Galactic vicinity of the Sun indicates several signatures of massive-star action \cite{1995SSRv...72..499F}: The Local Bubble is  a rather tenuous cavity around the Sun \cite{1998LNP...506.....B}, mapped in X-ray emission from hot gas. In addition to the Eridanus cavity, several loops/shells reminiscent of supernova remnants have been identified \cite{1992A&A...262..258D}. The stellar association of Scorpius-Centaurus (Sco OB2) and its subgroups are located at a distance of about 100--150~pc \cite{1992A&A...262..258D,1999AJ....117..354D,1999AJ....117.2381P}. Several subgroups of different ages (5, 16, and 17~My, with typical age uncertainties of 1--2 My; \cite{1989A&A...216...44D}, \cite{2008ApJ...688..377S}) have been identified. 

Stellar subgroups of different ages would result from a star forming region within a giant molecular cloud if the environmental effects of massive-star action of a first generation of stars (specifically shocks from winds and supernovae) would interact with nearby dense interstellar medium, in a scenario of  propagating or triggered star formation. Then later-generation ejecta would find the ISM pre-shaped by previous stellar generations. Such a scenario (figure~\ref{fig:triggeredSF_sco-cen}) was proposed based on the different subgroups of the Scorpius-Centaurus Association \cite{1989A&A...216...44D,2002AJ....124..404P} and the numerous stellar groups surrounding it \cite{2008hsf2.book..235P,2008A&A...480..735F}. 
This may be the kind of activity that shaped the local interstellar medium\cite{1995SSRv...72..499F,2003A&A...411..447L}, and specifically created the \emph{Local Bubble} \cite{1996SSRv...78..183B}. 

\begin{figure}
\centering
\includegraphics[width=0.48\textwidth]{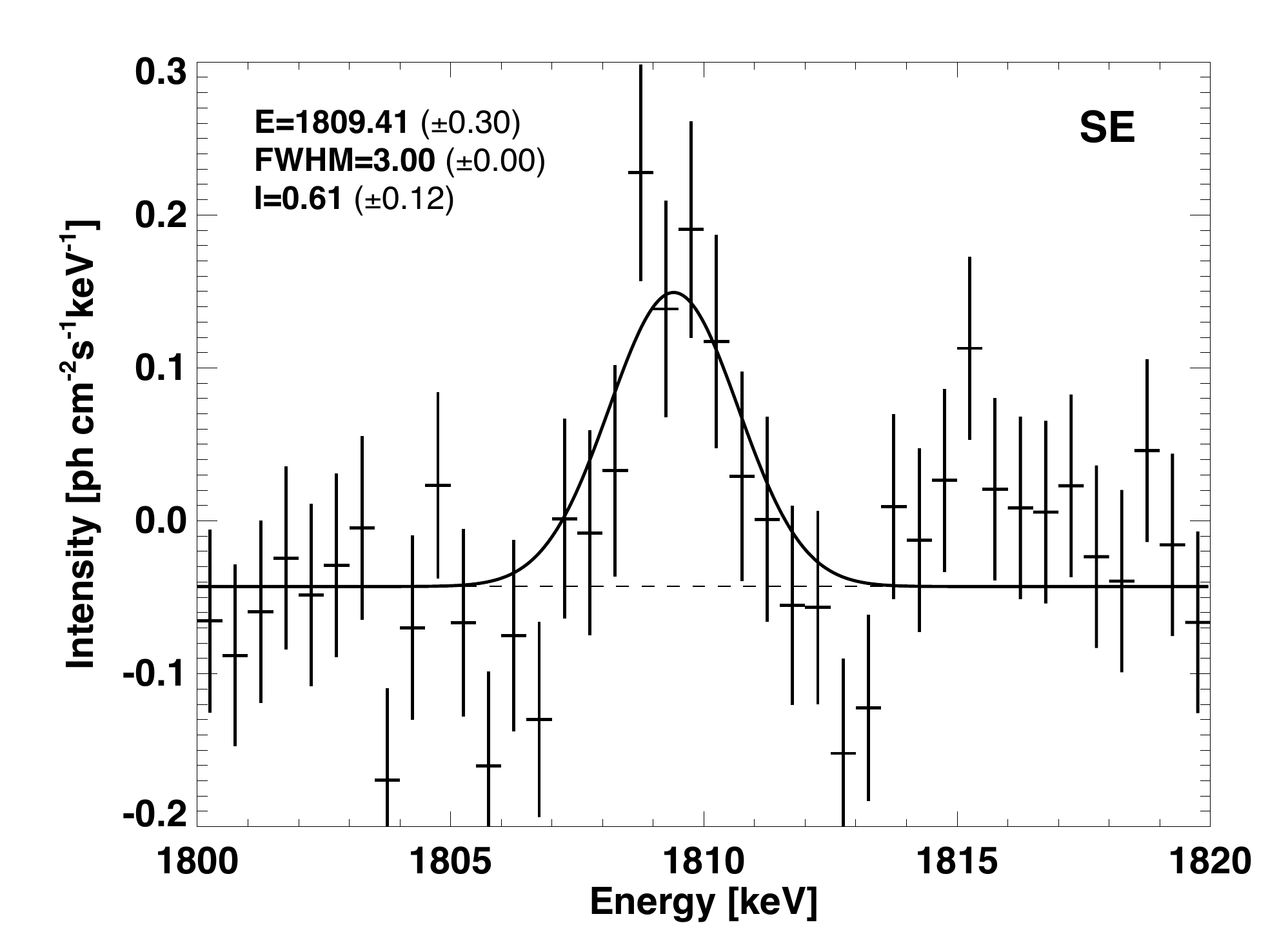}
\caption{The SPI measurement of \Al from the Scorpius-Centaurus association of stars, at nearby distance of 100--150~pc \cite{2010A&A...522A..51D}.}
\label{fig:26Al_ScoCen}
\end{figure}

With SPI, \Al emission from the direction overlapping the current location of the Upper-Sco group of stars could be discriminated against the large-scale Galactic \Al emission, due to its favorable location about 20\degree above the plane of the Galaxy (Fig.~\ref{fig:26Al_ScoCen}, \cite{2010A&A...522A..51D}). 
The \Al flux  of (6.1$\pm$0.11)~10$^{-5}$ph~cm$^{-2}$s$^{-1}$ can be converted to an observed amount of \Al, adopting a specific distance for the source; for a distance of 150~pc , a mass of  1.1~10$^{-4}$~\Msol of \Al has been derived  \cite{2010A&A...522A..51D}, consistent with another  independent  analysis \cite{2009A&A...506..703M} (who obtained a flux of (6.2$\pm$1.6)~10$^{-5}$~ph~cm$^{-2}$s$^{-1}$). 
Application of a population synthesis model for the Scorpius-Centaurus subgroups \cite{2009A&A...504..531V} showed that  an intensity in the \Al $\gamma$-ray line of 7~10$^{-5}$~ph~cm$^{-2}$s$^{-1}$ is predicted from 1.2~10$^{-4}$~\Msol of \Al, consistent with the measurement. 
This is based on the number of massive stars identified in the Upper Sco group of stars and an adopted age of 5~My. Unclear remains, however, if the adopted spatial distribution of \Al emission is appropriate for the extent of \Al dispersal, considering the proximity of the sources and the complex structure of the local interstellar medium  \cite{2009SSRv..146..235F,2009SSRv..143..241W}. Possibly, the apparent agreement between SPI measured and stellar-population predicted flux is fortuitous.

The \Al line from  the Scorpius-Centaurus region source may be slightly  blue-shifted:  a centroid energy of 1809.46~keV ($\pm$0.48 keV) has been found, and  implies a blue shift of $\sim$0.8~keV corresponding to bulk streaming towards the Sun at about (137$\pm$75)~km~s$^{-1}$. 
Interestingly, measurements of hot gas in the solar cavity also suggested gas inside the local cavity streaming from this general direction towards the Sun \cite{2009SSRv..146..235F}.

\subsubsection*{$^{26}$Al in the Orion and Carina Regions.}    
The Orion region is the most-nearby region of massive stars, at a distance of $\sim$450~pc \cite{2008hsf1.book..459B,1989ARA&A..27...41G}. Its location towards the outer Galaxy and at Galactic latitudes around 20\degree is favorable, as potential confusion from other Galactic sources is negligible. The dominating group of massive stars is the Orion OB1 association \cite{1994A&A...289..101B} with three major subgroups of different ages, one oldest subgroup \emph{a} at 8--12~My, and two possibly coeval subgroups \emph{b} (5--8~My) and \emph{c} (2--6~My); subgroup \emph{d} is the smallest and youngest at 1~My or below. Subgroup c holds the most massive stars, about 45 in the mass range 4--120~\Msol. These groups are located on the near side of the Orion A and B molecular clouds, which extend from 320 to 500~pc distance away from the Sun, and span a region of $\sim$120~pc perpendicular to our viewing direction.  

Since at least two subgroups of the Orion OB1 association are in the age range where one expects ejection of \Al (see Fig.~\ref{fig:popsyn_26Al60Fe}), it is plausible to search for \Al gamma-rays: at 400~pc, the output from ten massive stars at 10$^{-4}$~\Msol each would lead to a gamma-ray flux of 7.5~10$^{-5}$ph~cm$^{-2}$s$^{-1}$, which should be detectable with current telescopes. 

But with the COMPTEL imaging telescope, only faint hints for emission in the wider Orion region were found, at low surface brightness and apparently only at the level of typical noise. Upon a closer inspection, however, a clear line at 1.8~MeV seen in spectra from those areas of the sky (typically not seen for other low-brightness regions in the map) was encouraging, and a model fit found the total signal to be significant . 

 \Al emission was suggested from measurements with COMPTEL (a gamma-ray flux of 7~10$^{-5}$ph~cm$^{-2}$s$^{-1}$ was reported at  5$\sigma$ significance, \cite{2002NewAR..46..547D}), and appears quite extended and not concentrated near the OB1 association. As a huge interstellar cavity extends from the Orion molecular clouds towards the Sun,  which extends over almost 300~pc  as seen in X-ray emission \cite{1993ApJ...406...97B}, Orion OB association stars could be responsible for creating this cavity. Then,  \Al ejected from current-generation stars would find a pre-shaped cavity directing the flow of ejecta into it, rather than towards the dense remains of the parental molecular clouds on the far side of the OB association.
With INTEGRAL's spectrometer interesting measurements would be the centroid (bulk motion towards the Sun?) and width (\Al fractions moving within the hot cavity versus \Al deposited at the cavity walls?). With limited (\about~2Ms) exposure, no signal has been found yet. This is however not surprising, in particular since the line may be broadened;  more exposure would be needed to study this proposed scenario. 

From the \emph{Carina} region, COMPTEL data had shown a suggestive bright emission spot in their image \cite{1996A&AS..120C.327K}. INTEGRAL observations could not confirm the reported brightness, and finds only \about~1/2 of this at (1.5~$\pm$1.0)~10$^{-5}$~ph~cm$^{-2}$s$^{-1}$ from this region. Accounting for candidate massive-star groups in the region, even less \Al brightness is expected \cite{2012A&A...539A..66V}. Interestingly, this expected \Al should originate mostly from stellar winds of Wolf-Rayet stars rather than from supernovae. Hence, similar to the Cygnus region, one may be able to probe the main-sequence hydrogen-burning part of \Al synthesis through the specific and young age range of stars in Carina, with deeper observations of \Al emission.

\begin{figure}
\centering
\includegraphics[width=0.68\textwidth]{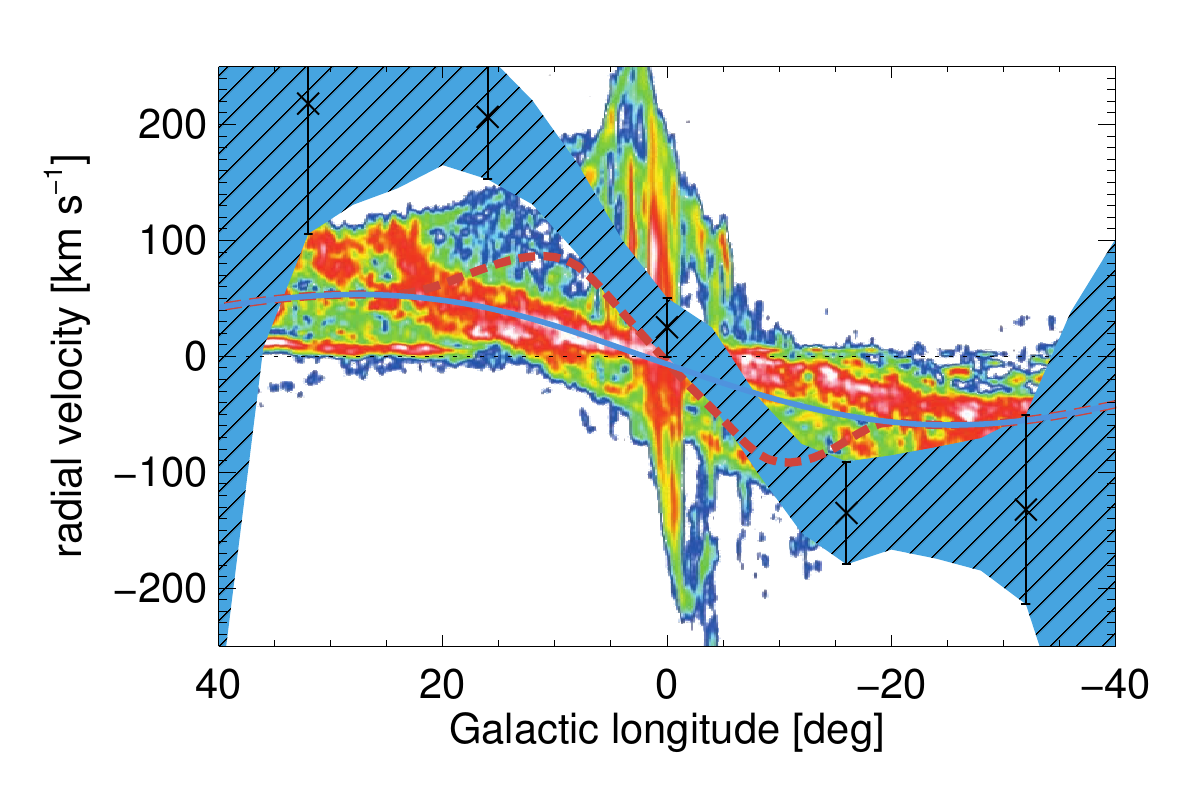}
\caption{The centroid of the $^{26}$Al line appears systematically Doppler-shifted along the inner Galaxy, as plausible from large-scale Galactic rotation. Note that data points as shown are not independent, but chosen to include sufficient $^{26}$Al signal and to maximize the longitudinal signature of $^{26}$Al line shifts (see \cite{2011SSRv..161..149W} and Kretschmar et al., in preparation (2012)).}
\label{fig:26Al_long-vel}
\end{figure}

\subsubsection*{The Inner Galaxy and Bar}
Along the inner Galaxy, the uniquely-high spectral resolution of SPI allows an interesting application: A kinematic shift of the $^{26}$Al line from the Doppler effect and due to large-scale rotation of sources about the center of the Galaxy could be recognized \cite{2006Natur.439...45D}.  This opens the possibility to study motion of hot interstellar gas as it is shaped around massive-star regions, and compare its dynamics to dynamics of stars and gas as we know it otherwise. For example, CO line emission has been used to trace the molecular gas in the Galaxy \cite{2001ApJ...547..792D}. Here, the Galactic ridge was recognized clearly, with peculiar and high cloud dynamics in the vicinity of the Galaxy's central supermassive black hole. As star formation occurs from dense clumps in molecular clouds, \Al ejected from recently-formed stars could show how closely its kinematic motion can be related to the current molecular cloud population.  

The velocity resolution obtainable from $^{26}$Al gamma-rays now has reached 100~km~s$^{-1}$, thereby becoming astrophysical significant for tracing Galactic gas streams, or potentially constraining turbulence within the wind- and supernova-blown  cavities of the interstellar medium.
The inner Galaxy $^{26}$Al measurements accumulated over \about~5 years have been analyzed in a way which aims to extract the signature of such large-scale motion. As sufficient \Al signal is required to determine the line centroid, longitude intervals between 6 and 20\degree\  have been chosen, and separately fitted from the Galaxy's large-scale remaining part of the sky, one by one; moving the selected longitude stretch across the Galactic plane in longitude, the line centroid shift can be traced. From overlapping longitude bins of smallest extent with weakest signals, to larger and independent longitude ranges, thus a longitude-velocity diagram has been derived as shown in Fig.~\ref{fig:26Al_long-vel}. Here, the blue-hatched area shows the range of uncertainty resulting from different choices of longitude bins in this analysis. The derived velocities extend from \about~-200 to \about~+200~km~s$^{-1}$ at the extremes (tangential directions to Galactic-ridge directions with largest relative motion with respect to the Sun, at longitudes $\pm$30\degree. 

It is apparent that the \Al velocities extend over higher velocities on large scale than known for other sources and in particular for cold, molecular gas (also shown in Fig.~\ref{fig:26Al_long-vel} in color). This appears surprising. If we consider \Al sources across the Galaxy traced by free electrons produced by ionizing starlight, which has been mapped through pulsar dispersion measurements, we would expect a longitude-velocity trend as shown by the thick (blue) line in Fig.~\ref{fig:26Al_long-vel}. But this representation of free electrons according to \cite{2002astro.ph..7156C} is deficient in the inner Galaxy, from an observational bias of available pulsar data. Thus it may not be surprising that this apparently does not represent \Al observations of motions, providing a hint that some inner-Galaxy sources may be missing in this free-electron model. Adding an additional \Al source population also within a Galactocentric radius of 2~kpc (where the free-electron map is \about~empty), and applying a density structure according to the Galaxy's inner bar \cite{2005ApJ...630L.149B} plus rotational behavior as extrapolated towards the inner Galaxy from available measurements \cite{2009PASJ...61..227S}, the \Al kinematic data are more closely reproduced (Kretschmer et al., submitted).  The generally-larger velocities seen in \Al sources suggest that preference towards the direction of Galactic rotation may be given to massive-star ejecta as they leave their sources. It remains to be shown if more detailed modeling of star formation along the Galaxy's bar and inner spiral arms  can provide an explanation of the observed \Al velocities.

\subsection{Novae}\label{sources_novae}
Nova nucleosynthesis is characterized by explosive hydrogen burning, according to the classical model for novae whereby nuclear ignition of the accreted surface layer  on a white dwarf causes this phenomenon. As the ignited envelope expands and cools, nuclear burning only characterizes a short, initial period, typically of duration of seconds. Nuclear burning occurs at temperatures around 2~10$^8$K, and processes seed nuclei dominated by accreted hydrogen and He, C, and O admixed from the underlying white dwarf. Nuclear-burning ashes should be enriched in unstable isotopes, preferentially on the proton-rich side of the isotopic valley of stability, and thus should lead to characteristic gamma-ray signals  \cite{1998ApJ...494..680J,2000MNRAS.319..350J}. Recurrent novae, which may occur on more-massive white dwarfs accreting at a higher rate, only process small amounts of material in their thermonuclear explosion; classical novae accrete total amounts of 10$^{-4}$-10$^{-5}$~\Msol at very slow rate of order 10$^{-9}$~\Msol~y$^{-1}$, thus are more relevant both for galactic nucleosynthesis and as potential gamma-ray sources. 

The major uncertainties in current nova models are the amount of admixed white-dwarf material, and the total amount of ejected material. 
Classes of novae are distinguished, both observationally and in view of plausible progenitors: \emph{Fast} novae decline in luminosity on the order of days, and \emph{slow} novae may take hundred or more days to do so; underlying white dwarf compositions may differ, depending on the mass and evolutionary history of the progenitor star, from He (M$_{initial}<$0.5~\Msol) through C/O (0.5~\Msol$<$M$_{initial}<$8-9~\Msol) and O/Ne/Mg (M$_{initial}>$8-9~\Msol up to the maximum mass avoiding core-collapse, probably 10--12~\Msol) stars. 
Nova nucleosynthesis proceeds by proton captures and successive $\beta^+$-decays through the hot CNO cycle, with break-out from the cycle towards production of heavier isotopes depending on the duration, temperature, and availability of seed nuclei. Novae from CO white dwarf stars are more frequent, and are thought to be significant producers of isotopes such as $^{18}$F. 
Short-lived radioactivities from several isotopes expected to be produced in explosive H burning typically undergo $\beta^+$~decay, thus leading to positron release and a corresponding, bright gamma-ray line at 511~keV from positron annihilation \cite{1998ApJ...494..680J}. The dominating unstable isotopes expected to be produced are $^{13}$N and $^{18}$F, and expected brightness of such gamma-ray line emission would be at the level of 10$^{-3}$ph~cm$^{-2}$s$^{-1}$ or above for a nova 1~kpc away, which should be detectable easily with gamma-ray telescopes such as SPI on INTEGRAL. However, this \emph{annihilation flash} would occur reveal days before the nova achieves its optical peak brightness, and thus would precede the detection and identification of the nova as a celestial source. Therefore, post-identification searches for serendipitous exposures of such novae are the way to measure this annihilation flash.
 
About 1/3 of all novae may be due to the more-massive and more-evolved white dwarf progenitors, which are enriched in heavier seed nuclei such as Na, Mg, and Ne. These novae have been found to eject significant amounts of heavier elements up to S. Therefore, it appears plausible that H burning also passes through the Ne-Na cycle and leaves behind significant amounts of $^{22}$Na, a radioactive isotope with decay time of 3.8 years, and thus a candidate emitter of gamma-rays at 1274.53 and 511~keV. 
The variety of gamma-ray signals which could arise from novae of the CO and O-Ne types have been summarized in \cite{2004NewAR..48...35H}.

\begin{figure}
\centering
\includegraphics[width=0.78\textwidth]{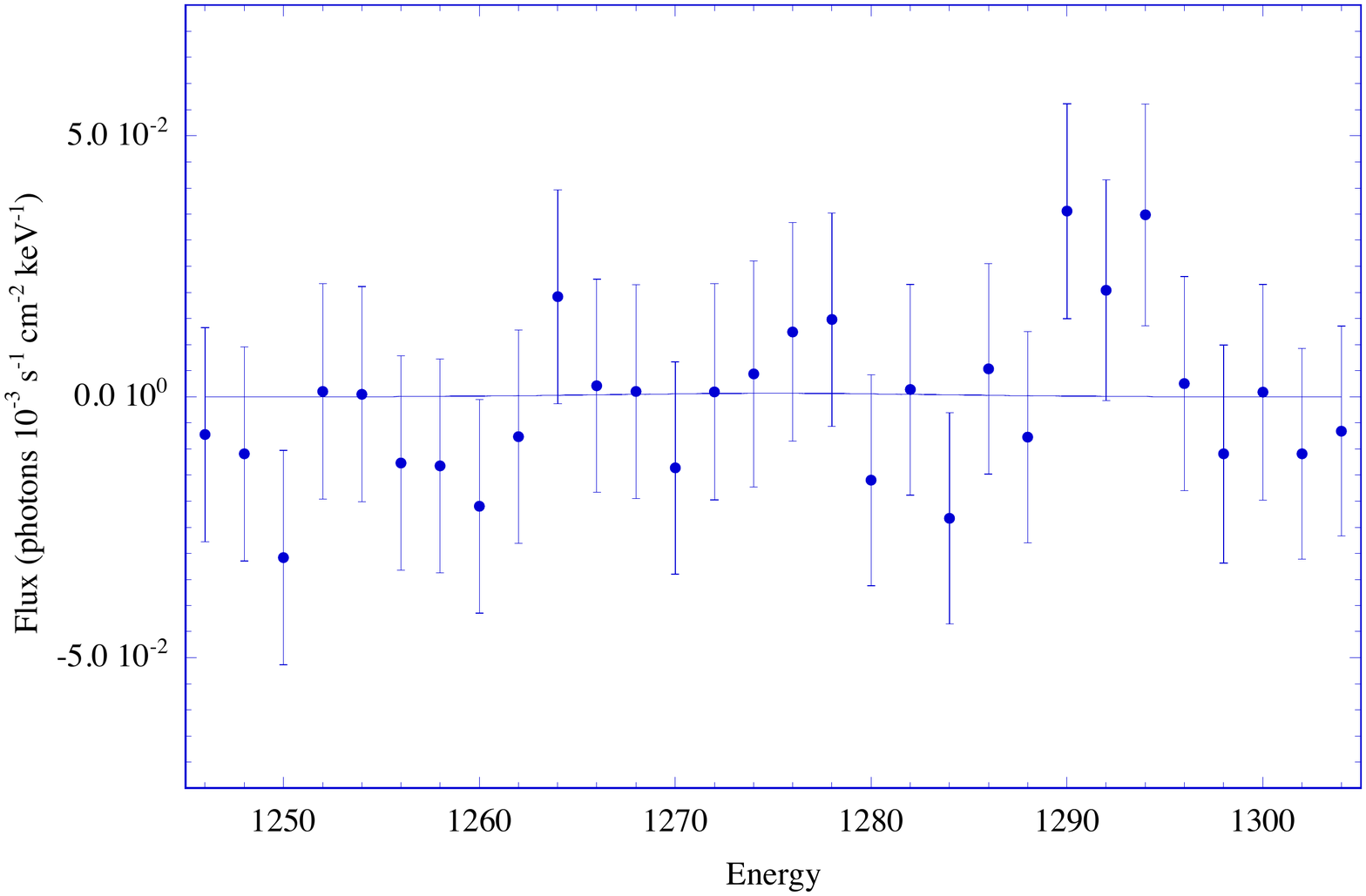}
\caption{SPI data from 3 years of data have been systematically searched  for a signal from the galactic distribution of novae \cite{2004ESASP.552..119J}. Here a spectrum shows the result in the vicinity of the $^{22}$Na line, derived testing a spatial source distribution as plausible for nova events \cite{1991ApJ...378..131K}. The fitted line at 1275~keV can hardly be seen, and corresponds to an (insignificant; 1$\sigma$) flux of 1.3~10$^{-5}$ph~cm$^{-2}$s$^{-1}$ (P. Jean, private communication).}
\label{fig:22Na_SPI}
\end{figure}

Even with several gamma-ray missions in operation, none of the searches performed so far (SMM, TGRS, COMPTEL, INTEGRAL) have resulted in any positive detection of any of the gamma-ray lines expected from novae; neither annihilation, nor $^{22}$Na nor $^7$Be decays could be reported. Several candidate novae have been targeted, and some hints seemed promising; but gamma-ray telescopes all suffer from intense activation and instrumental background in both the 511~keV and more so in the 1275~keV part of the measured spectrum.

The most sensitive study searching for $^{22}$Na emission from novae was done by \cite{2004ESASP.552..119J}. Here, the sky distribution was modeled from the Galactic nova distribution assumed to follow the 2.4~$\mu$m dust emission in the Galaxy \cite{1991ApJ...378..131K}, and was fitted to data from two years of SPI observations along the plane of the Galaxy, with 1.8~Ms of total exposure. Only an insignificant hint of the expected line at 1275 keV could be seen (figure~\ref{fig:22Na_SPI}). For the spatial model by  \cite{1991ApJ...378..131K}, which includes both a spheroidal and a significant disk population of novae, a 2$\sigma$ limit of 2.5~10$^{-4}$ph~cm$^{-2}$s$^{-1}$ was obtained. Translating this into a limit for the nova population attributed to this spatial distribution, adopting a nova rate of 20...40 per year and an O-Ne-Mg enriched nova fraction of 13...33\%, an average ejected-mass limit of 2.5--5.7~10$^{-7}$\Msol was obtained with INTEGRAL. This limit is still based on 3 years of data only, and thus a factor 3 above what could be obtained from COMPTEL; deeper SPI analysis is pending.

It appears that nova nucleosynthesis, while it certainly occurs, is beyond the reach of current instruments by at least an order of magnitude. Even the fortunate event of a nearby nova at few 100~pc distance, while confirming the general picture of nuclear runaway driven nova explosions, would still leave us with major uncertainty on how the path of nucleosynthesis in the p-rich part of the table of isotope really proceeds upward towards observed elemental enrichments such as Ca and S.

\subsection{Positron Annihilation}\label{positrons}
Nuclei such as the above-discussed $^{26}$Al, $^{44}$Ti, and $^{56}$Ni isotopes decay through $\beta^+$-decays, thus releasing positrons. These positrons eventually annihilate with electrons, converting the rest mass of the particle-antiparticle pair of 1.022~MeV into radiation. Momentum conservation requires emission of two or more photons in this annihilation, producing a bright and unique gamma-ray line at 511~keV. 
This line was the first cosmic gamma-ray line ever detected, first in 1972 with a low-resolution NaI detector instrument \cite{1972ApJ...172L...1J}, and later identified as annihilation line by a Ge detector measurement \cite{1978ApJ...225L..11L}. 

For decades apparently time-variable annihilation emission was pursued \cite{1982ApJ...260L...1L,1986ApJ...302..459L}. But it was found to appear as different instruments with different field-of-view sizes recorded different fractions of the diffuse galactic annihilation emission, once sufficiently sensitive instruments such as the SMM Gamma-Ray Spectrometer or OSSE on the Compton Observatory were able to observe large parts of the sky and for extended periods of time \cite{1990ApJ...358L..45S,1993ApJ...413L..85P}.

INTEGRAL has measured positron annihilation gamma-rays across the sky in great detail, confirming this diffuse nature of annihilation across our Galaxy \cite{2005A&A...441..513K}. The scientific surprise already apparent in OSSE results \cite{1997ApJ...491..725P} was consolidated by SPI images, and finds that the annihilation emission predominantly arises in the inner Galaxy in an extended region of size \about~10\degree. By comparison, the disk of the Galaxy is much fainter, wit a bulge-to-disk intensity ratio of 1.4 from a total luminosity of \about~2~10$^{-3}$ph~cm$^{-2}$s$^{-1}$ \cite{2005A&A...441..513K,2008Natur.451..159W}. 

\begin{figure}
\centering
\includegraphics[width=0.68\textwidth]{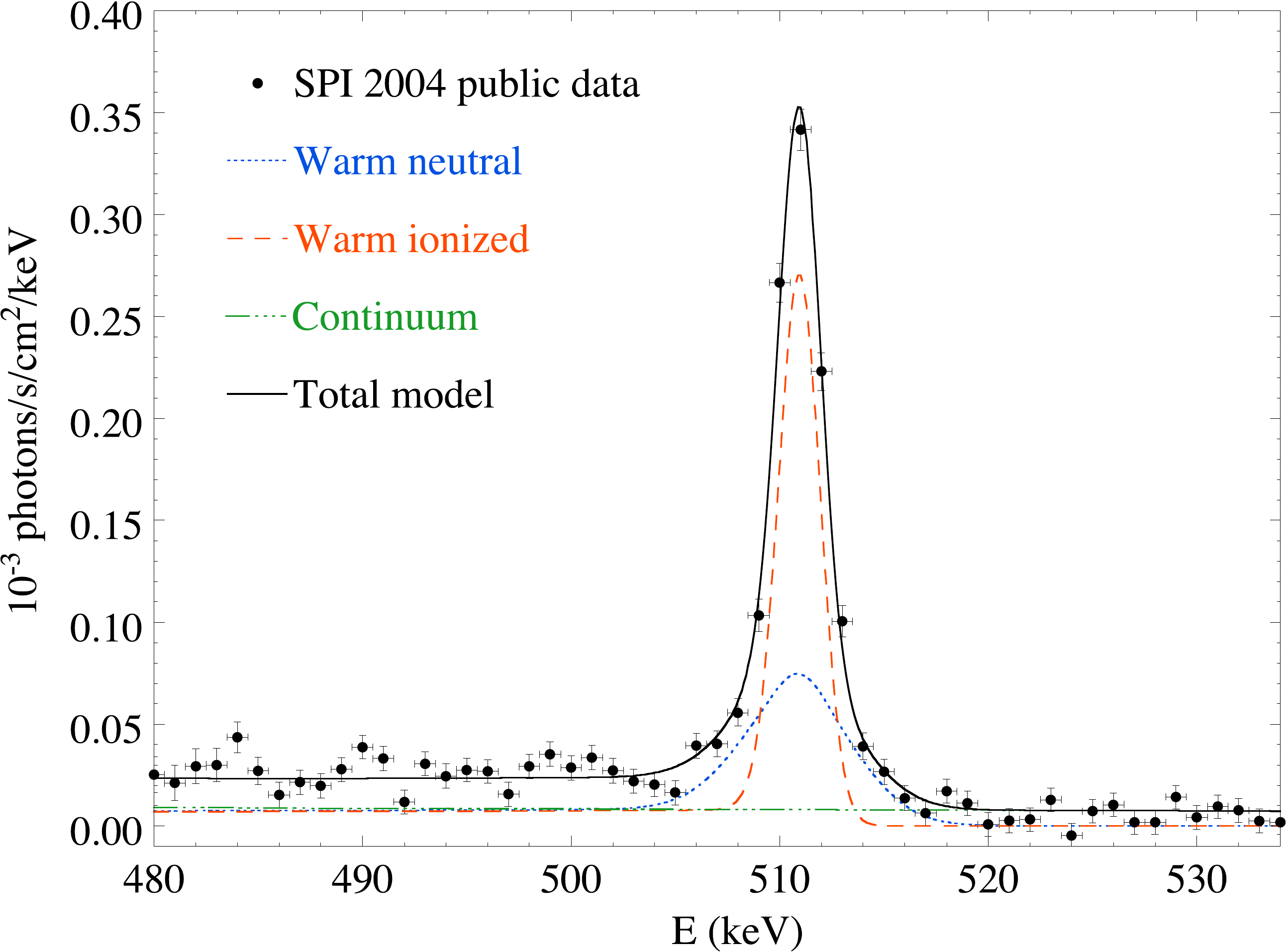}
\caption{The SPI spectrum of emission from positron annihilation in the Galaxy. The 511 keV line appears only moderately broadened, and, together with the high positronium fraction derived from the annihilation continuum $<$500~keV, supports annihilation to occur on average in moderately hot and only partially-ionized gas.}
\label{fig:511spectrum_SPI}
\end{figure}

The spectral shape of emission from positron annihilation conveys information about the conditions in the annihilation site: Doppler broadening is expected to increase the line width, when the electrons carry substantial velocities at the time of annihilation. This can either occur for direct annihilation with electrons of high temperatures, but, more importantly when charge exchange reactions with interstellar hydrogen occur on energized hydrogen atoms. But when annihilation occurs on electrons or hydrogen in cold, neutral interstellar matter, lines should be narrow with negligible broadening below \about~1~keV, as observed: SPI measurements find the main line component to be 1.3~keV widened beyond the instrumental resolution of 2.2~keV  \cite{2006A&A...445..579J}. Annihilation on interstellar grains would lead to an even narrower line, as momentum is mainly taken over by the massive grain at interstellar temperatures of \about~tens of K at most. This seems to be unimportant in view of the measurements. But neutral gas appears to be an important ingredient of annihilation sites. On average, annihilation environments are only moderately ionized (percent level), and moderately hot at \about~8000~K; this suggests that annihilation does not occur in hot interstellar medium such as characteristic for the environments of most candidate sources, but rather in the outer boundary layers of molecular clouds. Annihilation of positrons mostly involves formation of an atomic-state configuration of the positron with an electron, the \emph{positronium} atom. This occurs in two spin configurations, hence a singlet and a triplet state. The latter can only lead to annihilation emitting more than two photons, from momentum balance, and thus leads to a characteristic three-photon continuum spectrum that tells us about the importance of this positronium triplet state in the variety of positron annihilation processes. Neutral-gas annihilation environment thus would be expected to have positronium fractions 
above 0.9, and the SPI value of 0.95 supports that annihilation typically occurs in partially-neutral gas environments. Propagation of positrons from their sources to their annihilation sites seems a characteristic property, and thus makes any connections of annihilation gamma-rays to positron sources somewhat indirect.  

There is a variety of candidate sources of cosmic positrons \cite{2011RvMP...83.1001P}: Beyond radioactive decays, accreting compact sources are expected to produce jets of pair plasma, similarly the high magnetic field in pulsars leads to electromagnetic cascades involving pair creation. High-energy collisions above the pair threshold will always create pairs of electrons and positrons, so also in cosmic-ray interactions in interstellar space, possibly enhanced when cosmic rays interact with denser material in interstellar clouds. Finally, dark matter may decay, annihilate, or scatter, and several branches of these interaction sequences involve the release of positrons. 

This latter source, unlike all the others, is different in nature: Dark-matter particles are expected to fill a much larger volume of scale 100~kpc, and positron production should be extended and smooth compared to sources related to stellar objects, which always will have some sort of clumping or spatial variations according to locations of their parent objects. 
For this reason, the remarkably-symmetric emission of annihilation emission immediately attracted such interpretation as dark matter signature. 
The morphology of the emission can, in principle, be reproduced through annihilation or scattering of dark matter particles, as source intensity scales with density$^2$ and thus steepens in the inner Galaxy \cite{2012arXiv1201.0997V}. The rate of such dark matter interactions producing positrons should be near 3~10$^{-26}$cm$^3$s$^{-1}$ to obtain the correct relic density while not causing an observable signature in cosmic microwave background; both WIMP and MeV-type dark matter interactions are sufficiently uncertain to satisfy these constraints at present.

\begin{figure}
\centering
\includegraphics[width=0.48\textwidth]{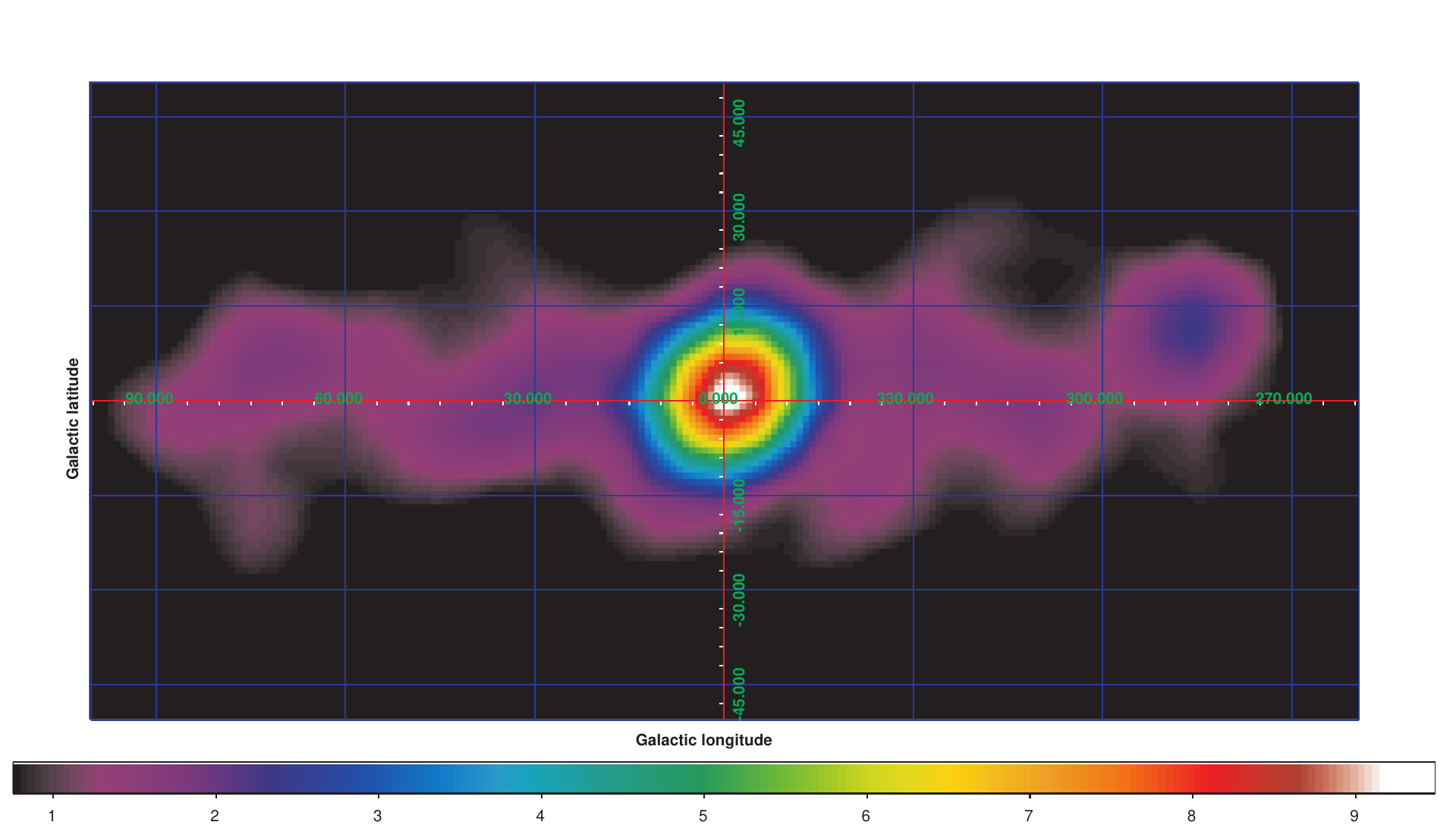}
\includegraphics[width=0.48\textwidth]{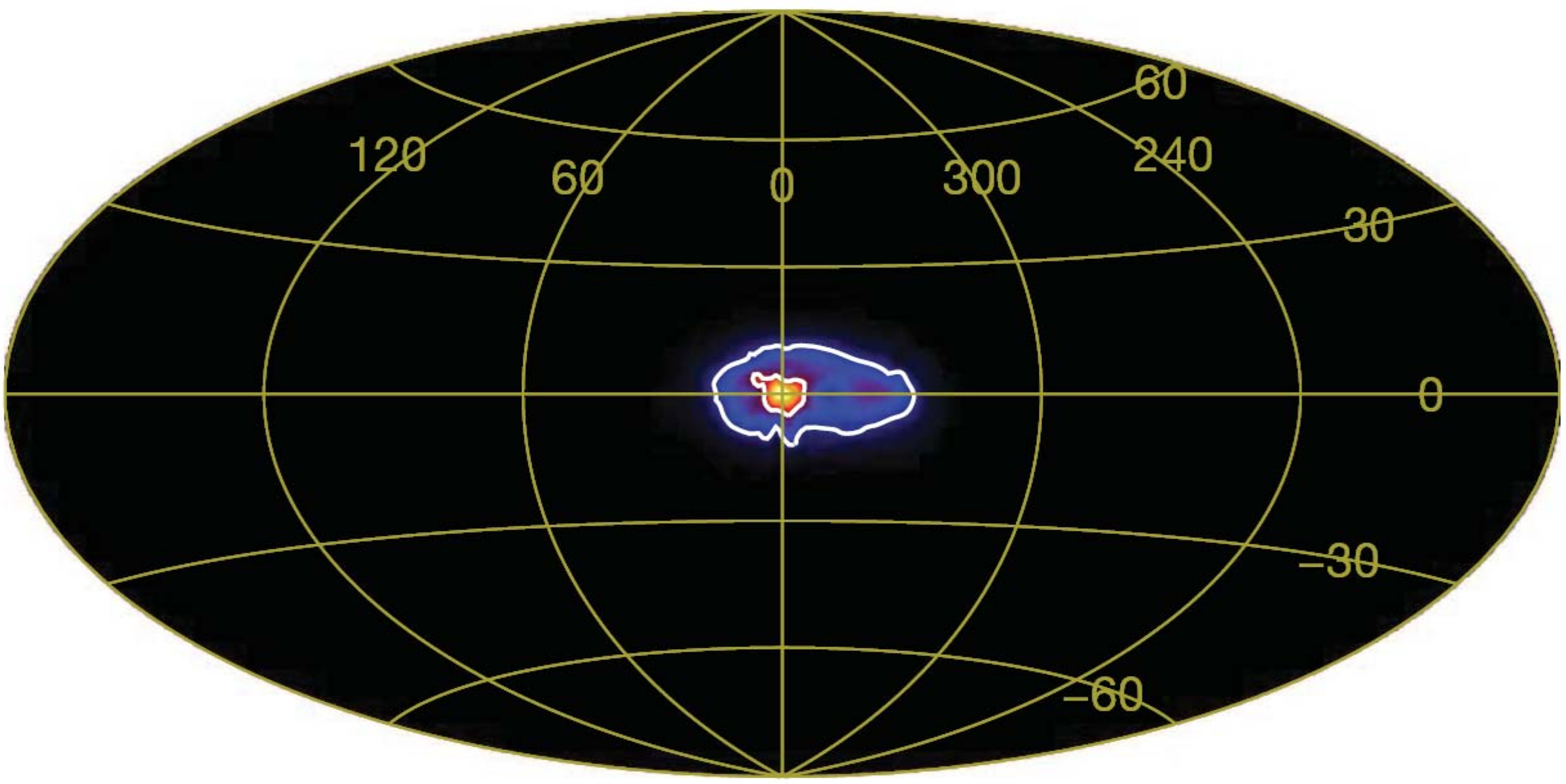}
\caption{The SPI images of Galactic annihilation emission show a dominating bright bulge-like (extended) emission centered in the Galaxy, with weak disk emission. \emph{left:} Significance map of a pixelized maximum-likelihood imaging deconvolution \cite{2010ApJ...720.1772B}.
\emph{right:} Intensity map of a multi-resolution expectation maximization imaging deconvolution \cite{2008Natur.451..159W}.
The centroid of the bulge emission may hold clues to the nature of the sources in the bulge, relating a small offset or an asymmetric inner-disk component (depending on analysis approach; see text) to spatial distributions as expected from stellar or, alternatively, dark-matter related sources.}
\label{fig:annihilation_image_SPI}
\end{figure}

SPI images (figure~\ref{fig:annihilation_image_SPI}) include hints for deviations from the ideal symmetry expected for dark-matter origins of the bulge emission. Asymmetric disk emission was reported, and interpreted as possibly related to a positron source population in X-ray binaries \cite{2008Natur.451..159W}. Microquasars are a subclass of such binaries, where accretion has been found to launch jets, which most likely include pair emission. But as the population of binaries and in particular micro quasars extends throughout the disk of the Galaxy, this interpretation is seen with some reservations. 

It turns out that the reported asymmetry can also be the result of an offset of the central bright emission in longitude by -0.6\degree, rather than originating in the disk itself \cite{2010ApJ...720.1772B}.  That may disfavor dark-matter interpretations at first glance, which should be centered on the Galaxy's gravitational potential only. Thus, models relating the mass-accretion activity of the supermassive black hole to positron production were pursued (e.g. \cite{2006ApJ...645.1138C}). 
Complementing these, the hints for past increased activity in our Galaxy's center derived from surprisingly-bright reflected off interstellar clouds led to speculations of a starburst creating a large population of massive stars and supernovae in the central few hundred parsecs of our Galaxy. The large-scale \emph{bubbles} extending into the Galaxy's halo and recently discovered in Fermi gamma-ray data may also be the result of blow-out from the Galaxy's disk towards the halo of such increased wind and supernova energy released some time ago  \cite{2010ApJ...724.1044S}. Then, the same nucleosynthesis positron sources that explain the current annihilation emission throughout the Galactic disk would also have filled the inner Galaxy with positrons, and the current annihilation gamma-ray brightness  may be the positron annihilation afterglow due to the long lifetime of positrons in interstellar space.

The annihilation emission of the disk could only be seen with INTEGRAL measurements when more than four years of data were integrated. Its surface brightness is low, and currently only identified in the inner ridge, within longitudes $\leq$50\degree. Different approaches have been taken to trace the annihilation gamma-ray signature from expected sources. Only $^{26}$Al is observationally established as a positron source, through measurement of its gamma-ray emission at 1809~keV (see above). But plausibly, the massive star groups producing such $^{26}$Al will also produce, though on different time scales, radioactivities in $^{44}$Ti and $^{56}$Ni, at roughly the same locations. Hence the $^{26}$Al gamma-ray emission map can be considered an observationally-founded tracer of positron sources across the Galaxy from nucleosynthesis. Even for high-mass binaries, pulsars, and microquasars, this spatial source distribution may be considered a plausible first-order approximation of a spatial model, considering that source catalogues for these objects suffer from major incompleteness biases. 

But positron annihilation itself is a complex process, as all sources release positrons with relativistic energies, \about~MeV for nucleosynthesis, and up to GeV energies for other candidate sources. The propagation and transport of such energetic positrons in interstellar space \cite{2009A&A...508.1099J,2011RvMP...83.1001P} is driven by the configuration of interstellar magnetic fields. These are expected to be rather irregular within the spiral arms in the disk of the Galaxy, but probably more regular between arms and at higher latitudes towards the halo. Path lengths between tens of pc and kpc are possible for positrons, depending on assumptions about energies and magnetic-field configurations. 

Positrons from nucleosynthesis as we know it to occur throughout the Galaxy (from $^{26}$Al, but also molecular-gas distributions, pulsars, supernova remnants, and massive stars) have been followed in such propagation studies, and have been related to the observed annihilation gamma-ray emission. 
While one approach finds that the entire emission with faint disk and bright bulge component could be explained by nucleosynthesis positrons alone \cite{2009ApJ...698..350H}, another study \cite{2012A&A...543A...3M} concludes that this cannot be the case, and nucleosynthesis positrons most likely only explain the disk component of the annihilation gamma-rays. Then, another component would be needed, but is not really obvious from astronomical data. Speculations discuss extreme positron propagation from disk through halo into the central bulge \cite{2006NewAR..50..553P}, or special (and past) activity around the central supermassive black hole \cite{2006ApJ...645.1138C}, or the dark matter signatures discussed above. Understanding positron propagation in the bulge region \cite{2009A&A...508.1099J}, and refined mapping of the annihilation emission morphology need to be combined towards discrimination of these alternatives \cite{2011RvMP...83.1001P}.

\section{Cosmic High-Energy Sources and Nuclear Emission}
\subsection{Low-Energy Cosmic Ray Interactions}\label{diagnostics_cr}
INTEGRAL/SPI data analysis for the inner Galaxy region with an extent of \about~60\degree covering the Galactic ridge and molecular-ring regions did not result in any nuclear-line signal such as expected from cosmic-ray interactions. A diffuse, power-law type continuum emission from this region is detected up to \about~2.5~MeV and with a power law slope of 1.79, above which only the \Al and \Fe lines from radioactive decays (and positron annihilation at and below 511~keV) are detected (see Fig.~\ref{fig:galridge_spec} \cite{2011ApJ...739...29B}. Beyond, at higher energies, the Galactic ridge, and in particular diffuse and extended line emission such as expected from $^{12}$C and $^{16}$O at 4.43 and 6.1~MeV, respectively, have not been found with INTEGRAL.

With respect to point sources, a nuclear line de-excitation spectrum  as predicted \cite{2011A&A...533A..13S} for the most-nearby candidate source, the Cas A supernova remnant, shows rather broad lines from C and O at 4.4 and 6.1~MeV. Estimated fluxes fall close to the sensitivity of COMPTEL, but due to the large line width these are well below INTEGRAL's capabilities.

The capture of neutrons on hydrogen is expected to produce a line at 2.223~MeV. Such line emission could arise from the surface of compact stars (white dwarfs, neutron stars), in particular in accreting binary systems. Line shape information could be valuable, as orbiting material in accretion disks could be identified, or gravitational redshifts be exploited wrt. neutron star accretion geometry. A search for this line and corresponding point sources was negative, and only provided upper limits\cite{2004NuPhS.132..396G}. This is in line with expectations for the candidate sources and their plausible accretion rates.

\subsection{Our Sun and Solar Flares}\label{sources_sun}
According to the standard model for solar-flare particle acceleration \cite{2001ApJ...548..492S}, reconnection events of magnetic field lines in the upper solar corona set up some kind of particle-acceleration environment, at significant altitude above the solar chromosphere. It remains to be understood which type of acceleration process occurs. Candidate processes are electrical fields, shock acceleration according to the Fermi mechanism, or stochastic acceleration from Alfven waves \cite{2005AdSpR..35.1825M}. It seems clear that the acceleration energy derives from the magnetic field, and that changes in its configuration set up the critical conditions for relativistic particle acceleration. As a result, protons, nuclei, and electrons are accelerated towards the deeper layers of the solar chromosphere, and produce a variety of observational signatures once interacting with denser gas towards the solar photosphere, at typical gas densities 10$^{14}$cm$^{-3}$ or above. Among those signatures, electron Bremsstrahlung produces continuum emission, nuclear excitations produce a variety of lines, most prominently $^{20}$Ne, $^{12}$C, and $^{16}$O de-excitation at 1.634, 4.439, and 6.129~MeV, respectively, and spallation-produced neutrons capture on hydrogen and produce a characteristic line at 2.223~MeV. 

Most-detailed gamma-ray spectra had been measured for the 4 Jun 1991 solar flare with the OSSE spectrometer on CGRO, confirming these spectral characteristics \cite{1997ApJ...490..883M}. The modest spectral resolution of OSSE (\about~8\%~FWHM) implied that the most-valuable diagnostics were temporal changes of spectral-line intensities. For example, the relative intensities of the $^{20}$Ne and $^{16}$O lines encode the particle spectrum, because the nuclear-excitation cross sections for these isotopes are very different, $^{20}$Ne activation having a much lower excitation threshold energy. Analyzing this line ratio, and comparing with electron Bremsstrahlung, it was found that the charged-particle energy is about equally distributed between electrons and nuclei, and that the ion energy spectrum is steep with typical power-law slopes between -3.5 and -5.5 \cite{2005AdSpR..35.1825M}. Other line ratios could be exploited to set constraints on the composition of the upper chromosphere, generally confirming that ion abundances are typical solar abundances except with a bias deriving from their first ionization potential, i.e. more-easily ionized species are enhanced.  
Important diagnostics could be expected from high-resolution spectroscopy, as Doppler shifts of gamma-ray lines or attenuation between source and observer modify the position and shape of the gamma-ray lines. Flare-accelerated particles propagate downward, spiraling around magnetic field lines. Depending on scatterings, and on convergence of field lines, particles may propagate deeper into the chromosphere, or get magnetically mirrored when pitch angles are large. Therefore the line shape details encode how accelerated particles lose their energy after the acceleration event and before thermalization, including possibly oscillations along magnetic loops. Therefore, it was clear that high-resolution spectroscopy as provided by cryo-cooled Ge detectors would be desirable for such study.

In February 2002, the \emph{Ramaty High-Energy Solar Spectrometer Instrument (RHESSI)} \cite{2002SoPh..210....3L} was launched to specifically study solar flares. Among its instruments, it featured a gamma-ray spectrometer based on 9 Ge detectors, similar to INTEGRAL's SPI instrument, but here located behind a rotation-modulation collimator for high-resolution imaging (at 2.3~arcsec resolution). RHESSI operations maintain pointing of the instrument towards the Sun, according to its main science goal. Details of the instrument are geared towards handling a large dynamic range of intensities, including remotely-operated attenuators which could be moved in front of detectors.

\begin{figure}
\centering
\includegraphics[width=0.65\textwidth]{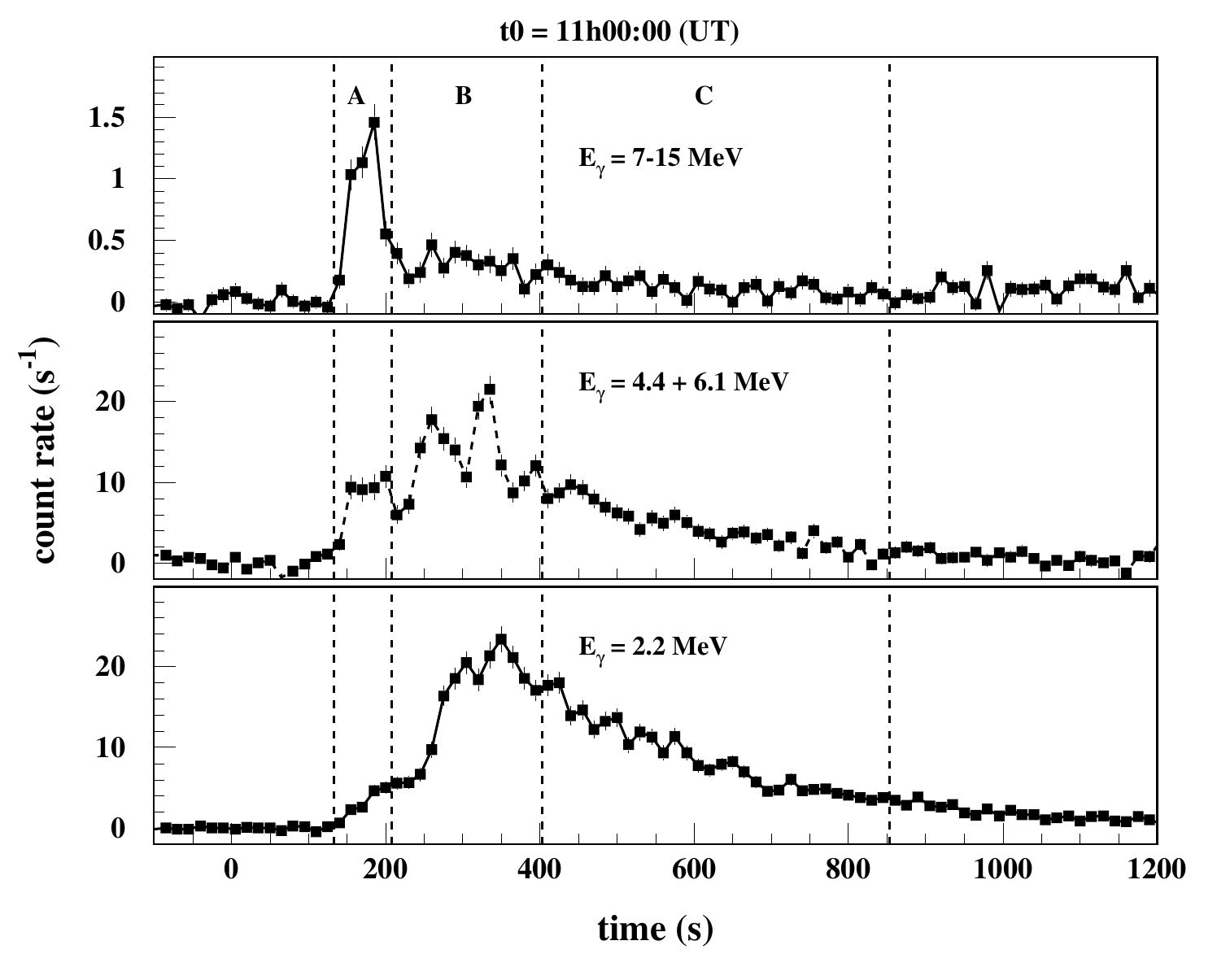}
\caption{The solar flare on 28 October 2003 showed an extended (15 min) period of gamma-ray emission. Different phases with their characteristic emission could be studied. \cite{2006A&A...445..725K}}
\label{fig:SolFlare_timeHist}
\end{figure}

\begin{figure}
\centering
\includegraphics[width=0.6\textwidth]{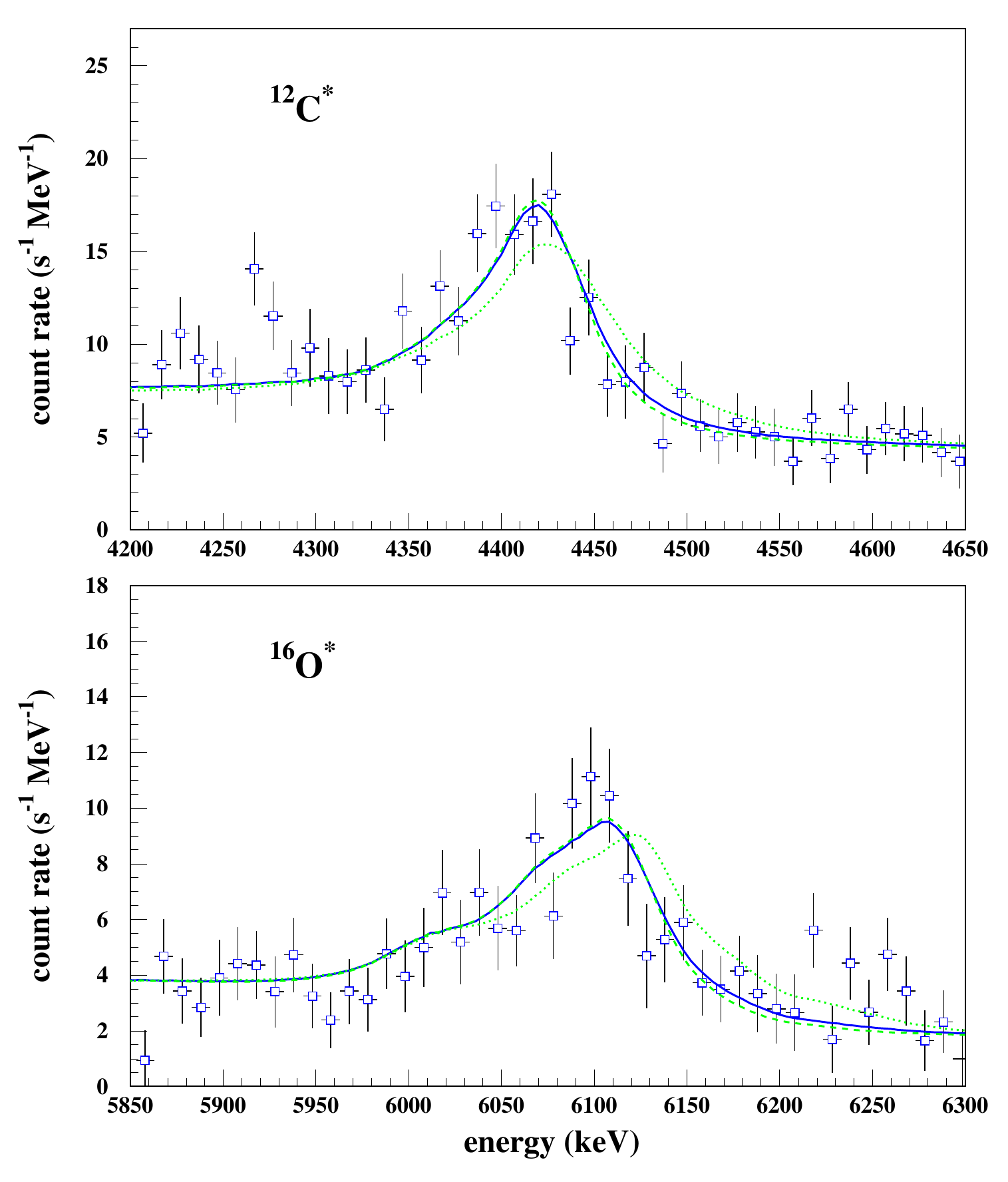}
\caption{The SPI measurements of the de-excitation emission from $^{12}$C \emph{(above)} and $^{16}$O
\emph{(below)} reveal low-energy tails. These arise from pitch-angle scattering of solar-flare particles, as they hit the solar atmosphere from above. See \cite{2006A&A...445..725K}  for details.}
\label{fig:SolFlare_C_O}
\end{figure}

INTEGRAL's spectrometer SPI provides 19 similar Ge detectors, optimized for a minimum of intervening material towards the sky as observed though the tungsten coded mask about 171~cm above the Ge array. Solar-flare observations thus typically occur with the instrument pointed away from the Sun, and gamma-rays entering Ge detectors through the side and penetrating the BGO scintillation detector which forms SPI's anti-coincidence system. 
Nevertheless, early in the INTEGRAL mission, several gamma-ray bright flares could be registered, as solar activity was still at a maximum. 

The flare event of 28 Oct 2003 produced gamma-ray emission over a period of \about~15 min \cite{2006A&A...445..725K} (see Fig.~\ref{fig:SolFlare_timeHist}). 
It started off with intense continuum emission, but within \about~a minute, line emission set in, with nuclear de-excitation lines preceding the neutron capture line at 2.223 MeV, as expected from the required slowing-down processes.

Nuclear lines were identified from neutron capture on hydrogen, and from excited $^{12}$C, $^{16}$O, $^{24}$Mg, $^{20}$Ne, and $^{28}$Si nuclei. The line intensity ratios evolve somewhat over the period of the flare. In particular the C and O line ratios are sensitive to the ratio of $\alpha$ particles over protons in the flare, and suggest that the $\alpha$/p ratio also evolves during the flare itself, reducing the $\alpha$-particle content in the later flare phase.

With SPI's high spectral resolution, line shape studies could be made, towards detections of kinematic Doppler shifts/distortions (see Fig.~\ref{fig:SolFlare_C_O}).  From the red wing which appears especially in the stronger signal from $^{12}$C, the beam geometry of accelerated particles as they hit the solar chromosphere could be constrained to be downward-directed (as expected), with a rather broad, fan-like widening, as it results from pitch-angle scattering in the solar atmosphere. 

With solar activity rising after 2010 towards its next maximum, new flare measurements are being analyzed. As RHESSI's energy resolution degraded over the mission years, SPI measurements will prove significant for the analysis of nuclear line ratios and line shapes.

\subsection{Nuclear Absorption of Background Gamma-Ray Sources}\label{diagnostics_abslines}
Absorption line spectroscopy has been successfully applied to a variety of astrophysical studies ever since Fraunhofer discovered characteristic spectral structure in the electromagnetic spectrum of the Sun and  identified it with absorption of the background thermal black body radiation of the Sun by atomic species in overlying gas. This astronomical principle has been employed in a variety of settings. For the study of sources at large, cosmological, distances, absorption against background emission from luminous quasars has become a key tool.

As with quasars, also with gamma-ray bursts, gas between the source and the observer will interact with the source emission, and imprint absorption effects.  While for the optical emission spectrum of quasars, atomic absorption due to electronic transitions is responsible for absorption features, for gamma-ray burst emission with typical energies of MeV,  \emph{nuclear} absorption and scattering properties of matter are capable of leaving \emph{nuclear} signatures in the observed gamma-ray burst spectrum. 
For the near-source part of the line of sight,  gamma-ray burst emission is so intense that gas should be instantaneously ionized by the burst, and hence cannot be studied through atomic spectroscopy.  This is also the case for the hot intergalactic medium which is believed to be at temperatures in excess of 10$^7$~K and fully ionized. 

 It is one of the most-interesting astrophysical questions, how in the early phases of the universe beyond redshifts \about~10 the probably very massive stars were formed before or in first galaxies. The composition of surrounding matter is believed to provide clues to the stellar population which created such early-generation stars. Gamma-ray bursts appear to be the best opportunity to measure such early star formation, from times where even UV emission from very massive stars would be redshifted beyond infrared windows of current and future telescopes. Nuclear absorption features are a candidate tool. Candidate nuclear absorption from early-universe cosmic matter could arise from nuclear excitation of abundant nuclei such as C and O (see above), but probably more significantly from He excitation and dissociation through the giant dipole resonance with a threshold energy of \about~25~MeV.   High redshifts of the most-interesting objects move such features into the \about~MeV regime. INTEGRAL could not find these signals, which should, however, be detectable with next-generation gamma-ray telescopes \cite{2011ExA...tmp..116G}. 
\section{Summary and Conclusions}
The INTEGRAL mission deepened the exploration of the nuclear-radiation sky since 2002, after the pioneering balloone-borne experiments (\about~1976) and the first sky survey with COMPTEL aboard the Compton Observatory (1991-2000).  Its imaging spectrometer instrument SPI has added high-resolution spectral information on the bright sources known from earlier observations. 

For core-collapse supernovae, details of $^{44}$Ti emission from the Cas A supernova remnant, and the absence of other $^{44}$Ti source detections, have supported the notion that  nucleosynthesis in such events occurs under a diversity of environmental conditions, which result from profound non-sphericity of the explosion. 
For Type-Ia supernovae, the exploitation of gamma-rays from $^{56}$Ni decay, as it powers supernova light, still awaits a sufficiently-nearby event; the independent information carried by gamma-rays from this decay chain could help to understand details of explosion physics, which happens in the dense and dynamic initial phase of the exploding star that is otherwise occulted to observations.

Diffuse gamma-ray emission from $^{26}$Al decay has been established as an astronomical window by itself from INTEGRAL measurements: Individual massive-star groups can now be distinguished, and the detailed multi-messenger comparison of their \Al emission with model predictions for each specific group of stars provides useful constraints for our models of massive-star interiors and their supernovae.
With the new detection of $^{60}$Fe decay from (probably) these same sources, an additional diagnostic of the interior structure of massive stars has become available, as the ratio of the gamma-ray lines from $^{60}$Fe and $^{26}$Al eliminates many uncertainties about the source numbers and distances. The discovery of $^{60}$Fe also in  ocean crust material underlines the role of such long-lived radioactive isotopes as tools to study  cosmic nucleosynthesis, in particular also addressing the aspect of how core-collapse supernovae spread their ejecta in their surroundings. \Al emission from nearby regions is bright enough so it can be located relative to its sources. 
\Al line astronomy in the inner Galaxy shows Doppler shifts from the large-scale motion of galactic rotation. This unexpected kinematic information about diluted hot interstellar gas is capable to provide new insights into the massive-star population in the inner parts of our Galaxy and how such short-lived stars impact on their surrounding interstellar medium through the feedback of mechanical energy.

Positron annihilation throughout the Galaxy has been mapped in detail with INTEGRAL for the first time. The puzzling brightness of the bulge region of our Galaxy may find explanations in the complex propagation of positrons in interstellar space. It may also shed light on past activity in the center of our Galaxy, or even interactions of dark matter, which could possibly have been directly detected through annihilation gamma-rays.
The origins of cosmic-rays are still unclear a hundred years after their discovery. How the acceleration of particles in cosmic sites is initiated, and how it proceeds towards relativistic energies,  remains to be one of the big astrophysical questions. Gamma-ray line observations can address the low-energy part of physical processes herein, as acceleration begins from a seed of thermal particles, and supra-thermal energies in the range of nuclear excitation energies provide an observational signature, from first steps of particle acceleration. The expected de-excitation lines from nuclei of interstellar gas have however not yet been seen. 
The detailed spectra of these same lines have been measured from solar flares, and confirm this astrophysical connection. 

INTEGRAL and other gamma-ray telescopes have only seen the brightest sources of nuclear lines. Nuclear fusion reactions have a key role in astrophysics, as they stabilize the existence of stars and power supernova light. There must be nuclear emission from many more sources, and from other isotopes. Examples are the nova explosions or nuclei on the surfaces of compact stars in binary systems. The unique and different processes which generate nuclear line features are only partly exploited in astrophysics. Ideas for next-generation telescopes have been presented, both to the Decadal Survey \emph{Astro2010} in the US, and to the \emph{Cosmic Vision} program of ESA. The evaluating committees have not given priority to these projects. Therefore, the INTEGRAL mission will form a legacy database for cosmic nuclear emission, which will provide the base to relate such emission to other observables. Continuing to operate this mission will not only exploit at best the investments, it also will ensure that scientists will be ready for measurements of a rare sufficiently nearby supernova or other potential surprises of the nuclear-radiation sky. 
Its   nuclear emission will be essential to understand physical processes in their interior. 
At the same time, deep observations in carefully-selected regions are capable to advance nuclear-line data beyond thresholds which  limit astrophysical interpretations, even with INTEGRAL.

\section*{References}
\bibliographystyle{jphysicsB} 

\end{document}